\documentclass[12pt]{article}
\usepackage{latexsym}

\newcommand{\xxx}[1]{ [#1]}
\newcommand{\mysection}[1]{\section{#1}
   \hspace{0.8cm}\setcounter{equation}{0}}
\renewcommand{\theequation}{\arabic{section}.\arabic{equation}}
\newcommand{\myappendix}{\appendix
   \renewcommand{\theequation}{\Alph{section}.\arabic{equation}}
   \vspace{30pt} \noindent {\Large \bf Appendices}}
\def\thefootnote{\fnsymbol{footnote}}
\def\notd{\not{\hspace{-.03in}\pp}}

\newcommand{\vev}{{\it vev}}
\def\tG{{\tilde G}}
\def\tg{{\tilde g}}

\def\hp{{\hat p}}

\def\tk{{\tilde k}}

\def\ts{{\tilde s}}
\def\tL{{\tilde L}}

\def\hE{{\hat E}}

\def\tK{{\tilde K}}

\def\bW{{\bar W}}

\def\hL{\hat{L}}

\def\hS{\hat{S}}

\def\hK{\hat{K}}

\def\hb{\hat{b}}

\def\dx{\delta_X}

\def\r{\right|}
\def\l{\left.}
\def\[{\left [}
\def\]{\right ]}
\def\({\left (}
\def\){\right )}
\def\lbr{\left\{}
\def\rbr{\right\}}
\def\pp{\partial}

\def\R{\bar{R}}

\def\ph{\bar{\phi}}
\def\Ph{\bar{\Phi}}

\def\bj{\bar{\jmath}}

\def\bTh{\bar{\Theta}}
\def\ux{$U(1)_X$}
\def\ua{$U(1)_a$}

\def\B{\bar{B}}

\def\t{\bar{t}}
\def\T{\bar{T}}
\def\S{\bar{S}}

\def\M{\bar{M}}

\def\Det{{\rm Det}}

\def\re{{\rm Re}}
\newcommand{\beq}{\begin{equation}}
\newcommand{\eeq}{\end{equation}}
\newcommand{\bea}{\begin{eqnarray}}
\newcommand{\eea}{\end{eqnarray}}
\def\bF{\bar{F}}
\def\G{{\cal G}}

\def\bA{\bar{A}}

\def\L{{\cal L}}
\newcommand{\superint}{\int \diff^{4}\theta \, }

\newcommand{\diff}{\mbox{d}}

\def\D{{\cal D}}
\def\bD{{\bar{\cal D}}}

\newcommand{\Da}{{\cal D}^{\alpha}}

\newcommand{\Dc}{{\cal D}_{\alpha}}

\newcommand{\Wa}{{\cal W}^{\alpha}}

\newcommand{\Wc}{{\cal W}_{\alpha}}

\def\CA{|C_A|^2}

\def\vx{V_X}

\def\hx{h_X}
\def\pp{\partial}
\def\rvev{\right\rangle}
\def\lvev{\left\langle}

\def\hel{\hat\ell}

\def\Del{\Delta}
\def\bDel{\bar{\Del}}
\def\bdel{\bar{\del}}
\def\del{\delta}

\def\u{\bar{u}}
\def\ep{\epsilon}

\newcommand{\bbar}[1]{{\overline{#1}}}

\newcommand{\myref}[1]{(\ref{#1})}
\newcommand{\beqa}{\begin{eqnarray}}
\newcommand{\eeqa}{\end{eqnarray}}
\newcommand{\nnn}{ \nonumber \\ }
\newcommand{\p}{\partial}
\newcommand{\W}{{\cal W}}
\newcommand{\Wb}{{\bbar{{\cal W}}}}
\newcommand{\chiproj}{(\bD^2 - 8R)}

\newcommand{\bchiproj}{(\D^2 - 8 \bar R)}

\newcommand{\Zbf}{{{\bf Z}}}

\newcommand{\uone}{$U(1)$}

\newcommand{\myvev}[1]{{\langle #1 \rangle}}
\newcommand{\bigvev}[1]{{\left\langle #1 \right\rangle}}

\newcommand{\Dk}{\Delta k}

\newcommand{\tr}{\mathop{{\hbox{Tr} \, }}\nolimits}

\newcommand{\half}{{1 \over 2}}

\newcommand{\Az}{{A_0}}
\newcommand{\DA}{\Del_A}

\newcommand{\ddd}{\nnn && \quad}
\newcommand{\mmm}{\nnn & &}

\def\hSig{\hat\Sigma}
\def\hTh{\widehat\Theta}
\def\bhTh{\widehat{\bTh}}
\def\hG{\hat G}

\newcommand{\F}{{\cal F}}
\def\eee{\nonumber \\ &=&}
\def\edk{e^{-\Dk/3}}
\def\lhel{\ell_{\hel}}
\def\lhela{\ell_{\hel A}}

\def\tc{{\tilde c}}

\newcommand{\mysec}[1]{Section \ref{#1}}
\newcommand{\myapp}[1]{Appendix \ref{#1}}

\begin{document}

\begin{titlepage}

\hfill   LBNL-54203

\hfill   UCB-PTH-03/29

\hfill   hep-th/0312125

\hfill (rev.) May 2004

\begin{center}

\vspace{18pt}
{\bf \Large Modular Invariant Gaugino Condensation
in the Presence of an Anomalous ${\bf U(1)}$}\footnote{This
work was supported in part by the
Director, Office of Science, Office of High Energy and Nuclear
Physics, Division of High Energy Physics of the U.S. Department of
Energy under Contract DE-AC03-76SF00098, in part by the National
Science Foundation under grants PHY-0098840 and INT-9910077,
and in part by National Science and Engineering Research Council of
Canada.}

\vspace{18pt}

Mary K. Gaillard,\footnote{E-Mail: {\tt MKGaillard@lbl.gov}}
\vskip .01in
{\em Department of Physics, University of California 
and \\ Theoretical Physics Group, Bldg. 50A5104,
Lawrence Berkeley National Laboratory \\ Berkeley,
CA 94720 USA}\vskip .03in
 Joel Giedt\footnote{E-Mail: {\tt giedt@physics.utoronto.ca}}
\vskip .01in Department of Physics, {\em University of Toronto, 60
Saint George Street,\\ Toronto, ON M5S 1A7, Canada}
\vskip .03in
{\em and}
\vskip .03in
Aleksey L. Mints\footnote{E-Mail: {\tt mints@socrates.berkeley.edu}}
\vskip .01in
{\em Department of Physics, University of California,Berkeley,
CA 94720 USA}

\vspace{18pt}

\end{center}

\begin{abstract}
Starting from the previously constructed effective supergravity theory
below the scale of $U(1)$ breaking in orbifold compactifications of
the weakly coupled heterotic string, we study the effective theory
below the scale of supersymmetry breaking by gaugino and matter
condensation in a hidden sector. Questions we address include vacuum
stability and the masses of the various moduli fields, including those
associated with flat directions at the $U(1)$ breaking scale, and of
their fermionic superpartners.  The issue of soft
supersymmetry-breaking masses in the observable sector presents a
particularly serious challenge for this class of models.
\end{abstract}

\end{titlepage}

\newpage
\renewcommand{\thepage}{\roman{page}}
\setcounter{page}{2}
\mbox{ }

\vskip 1in

\begin{center}
{\bf Disclaimer}
\end{center}

\vskip .2in

\begin{scriptsize}
\begin{quotation}
This document was prepared as an account of work sponsored by the United
States Government. Neither the United States Government nor any agency
thereof, nor The Regents of the University of California, nor any of their
employees, makes any warranty, express or implied, or assumes any legal
liability or responsibility for the accuracy, completeness, or usefulness
of any information, apparatus, product, or process disclosed, or represents
that its use would not infringe privately owned rights. Reference herein
to any specific commercial products process, or service by its trade name,
trademark, manufacturer, or otherwise, does not necessarily constitute or
imply its endorsement, recommendation, or favoring by the United States
Government or any agency thereof, or The Regents of the University of
California. The views and opinions of authors expressed herein do not
necessarily state or reflect those of the United States Government or any
agency thereof of The Regents of the University of California and shall
not be used for advertising or product endorsement purposes.
\end{quotation}
\end{scriptsize}

\vskip 2in

\begin{center}
\begin{small}
{\it Lawrence Berkeley Laboratory is an equal opportunity employer.}
\end{small}
\end{center}

\newpage
\renewcommand{\thepage}{\arabic{page}}
\setcounter{page}{1}
\def\thefootnote{\arabic{footnote}}
\setcounter{footnote}{0}

\mysection{Introduction} In previous articles \cite{GG02a,GG02b} the
effective supergravity theory obtained in the presence of an anomalous
\uone, which for the remainder of this article we will denote \ux, was
studied.  In the cases investigated, scalar fields charged under \ux\,
as well as other \uone's, acquired nonvanishing vacuum expectation
values ({\it vev}'s).  Associated chiral multiplets were ``eaten'' by
the \uone\ vector multiplets, including the \ux\ multiplet, to form
massive vector multiplets.  Tree-level exchange of these massive
vector multiplets were eliminated by redefinitions that removed linear
couplings between the heavy and light fields.  It was demonstrated
that these redefinitions can be made at the superfield level, while
maintaining manifest {\it modular invariance}\footnote{The modular
transformations on the fields of the effective theory are defined in
\myref{mdtr} below.}, local supersymmetry and the (modified) linearity
conditions,
\beq
\chiproj L = - \sum_a (\W \W)_a, \qquad
\bchiproj L = - \sum_a (\Wb \Wb)_a,
\label{abc}
\eeq
for the linear superfield $L$, whose
lowest component is the real scalar associated
with the dilaton.  A comparison with redefinitions
at the component field level provided assurances
that the superfield approach was reliable \cite{GG02a}.

Our motivation for studying these theories stems from the prevalence
of a \ux\ factor in the string scale gauge group of semi-realistic
string compactifications; for example, in a recent study \cite{Gie02a}
of a certain class of standard-like heterotic $Z_3$ orbifold models,
it was found that 168 of the 175 models in the class had an anomalous
\ux.  Thus the additional ingredient of a \ux\ factor is an important
modification of the string-inspired effective supergravity worked out
by Bin\'etruy, Gaillard and Wu (BGW) \cite{BGWa,bgw2}.  Indeed, we
expect low energy phenomenological aspects of these models---general
features of the superpartner spectrum \cite{gn}--\cite{BGWsp}, cosmology
of the models \cite{lyth,BGWcos}, implications for accelerator searches
\cite{BGWacc}---to be modified by the presence of a \ux\ at the high
scale.  In \cite{GG02b} complications were addressed that arise in
the semi-realistic models that we seek to understand, since the
scalars that get \vev's due to the \ux\ are typically charged under
several \uone\ factors and multiple scalars must generally get \vev's
in order for the {\it D-terms} of the several $U(1)$'s to
(approximately) vanish.  However in that article only the
supersymmetric phase was examined; the nonperturbative dynamics in a
hidden sector---which ultimately leads to supersymmetry breaking by
gaugino condensation---was not addressed.  The purpose of this paper
is to examine the effective supergravity theory when these important
effects are accounted for; {\it i.e.}, we intend to study the
effective theory below the scale of gaugino condensation.  As has been
noted previously, the supersymmetric vacuum is approximately the
stable vacuum in the case where dynamical supersymmetry breaking {\it
via} gaugino condensation occurs in an effective supergravity context
\cite{Bar98}.  Thus the tools developed already in \cite{GG02a,GG02b}
will prove useful here.

In Section 2 we review those aspects of References \cite{GG02a} and
\cite{GG02b} that are needed for the present discussion.  The string
scale theory is first defined; it is the same as in the BGW models
\cite{BGWa,bgw2} except that now a \ux\ is present.  As a consequence
large \vev's are induced and some fields get masses of order $10^{-1}$
to $10^{-2}$ $m_P$, where $m_P=1/\sqrt{8 \pi G} \approx 2.44 \times
10^{18}$ GeV is the reduced Planck mass.  (In the remainder of this
article we work in units where $m_P=1$.)  These large fields are
integrated out from the one-loop effective action by a sequence of
field redefinitions that are chosen so as to obtain an effective
theory below the \ua-breaking scale that is manifestly modular
invariant and locally supersymmetric and preserves the modified
linearity condition for the linear multiplet $L$.  We conclude Section
2 by summarizing these redefinitions which were worked out in
Refs.~\cite{GG02a,GG02b}.

In Section 3 we add effective terms to the Lagrangian that describe
the leading contributions from gaugino condensation in a hidden
sector.  Here again the description is basically that of the BGW
models, except that we must be careful about the \ux\ when anomaly
matching is considered.  This leads to important and interesting
constraints.  We construct the effective bosonic Lagrangian and
conclude Section 3 with discussions of the potentials for the T-moduli
and the dilaton and the masses of their superpartners.

In Section 4 we discuss scalar masses in the observable sector under
different assumptions for the K\"ahler potential for matter.  Our
results confirm those of earlier analyses~\cite{kk}--\cite{dp}, that
did not specify the mechanism for supersymmetry breaking, in that the
D-term contribution to these masses is generically dominant, resulting
in an unacceptably large scalar/gaugino mass hierarchy as well as the
possibility of large charge and color breaking \vev's.  In Section 5 we
discuss parameterizations of string nonperturbative effects and their
influence on the scales of coupling constant unification and
condensation and on the gravitino and scalar masses.  D-moduli masses
are addressed in Section 6, and in Section 7 we summarize our results
and discuss future lines of investigation.  Detailed calculations are
relegated to appendices.

Throughout this article we use the linear multiplet formulation
\cite{LMULT,BGG89} for the dilaton and the $U(1)_K$
superspace formalism \cite{Mul86,BGG89,BGG01} of supergravity, except
that, for reasons explained in \cite{GG02a}, we do not use $U(1)_K$
superspace for the Abelian gauge groups that are broken at the string
scale by the anomalous \ux.  (For a review of the $U(1)_K$ superspace
formalism see \cite{BGG01}; for a review of the linear multiplet
formulation see \cite{GG99}.)

\mysection{Review}
\label{rev}
In this section we review elements of Refs.~\cite{GG02a} and
\cite{GG02b}.  We unify our notation and present enough details to
render the present article reasonably self-contained.

\subsection{String scale effective theory}
We start with the effective theory at the string scale defined
as in \cite{GG02a}:
\beq
\L = \superint\tL + \L_Q + \L_{th},
\label{tlg}
\eeq
where $\tL$ is the real superfield functional
\beq
\tL = E\[- 3 + 2Ls(L) + L\(bG - \dx\vx\)\]
= E\[- 3 + 2LS\],
\label{oit}
\eeq
and the K\"ahler potential is given by
\beqa
K &=& k(L) + G + \sum_A K_{(A)}, \qquad
K_{(A)} = e^{G^A + 2 \sum_a q_A^a V^a}|\Phi^A|^2,
\nnn
G &=& \sum_I g^I, \qquad G^A = \sum_I q^A_I g^I,
\nnn
g^I &=& -\ln(T^I + \T^I), \qquad k(L) = \ln L + g(L).
\label{kpt}
\eeqa
In the dual chiral formulation $s(L) \to {\rm Re}(s);$ the \vev\,
$\lvev s(L)\rvev = g_s^{-2}$ determines the coupling at the string
scale.  Canonical normalization of the Einstein term requires:
\beq
k'(L) = - 2L s'(L).
\label{canon}
\eeq

Since the underlying
theory is anomaly free, it is known
that the apparent anomalies are canceled by
a four-dimensional version~\cite{UXR,gsterm} of the
Green-Schwarz (GS) mechanism \cite{GS84}.  This
leads to a Fayet-Illiopoulos (FI) term in
the effective supergravity Lagrangian.  Ignoring
nonperturbative corrections\footnote{The modification
in the presence of nonperturbative corrections
will be noted below.} to the dilaton K\"ahler potential, the
D-component of the \ux\, vector supermultiplet is given by
\beq
D_X = \sum_A {\p K \over \p \phi^A} q^X_A \phi^A + \xi,
\qquad \xi = {g_s^2 \tr T_X \over 192 \pi^2},
\label{eq1}
\eeq
where $K$ is the K\"ahler potential, $q_A^X$ is the
\ux\ charge of the scalar matter field $\phi^A$,
$\xi$ is the FI term, $T_X$ is the charge generator
of \ux, $g_s$ is the unified (string scale) gauge coupling.

Up to perturbative loop effects,
the chiral dilaton formulation has $g^2_s = 1/\re \myvev{s}$,
where $s= S|$ is the lowest component of the
chiral dilaton superfield $S$.  However, once
higher order and nonperturbative corrections
are taken into account the chiral dilaton
formulation becomes inconvenient.  The dual
linear multiplet formulation---which relates
a (modified) linear superfield $L$ to $\{ S,\bar S\}$
through a duality transformation---provides
a more convenient arrangement of superfield degrees
of freedom due to the neutrality of $L$
with respect to target-space duality transformations
(hereafter called {\it modular transformations}):
\beqa
T^I &\to& {a^I T^I - i b^I \over ic^I T^I + d^I}, \nnn
\Phi^A &\to& e^{-\sum_I q_I^A \F^I} \Phi^A, \nnn
a^I d^I - b^I c^I &=& 1, \qquad a^I,b^I,c^I,d^I \in \Zbf
\qquad \forall \quad I=1,2,3, \nnn
\F^I &=& \ln \( i c^I T^I + d^I \).
\label{mdtr}
\eeqa
The parameters $a^I$, {\it etc.}, may be taken as independent, or
subject to additional constraints, depending on the details of the
string construction.  In the limit of vanishing nonperturbative
corrections to the dilaton K\"ahler potential, $g^2_s = 2
\myvev{\ell}$, where $\ell = L|$.

In the linear multiplet formulation,
including nonperturbative corrections to the dilaton
K\"ahler potential, the FI term becomes
\beq
\xi(\ell) = {2 \ell \tr T_X \over 192 \pi^2}.
\label{eq2}
\eeq Consequently, the background dependence of the FI term in
\myref{eq2} arises from $\myvev{\ell}=\myvev{L|}$.  The FI term
induces nonvanishing {\it vev}'s for some scalars $\phi^A$ as the
scalar potential drives $\myvev{D_X} \to 0$, if supersymmetry is
unbroken.  In general a total number $m$ of \ua's are broken at
the same time. The nonvanishing \vev's in the supersymmetric vacuum phase
can be related to the FI term.  Then $\myvev{L|}$ serves as an order
parameter for the vacuum and all nontrivial \vev's can be written as
some function of $\myvev{L|}$. However the vacuum value
$\myvev{L|}$ is not determined at the \ua-breaking scale. The
conditions
\beq \lvev D_a\rvev = 0 \label{da=0}\eeq
require only that $m-1$ linear combinations of the modular invariant
functions $\myvev{K_A|}$ vanish and that one linear combination is
equal to (2.7), that is, proportional to $\ell$, which, like the
T-moduli, remains a dynamical field of the effective supergravity
theory below the \ua-breaking scale.  To account for this fact,
following \cite{GG02a,GG02b} we promote \myref{da=0} to a superfield
relation.  Thus we impose the superfield identity
\beq
\( {\pp K\over\pp V_a} + 2L{\pp S\over\pp V_a} \)_{\DA=0} =
\( {\pp K\over\pp V_a} \)_{\DA=0} - L\dx\del_{Xa}  = 0,\label{sfid}
\eeq
where $\DA$ are superfields, to be defined below,
that vanish in the supersymmetric vacuum.
This assures vanishing of
the D-terms at the \ua\, symmetry breaking
scale while maintaining manifest local supersymmetry below that scale.  The
latter point was demonstrated in detail, at both the superfield and the
component field levels, for the toy model studied in~\cite{GG02a}.

$\L_Q$ is the quantum correction~\cite{gt,kl,gnw}
that contains the field theory anomalies canceled by the GS terms:
\bea
\L_Q &=& - \int d^4\theta
{E\over 8R}\sum_a \W_a^\alpha P_\chi B_a
\W_\alpha^a + {\rm h.c.},
\label{lqa} \\
B_a(L,V_X,g^I) &=& \sum_I (b-b^I_a) g^I - \dx\vx + f_a(L),
\label{lqb}
\eea
where $P_\chi$ is the chiral projection operator~\cite{1001}:
$P_\chi \W^\alpha = \W^\alpha$, that reduces
in the flat space limit to $(16\Box)^{-1}\bD^2\D^2$, and
the $L$-dependent piece $f_a(L)$ is the ``2-loop'' contribution~\cite{gt}.
The string-loop contribution is~\cite{thresh}
\beq
\L_{th} = -\superint\,{E\over 8R}
\sum_{a,I} b_a^I (\W \W)_a
\ln\eta^2(T^I) + {\rm h.c.}
\label{lth}
\eeq
For each $\Phi^A$, the \ux\ charge is denoted
$q_A^X$ while $q^A_I$ are the modular weights.
The conventions chosen here imply \ux\ gauge
invariance under the transformation
\beq
\vx \to  \vx' = \vx + {1\over2}\(\Theta + \bTh\),
\quad \Phi^A \to \Phi'^A = e^{-q_A^X \Theta}\Phi^A.
\label{baa}
\eeq
The GS coefficients $b$ and $\dx$
must be chosen to cancel the quantum field
anomalies under modular and \ux\ transformations
that would be present in the absence of
the GS counterterms \cite{UXR,gsterm}.
It is not hard to check that the correct
choices are given by:
\beqa
\dx &=& -{1\over2\pi^2}\sum_AC^A_{a\ne X}q_A^X
= -{1\over48\pi^2} \tr T_X ,
\label{dxnorm} \\
8 \pi^2 b & = & 8 \pi^2 b_a^I+ C_a - \sum_A(1-2q^A_I)C_a^A
\qquad\forall\; I = 1,2,3\quad {\rm and} \quad\forall\; a.
\label{bdefs}
\eeqa

\subsection{Field redefinitions}\label{redef}
In this section we review the field redefinitions
of \cite{GG02a,GG02b}, phrased in the notation used
in the present article.  We state here the general case,
which was treated in Section 3.2 of \cite{GG02b}.

We introduce a vector superfield $V_a \; (a=1,\ldots,m)$ for each of
the \ua\, gauge groups that are broken by the presence of the FI term
$ - \dx L V_X$ in \myref{oit}.  One of these is assumed to be
anomalous; we denote it by \ux\ and the corresponding vector
superfield by $V_X$.  In addition, there are a number of chiral
superfields $\Phi^A$ that carry nontrivial charge under the \uone's.
Define the modular invariant \vev's\footnote{By the \vev\, of a
superfield we mean that the component fields in the expansion should
be evaluated at their \vev's.}  
\beq \myvev{ e^{G^A}|\Phi^A|^2 } =
\CA,
\label{defca}
\eeq
where $C_A$ is a complex constant.  For $C_A \ne 0 \;
(A=1,\ldots,n)$ we may define chiral superfields $\Theta^A$
through the identification
\beq
\Phi^A = C_Ae^{\Theta^A}.
\eeq
Then the (composite) superfield whose \vev\, appears in
\myref{defca} can be written
\beq
|\Phi^A|^2 e^{G^A} = \CA \exp \( G^A + \Theta^A + \bTh^A \) .
\eeq
This motivates the definition of the modular invariant real superfield
\beq
\Sigma^A = \Theta^A + \bTh^A + G^A ,
\label{jjrt}
\eeq
that satisfies $\myvev{ \Sigma^A } = 0.$
By contrast, we generically have $\myvev{G^A} \not= 0$ and
$\myvev{\Theta^A} \not= 0$.
The basis $(g^I,V^a,\Sigma^A)$ is equivalent
to the basis $(g^I,V^a,|\Phi^A|)$, but the
fields $\Sigma^A$ that have replaced $|\Phi^A|$
are modular invariant superfields with vanishing
\vev's; the usefulness of this feature is apparent
when we expand about a given vacuum.

The corresponding contribution
to the K\"ahler potential \myref{kpt} now takes the form
\beq
K_{(A)} = \CA \exp \( \Sigma^A + 2 \sum_a q_A^a V^a \).
\label{pies}
\eeq
Note that $\myvev{K_{(A)}} = |C_A|^2$ if we take $V^a$ in Wess-Zumino
gauge. However for the condition \myref{da=0} to hold in the effective
theory that is operative between the \ua-breaking scale and the
condensation scale where the \vev\, of the dilaton $\ell$ is
determined, \myref{pies} is not fixed at a constant value but rather as a
functional of $L$.  This is most easily achieved by absorbing a
dependence on $L$ in the vector fields $V_a$, as will be done below.

From \myref{pies} it is evident that each $V^a$ will generically
``eat'' some combination of the $\Sigma^A$ when we go to unitary
gauge, since linear couplings between $V^a$ and $\Sigma^A$ are
implied.  It is possible to identify a set of vector superfields that
do not couple linearly to the massless matter superfields.  To this
end we make the following field redefinitions:
\beq
V_a = U'_a - \Sigma_a, \qquad \Sigma_a = \sum_A T_{aA} \Sigma^A,
\qquad \Sigma'^A = \Sigma^A - 2 \sum_a q^a_A \Sigma_a,
\label{ppos}
\eeq
where $T_{aA}$ is a projection from the $n$-dimensional space of
chiral superfields with nonvanishing \vev's onto the $m$-dimensional
$U(1)^m$ space of linearly independent generators of the spontaneously
broken \ua's. It is defined by:
\beq
T_{aA} = {1\over2} B_A \sum_b M^{-1}_{ab} q^b_A, \qquad
M_{ab} = \sum_Aq^a_Aq^b_A B_A.
\label{ppot}
\eeq
Eq.~\myref{pies} then becomes:
\beq
K_{(A)} = |C_A|^2 \exp \[2 \sum_a q_A^a (U_a'-\Sigma_a) + \Sigma^A \]
= |C_A|^2 \exp \[2 \sum_a q_A^a U_a' + \Sigma'^A \] .
\label{piet}
\eeq
The linear dependence of $m$ of the uneaten matter fields is apparent
in the $\Sigma'^A$ basis, for it is easy to check that
\beq
\sum_Aq^a_AB_A\Sigma'^A = 0 \qquad \forall \; a .
\label{gb2}
\eeq
Thus, only $n-m$ of the $\Sigma'^A$ are linearly
independent.\footnote{For this reason the $V',\Sigma'$ basis was
referred to as ``quasi-unitary'' gauge in \cite{GG02b}; the
conventional unitary gauge will be recovered below when certain
conditions are imposed on the constant, real parameters $B_A$.}

While \myref{ppos} is not a gauge transformation, it can be
related to one.  To arrive at this result, for the
fields $\Sigma'^A$ that appear in \myref{ppos} we
make the identification
\beq
\Sigma'^A \equiv G'^A + \Theta'^A + \bTh'^A,
\eeq
where $\Theta'^A$ is a chiral superfield and
$G'^A$ is a function of the $g^I$.  From the
transformation \myref{ppos} we read off:
\beqa
\Theta'^A &=& \Theta^A - 2 \sum_a q^a_A \Theta_a, \qquad
G'^A = G^A - 2 \sum_a q^a_A G_a .
\label{jjui} \\
\Theta_a &=& \sum_A T_{aA} \Theta^A , \qquad
G_a= \sum_A T_{aA} G^A .
\qquad
\label{jjut}
\eeqa
This leads us to rewrite the vector superfield
shift that appears in \myref{ppos} in the following way:
\beq
V_a = (U'_a - G_a) - (\Theta_a + \bTh_a)
\equiv V'_a - (\Theta_a + \bTh_a) .
\label{hhrr}
\eeq
The shift $V_a \to V'_a$ {\it is} a gauge transformation,
provided we simultaneously shift {\it all} the gauge-charged
fields correspondingly:
\beq
\Phi'^A = \Phi^A \exp \(-2 \sum_a q^a_A \Theta_a \) \quad \forall\; A .
\label{hhre}
\eeq
Indeed with the identification
\beq
\Phi'^A = C_A e^{\Theta'^A}, \quad A=1,\ldots,n,\label{phi'0}
\eeq
the shift in \myref{jjui} is precisely the change of
variables \myref{hhre}, for the fields that get \vev's.

From \myref{hhrr} we have $U'_a =V'_a + G_a$.  Thus $U'_a$ ``eats''
the combination of K\"ahler moduli $G_a$; it is this shift $V'_a \to
U'_a$ that is not a gauge transformation.  With this redefinition we
obtain corrections to the effective action that do not cancel between
the GS term and the one-loop quantum correction; we include these
explicitly in our total effective Lagrangian.  (As noted in
\cite{GG02b}), once the nonperturbative dynamics of the hidden sector
stabilizes the $T^I$, modular invariance is broken and we are free to
instead take $U'_a =V'_a + \myvev{G_a}$, which is just a gauge
transformation if $\myvev{G_a}$ is a homogeneous
background field.)

To account for the required $L$-dependence of $K_{(A)}$, we shift to a
new (unprimed) vector superfield basis
\beq
U'_a = U_a + h_a(L) + \sum_B b_{aB}(L) \Sigma'^B ,
\label{ugtr}
\eeq
where $\myvev{U_a}=0$ by definition.
We determine the functions $h_a(L)$
from the requirement that the D-term $vev$'s vanish:
\beq \sum_A \lvev q_A^b e^{G'^A}|\Phi'^A|^2 \exp \( \sum_a 2q_A^a\[U_a
+ h_a(L) + \sum_B b_{aB}\Sigma'^B \]\) \rvev_{(L,T)} =
{\dx\over2}L\del_{bX},
\label{vevs}\eeq 
where the subscript $(L,T)$ indicates that the dilaton and moduli
superfields $L$, $T^I$ are left as quantum variables; that is, the
superfield ``\vev's'' $\myvev{|\Phi^A|^2}_{(L,T)} = \CA e^{-G^A}$ and
$\myvev{V_a}_{(L,T)}= h_a(L)$ are defined as functionals of the superfields
$(L,T)$.  Since by assumption $\myvev{U_a}_{(L,T)} =
\myvev{\Sigma'^A}_{(L,T)} = 0$, \myref{vevs} gives a set of equations
for the functionals $h_a$:
\beq \sum_A q_A^b \CA \exp \( 2 \sum_a q_A^a h_a(L) \) \equiv \sum_A
q_A^b x^A = {\dx\over2}L\del_{bX}.\label{funct}\eeq
Evaluated at the vacuum values $U = \Sigma = 0$, the shifts \myref{ugtr}
in the $U'_a$ modify the functions that appear in \myref{oit} and
\myref{kpt}:
\bea k(L) &\to& \tk(L) = k(L) + \delta k(L), \nnn 2Ls(L)
&\to& 2L\ts(L) = 2Ls(L) + 2L\delta s(L),  \eea 
with \cite{GG02b}, using \myref{funct},
\beq \delta k(L) = \sum_Ax^A, \quad {\pp\over\pp L}\delta k(L) =
2\sum_{A,b}q^b_A h'_b x^A = h'_X\dx L,
\label{delk}\eeq 
and\footnote{A factor $\dx/2$ is missing from the right hand side of 
both equations in (3.9) of~\cite{GG02b}.}
\beq 2L\delta s = - \dx L h_X, \quad
2L{\pp\over\pp L}\delta s = - \dx L h'_X = - {\pp\over\pp L}\delta
k,\label{dels}\eeq 
so the Einstein condition \myref{aab} is satisfied
for $U=\Sigma=0$. The K\"ahler potential for matter is 
\bea K(\Phi)
&=& \sum_A e^{G'^A + 2\sum_aq^a_A
\[U_a + h(L)_a + \sum_Bb_{aB}\Sigma'^B\]}|\Phi'^A|^2
\nonumber \\
&=& \delta k(L) + L\dx U_X
+ \sum_A\Sigma'^A\(x^A + b_{XA}\dx L\) \nnn
&& + 2\sum_{A,a}\Sigma'^AU_a
\(q^a_Ax^A + 2\sum_{B,b}b_{bA}x^Bq^b_Bq^a_B\)
+ O(U^2,\Sigma'^2,|\Phi'^{A>n}|^2).
\eea
The linear coupling of $\Sigma'^A$ to
the vector multiplets $U_a$ is given by:
\beq
K \ni 2\sum_a U_a \[ \sum_{A}\Sigma'^A
\(q^a_Ax^A + 2\sum_{B,b}b_{bA}x^Bq^b_Bq^a_B\) \].
\label{hhyt}
\eeq
We can exploit \myref{gb2} to eliminate $U,\Sigma'$ mixing.
We choose the constants $b_{aA}$ in \myref{ugtr} such 
that\footnote{The functional $f_a(L)$ introduced here is
not to be confused with the one in \myref{lqb}.}
\beq
f_a(L)q^a_AB_A = q^a_Ax^A + 2\sum_{B,b}b_{bA}x^Bq^b_Bq^a_B .
\label{vltm}
\eeq
Then the term in brackets in \myref{hhyt} vanishes identically for
each $a=1,\ldots,m$.  Since the \ua\, are assumed to be
broken, the vectors $q^a=(q^a_1,\cdots, q^a_n)$ are linearly
independent, and the matrix
\beq
N_{a b} = \sum_Bx^Bq^a_Bq^b_B \label{nab}
\eeq
has an inverse. This allows us to uniquely determine the
required constraints:
\beq
b_{aA}(L) = {1\over2}\sum_bq^b_AN^{-1}_{ab}\[f_b(L)B_A - x^A(L)\].
\eeq
From \myref{gb2} we have the sum rule
\beq
\sum_Ab_{aA}\Sigma'^A
=  - {1\over2}\sum_{b,A} N^{-1}_{ab}x^Aq^b_A\Sigma'^A \label{sr}.
\eeq
It is convenient to identify the part of $b_{aA}$ that
actually contributes to the right-hand side of \myref{sr}:
\beq
\hb_{aA} = - \half \sum_b N^{-1}_{ab} x^A q^b_A, \quad
\sum_A q^c_A\hb_{aA} = - \half \del^c_a.
\label{baA}\eeq
Then the kinetic terms \myref{pies} can finally be written as
\beq K_{(A)} = \CA\exp\(\Sigma'^A + \sum_a 2q_A^a\[U_a + h_a(L) +
\sum_B\hat b_{aB}(L)\Sigma'^B\]\).\label{final}\eeq
The set of gauge transformations and field redefinitions leading to
\myref{final} defines a gauge that is closely related to the ``true''
unitary gauge, as can be seen~\cite{GG02b} by writing \myref{final}
in the form
\bea K_{(A)} &=& x^A(L)\exp \(\hSig^A + \sum_a 2q_A^a U_a\), \nnn
 \hSig^A &=& \Sigma'^A + 2\sum_{aB}q_A^a\hat b_{aB}\Sigma'^B =
 \Sigma^A - \sum_{abB}q_A^a x^Bq^b_BN^{-1}_{ab}\Sigma^B, \\ 0 &=&
 \sum_A q^a_Ax^A(L)\hSig^A(L). \label{sigcond}\eea
Here the $m$ massive vectors $U_a$ and the $n-m$ linearly independent
uneaten supermultiplets $\hSig^A$ appear as the physical states.  This
reduces to the conventional unitary gauge when $L$ is replaced by the
\vev\, $\ell_0$ of the dilaton $\ell = L|$, which is determined only
at the condensation scale.

However, there are two simple examples where
the constraint \myref{vltm} leads directly to
``true'' unitary gauge.  First suppose that
\beq
q^a_A = q^a_\Az, \quad \forall \; A=1,\ldots,n;
\eeq
that is, all fields acquiring \vev's have the
same charges.  Then the condition \myref{funct}
reads
\beq q^{a\ne X}_\Az = 0, \quad\sum_A \CA = {\dx\over2q^X_\Az}L\exp \(-
2 q_\Az^X h_X(L) \) = C, \quad x^A =
L\CA{\dx\over2q^X_\Az C}, \eeq
where $C$ is a constant, so that $\hSig$ is independent of $L$:
$\hSig(L) = \hSig(\ell_0)$.

Next suppose that $m=n$; {\it i.e.,} the number of fields
acquiring \vev's corresponds to the number of spontaneously
broken \uone\ generators.  Then there are
no uneaten fields among the $\Sigma^A$.  The charge matrix
$q^a_A$ is invertible, and we can uniquely define
solutions to the equations
\beq \sum_a q_A^a Q_a^B = (q Q)^B_A = \delta_A^B,\qquad \sum_A q_A^a Q_b^A =
(Q q)^a_b = \delta^a_b .\label{inft}\eeq
In this case we can invert \myref{funct} to obtain
\beq
x^A(L) = \half \dx L Q_X^A, \quad \hat b_{aA} = -\half Q_a^A,
\quad\hSig^A = 0\label{xmin} .
\eeq
In general $m\le n$; if $m<n$ it is not possible
to invert the conditions \myref{funct}, and $\hb_{aA}$
depends on $L$ if the \ua\, charges are not degenerate.

Note that the constant parameters $B^A$ do not appear in
\myref{final}.  As noted in \cite{GG02b}, these parameters are
unphysical and must simply be chosen such that the constraint
\myref{vltm} is satisfied.  The constants $C_A$ are also not physical
since they can be shifted by gauge transformations with constant
parameters~\cite{GG02b} that leave the physically relevant variables
$x^A$ unchanged; in particular one could choose $B_A = \CA$. On the
other hand, with the choice\footnote{The first equality in (3.32) of
\cite{GG02b} should read $\lvev b_{aA}\rvev = 0$.}~\cite{GG02b}
\beq b_{aA}(\ell_0) = 0, \quad f_a(\ell_0) = f, \quad
B_A = x^A(\ell_0)/f,\eeq
the gauge defined by \myref{final} with $L\to\ell_0$ is the same as
the one defined by \myref{ppos} up to the shift $h(\ell_0) = U' -
U(\ell_0)$. For the special case of degenerate \ua\, charges, this is
satisfied by $B_A = \CA,\; f_a(L) = L\dx/2a_{A_0}C$, while for the
case of minimal \vev's this requires $B_A = cQ^A_X,\;f_a(L) =
L\dx/2c,\; c=$ constant, suggesting $\CA = cQ^A_X$ as a convenient
choice.

The above field redefinitions also modify the kinetic terms for all
\ua-charged chiral fields $\Phi^B$.  We have 
\bea V_a &=& U_a + h_a(L) + \hSig_a, \quad \hSig_a =
\sum_A\hb_{aA}\Sigma^A = \hTh_a + \bhTh_a + \hG_a,\nnn \hTh_a &=&
\sum_A\hb_{aA}\Theta^A, \quad \hG_a =
\sum_A\hb_{aA}G^A.\label{newu}\eea
The kinetic term for $\Phi^B$ takes the form\footnote{The
presence of terms linear in $U_a$ in the exponent in \myref{shifts}
induces terms of $O(|\Phi^B\Phi^C|^2)$ in the effective low energy
K\"ahler potential defined by integrating out the massive vector
fields.} 
\beq K_{(B)} = e^{G^B + 2\sum_aq^a_B\(h_a(L) + \hSig_a\)}|\Phi^B|^2 +
 O(U_a). \label{kb}\eeq
First consider the case where $\hb_{aA}$ is independent of
$L$.  Then the term proportional to $\hTh_a + \bhTh_a$ in the shift
\myref{newu} in the vector field is just a gauge transformation
and the corresponding term in the exponent in \myref{kb}
is absorbed in the redefinition of $\Phi^B$ under the gauge 
transformation:
\beq \Phi'^B = \Phi^B\exp\(2 \sum_a q^a_B\hTh_a\).\label{phi'}\eeq
Since the exponent in \myref{phi'} is not modular invariant, the
modular weights of $\Phi'$ are modified with respect to
those of $\Phi$, as reflected by the $T$-dependence of the 
exponent in \myref{kb}:
\bea K_{(B)} &=& e^{G'^B+ 2\sum_aq^a_B h_a(L)}|\Phi'^B|^2 + O(U_a),
\quad G'^B = \sum_I q'^B_I g^I, \nnn q'^B_I &=& q^B_I +
2\sum_{aA}q^a_B\hb_{aA}q^A_I = q^B_I + \del q^B_I.
\label{shifts}
\eea
In the more general case where $\hb_{aA}$ depends on $L$ and
unitary gauge is defined {\it a postiori} after the \vev\,
$\myvev{L|} = \ell_0$ is fixed we have
\beq  \hb_{aA}(L) = \hb_{aA}(\ell_0) + O(\Del L), \quad
\Del L = L - \ell_0.\eeq
Then the gauge transformation \myref{phi'} and the redefinition
\myref{shifts} of modular weights are defined by the replacement
$\hb_{aA}\to \hb_{aA}(\ell_0)$. The additional terms of order $\Del L$
generate higher dimension couplings of the $\Phi'^B$ to the dilaton
and to the eaten superfields $\hSig_a$, whose components in unitary
gauge are identified with the longitudinally polarized components of
the massive vector supermultiplet $U_a$.  When the $U_a$ are
integrated out they generate operators of dimension eight ({\it e.g.,}
$|\Phi|^6\hel^2$) and higher in the low energy theory, and we may
neglect them. As noted above, the $m$ ``Goldstone modes''
$\hat\Sigma_a$ disappear from the Lagrangian due to overall gauge
invariance.  The $n-m$ uneaten physical states\footnote{There is a
term in $K$ that is linear in $\hat\Sigma$ whose only effect is to
slightly modify the K\"ahler metric for the $T$-moduli: see
(2.44)-(2.47) and (3.33) of~\cite{GG02b}. $D^A$ is a singlet of the
surviving gauge group, and terms in the K\"ahler potential that are
linear in an uncharged chiral superfield do not contribute to the
Lagrangian. Shifting this term from $K$ to $S$ {\it via} the Weyl
transformation has no effect because linear terms in the effective
K\"ahler potential $K - 2LS$ are Weyl invariant.} $\hat\Sigma^A\,;
<\hat\Sigma> = 0,$ introduced in \myref{sigcond} may be
expressed~\cite{GG02b} in terms of chiral and anti-chiral fields
$D^A,\bar D^{\bA}$:
\bea \hSig^A &=& \hTh^A + \widehat{\bTh}^A + \hG^A = D^A+ \bar D^{\bA}
+ O\(\[\hat T^I + \hat{\T}^I\]^2/\myvev{t^I + \t^I}^2\),\nnn D^A
&=& \widehat{\Theta}^A + \lvev\hG^A\rvev + \lvev{\pp\hG^A\over\pp
t^I}\rvev\hat T^I, \qquad \lvev{D^A}\rvev = 0,\nnn
0 &=& \sum_A q^a_Ax^A(\ell_0)\hSig^A = \sum_A q^a_Ax^A(\ell_0)\hTh^A =
\sum_A q^a_Ax^A(\ell_0)\hG^A.\label{ddef}\eea
where $T^I = \myvev{t^I} + \hat T^I$.

\subsection{Weyl transformation}
Of chief concern in \cite{GG02a,GG02b} was the
maintenance of the canonical normalization for the Einstein
term---concurrent to field redefinitions.
Here we recall the general prescription given in those papers
for determining the necessary Einstein condition
from $\L$ rewritten in a new field basis.
The relevant part of the Lagrangian is \myref{oit}.  We define $M$ to
stand collectively for the fields that are to be
regarded as independent of $L$ in a given basis.  We
then define the functional $S$ by the identification
\beq
\tL \equiv E[-3 + 2L S(L,M)].
\label{jut}
\eeq
The Einstein condition holds provided
\beq
\( {\p K \over \p L}\)_M + 2 L \( {\p S \over \p L}\)_M = 0.
\label{aab}
\eeq
Here the subscripts on parentheses
instruct us to hold constant
under differentiation the fields denoted
collectively by $M$.

As explained in \cite{GG02b}, since the redefinition \myref{ugtr}
involves $L$, some care is required if we are to retain a canonical
Einstein term.  In \myref{dels} we saw that the condition \myref{aab}
is automatically satisfied to zeroth order in the fields $\Del =
U,\Sigma$ with vanishing \vev's after the field redefinitions of the
previous subsection, provided \myref{canon} is satisfied.  To ensure
\myref{aab} holds at nontrivial orders in $\Del$ requires a Weyl
transformation that redefines the linear multiplet $L\to \hL(L,\Del)$
such that the linearity condition \myref{abc} holds for $\hL$ in the
new Weyl basis; this transformation eliminates the lowest dimension
terms linear in $\Del$, so that tree exchange of these fields may
be neglected when they are integrated out.  These results apply to the
supersymmetric phase.  When we introduce supersymmetry breaking
through gaugino condensation, we do not expect \myref{da=0} to remain
strictly true.  The required Weyl transformation to assure \myref{aab}
in the case of small nonvanishing D-terms is worked out in \myapp{weyl}.

\mysection{Gaugino condensation} After we make the gauge
transformations and field redefinitions of the previous section, as
summarized in \myref{newu}, as well as the requisite Weyl
transformation, the density $\tL$, Eq. \myref{oit}, is modified to
read\footnote{In this section we study the vacuum in the parameter
space of the dilaton, $T$-moduli and static condensates, and set
to zero all fields with vanishing \vev's.  Therefore the term linear
in $\hSig^A$ noted in the previous footnote is irrelevant here.
However when we shift the \vev's $\myvev{\Phi^A}$ by small amounts
$\Del^A$ to allow $D_a\sim |u|^2\ne0$, there are terms linear in
$\Del^A$; these are treated exactly in \mysec{cond} and \myapp{weyl}.}

\bea \tL &=& E \[- 3 + 2L\ts(L) + L \( bG - \dx\hG_X - \dx U_X \) \] 
= \hE\[- 3 + 2\hL\ts(\hL) + \tL_{GS}\] + O(U^2), \nnn
\tL_{GS} &=& \hL\(bG - \dx\hG_X \), \quad \ts(L) = s(L) - {\dx\over2}
h_X(L),  \quad \label{lgs} \eea
where here $U$ refers collectively to all the heavy modes with
vanishing \vev's that we have integrated out, and $\hG_X$ is defined in
\myref{newu}.  The quantum Lagrangian \myref{lqa} is also modified.
For the term quadratic in the field strength of the unbroken gauge
group factor $\G_u$, it is of the form \myref{lqa} with
\beq B_u = \sum_I(b - b^I_u)g^I - \dx\hG_X - \dx \hx(\hL) + f_u(\hL)
+ O(U).\label{lq2}\eeq
The shift $\hx(L)$ restores the gauge field kinetic energy term to its
original form; {\it i.e.} just multiplied by the original function
$s(L)$.  The anomalous modular transformation of the $\dx\hG_X$ term in
$B_u$ is canceled by the corresponding shift in $\tL_{GS}$:
\beq bG - \dx\hG_X = \sum_Ig^I\(b + \del b_I\)\equiv\sum_Ig^I b_I 
.\label{delgs}\eeq
The shift in the moduli dependence of $B_u$ corresponds to the shifted
modular weights given in \myref{shifts} of the \ua-charged fields in
the loops that contribute to the $\beta$-function for $\G_u$.  That
is, referring to \myref{dxnorm} and \myref{bdefs}, the shift in $B_u$ 
is given by
\bea 4\pi^2\del B_u &=& \sum_BC_u^B\sum_I\del q^I_Bg^I = 
2\sum_{B,a}C_u^Bq^a_B\sum_{A,I}\hb_{aA} q^A_Ig^I = 2
\sum_{B,a,A}C_u^Bq^a_B\hb_{aA}G^A \eee  2
\sum_{B,a}C_u^B q^a_B\hG_a = - 4\pi^2\dx\hG_X = 4\pi^2\sum_I\del b_I g^I,
\label{delb}\eea
where $C_u^B$ is the quadratic Casimir for the representation $B$
of ${{\cal G}}_u$, and we used the fact that
\beq \sum_B C_u^B q^{a\ne X}_B = 0, \eeq 
since by assumption only \ux\, is anomalous.  The terms in \myref{lqa}
that are quadratic in the field strengths of the broken \ua's generate
terms in the low energy theory that are of very high dimension in
derivative of fields and in auxiliary fields, as discussed in
\cite{GG02a}, and we neglect them.  We next construct the effective
Lagrangian for condensation at the scale where one gauge group $\G_c$
becomes strongly coupled, extending the approach of \cite{bgw2}. A
number of new features arise, which we describe below. 

\subsection{Construction of the effective theory at the condensation 
scale}\label{cond}

Here we follow the construction of \cite{bgw2}.  As shown there the
physics of condensation is dominated by the group $\G_c$ with the
largest $\beta$-function coefficient $b_c$ unless there are two groups
with nearly equal $\beta$-functions.  Therefore, for simplicity, we
consider here the case with just one condensing simple group $\G_c$ in
the hidden sector.  Here we set to zero fields with vanishing \vev's,
in which case the new Weyl basis discussed in the previous section is
equivalent to the original one.  The term in \myref{lqa} that is
quadratic in the strongly coupled gauge field strength $\Wa_c$ is
replaced by an effective VYT~\cite{vyt} action,
generalized~\cite{tom,bg89} to the case of local supersymmetry, that
is manifestly invariant under the nonanomalous symmetries of the
underlying quantum field theory:
\bea \L_{VYT} = {1\over8}\superint\,{E\over
R}U_c\[b'_c\ln(e^{-K/2}U_c) + \sum_\alpha b^\alpha_c\ln\Pi^\alpha\] +
{\rm h.c.},\label{vyt} \eea
where $U_c$ and $\Pi^\alpha$ are nonpropagating\footnote{The dynamical
condensate case was studied in ref.~\cite{yy} with just an $E_8$ gauge
condensate.  After correctly integrating out the heavy bound state
degrees of freedom, that have masses larger than the condensation
scale $\Lambda_c$, one recovers the theory with a static $E_8$
condensate~\cite{BGWa}.  We expect this result to be
generic~\cite{BGWsp}.} gauge and matter condensate chiral superfields,
with K\"ahler chiral weights 2 and 0, respectively:
\beq U_c\simeq {\Wa_c\Wc^c},\quad \Pi^\alpha\simeq
\prod_B\(\Phi_c^B\)^{n^B_\alpha}.\label{condef} \eeq 
The chiral superfields $\Phi_c^B$ that condense are charged under
the strongly coupled gauge group $\G_c$.  The effective theory is modular invariant; the modular anomaly
matching condition between the effective Lagrangian \myref{vyt} and
the underlying quantum Lagrangian \myref{lqa} reads~\cite{bgw2}
\beq b'_c + \sum_{\alpha,B} b^\alpha_cn^B_\alpha q_I^B =
{1\over8\pi^2}\[C_c - \sum_B C^B_c\(1 - 2q^B_I\)\]\;\;\;\; \forall
\;\;I.\label{mod} \eeq
The strongly coupled Yang-Mills sector also possesses a residual
global \ua\, invariance that is broken only by superpotential
couplings, such as \myref{pipot} below, that involve those chiral
superfields that get \vev's at the \ua-breaking scale.  They enter the
$\G_c$ gauge coupling RGE only through chiral field wave function
renormalization, which is a two-loop effect that is encoded in the
expression \myref{condeoms} for the gaugino condensate through the
appearance\footnote{See (2.29)--(2.33) of~\cite{bgw2} and (14)--(16)
of~\cite{gn} and the related discussions.} of the superpotential
coefficients $W_\alpha$.  We therefore impose the \ua\, anomaly
matching conditions
\beq \sum_{\alpha,B} b^\alpha_c n^B_\alpha q_B^a = \del_{aX}
\sum_B{C^B_c\over4\pi^2}q_B^X,\label{ua}\eeq
where again the sum over $B$ includes only the chiral superfields
$\Phi^B_c$ that are charged under $\G_c$.  Finally, assigning canonical
dimensions $({3\over2},1)$ to $\Wa_c,\Phi^B_c$, the standard trace
anomaly requires~\cite{bgw2}
\beq 3b'_c + \sum_{\alpha,B}b^\alpha_cn^B_\alpha = 3b'_c +
\sum_{\alpha}b^\alpha_c d_\alpha = {1\over8\pi^2}\(3C_c -
\sum_BC^B_c\) + O(\Lambda_c/m_P), \label{trace}\eeq
where $d_\alpha$ is the dimension of $\Pi_\alpha,$ and $\Lambda_c$
is the condensation scale. The matter
condensates are invariant under all the unbroken, nonanomalous
symmetries, and therefore the same monomials can appear in the
superpotential\footnote{The anomaly matching conditions are satisfied
if one generalizes the $\Pi^\alpha$ in \myref{condef} to a linear
combination of monomials: $\Pi^\alpha = \sum_i c^i_\alpha\Pi^\alpha_i$
with the same dimension, \ux\, charge and modular weights, for fixed
$\alpha$, and superpotential $W(\Pi) = \sum_{\alpha,i}
W^i_\alpha\Pi^\alpha_i$.  The only change is that the second equation
in \myref{condeoms} is replaced by $\pi^\alpha = - {b^\alpha_c
c^i_\alpha/4W^i_\alpha}u,$ $\forall\;\alpha,i$, requiring
$c^i_\alpha/W^i_\alpha$ independent of $i$ for fixed $\alpha$ if
condensation is to occur.}
\beq W(\Pi) = \sum_\alpha W_\alpha\Pi^\alpha,\label{pipot}\eeq
where $W_\alpha$ is a function of the $\G_c$-neutral unconfined chiral
multiplets.  Solving for the condensates gives~\cite{bgw2}
\bea |u|^2 &=& \l\bar U_cU_c\r = e^{-2b'_c/b_c}e^Ke^{-2s(\ell)/b_c -
b\sum_Ig^I/b_c} \prod_I|\eta(t^I)|^{4(b -b_c)/b_c}
\prod_\alpha|b^\alpha_c/4W_\alpha|^{-2b_c^\alpha/b_c}, \nonumber \\
\pi^\alpha &=& \l\Pi^\alpha\r = - {e^{-K/2}b^\alpha_c\over4W_\alpha}u,
\qquad b_c \equiv b'_c + \sum_\alpha b^\alpha_c.\label{condeoms} \eea
The expression for $|u|^2\simeq\Lambda_c^6$ is consistent~\cite{bgw2,gn}
with instanton calculations in supersymmetric Yang-Mills theories
provided $b_c$ is the coefficient of the $\beta$-function:
\bea b_c &=& -{2\over3g_c^3(\mu)}\mu{\pp g_c(\mu)\over\pp\mu} =
{1\over8\pi^2}\(C_c - {1\over3}\sum_BC^B_c\) \eee b'_c + \sum_\alpha
b^\alpha_c = {1\over8\pi^2}\(C_c - {1\over3}\sum_BC^B_c\) +
\sum_\alpha b^\alpha_c\(1 - {d_\alpha\over3}\) +
O(\mu/m_P),\label{beta}\eea
which is satisfied if only matter condensates of dimension three have
$b_c^\alpha\ne0$. This is consistent with the fact that operators of
dimension $3+ \del_\alpha$ are suppressed in the superpotential
\myref{pipot} by a factor\footnote{If \ua-breaking generates masses
$M$ such that $\Lambda^2_c\ll M^2< m^2_P$ the lowest of these would be
expected to replace $m_P$ everywhere.} $W_\alpha \sim
m_P^{-\del_\alpha}$: the second equation in \myref{condeoms} shows
that condensation can occur only if $b_c^\alpha\to0$ when
$W_\alpha\to0$.  Operators $\Pi^\alpha_2$ of dimension two may be
generated by the \vev's of fields $\Phi^A$ that break the \ua's
through superpotential couplings of the form
\beq W_M = c_R(T^I)\left<\prod_A\Phi^A\right>\Phi^R_c\Phi^{\R}_c=
M_R\Phi^R_c\Phi^{\R}_c\simeq M_R\Pi^R_2.\label{masspot}\eeq
where $R,\R$ denote a representation of $\G_c$ and its conjugate.
In this case, for finite $b^R_c,u$, \myref{condeoms} requires
\beq \pi_2^R = - e^{-K/2}b_c^R u/M_R \to 0, \quad M_R
\gg \Lambda_c = |u\u|^{1\over6},\eeq
in conformity with conventional wisdom. 

The most straightforward solution to \myref{trace} is\footnote{This
expression for $b'_c$ also produces the correct anomaly under the
phase transformation $\theta\to e^{i\alpha}\theta$ on the fermionic
superspace coordinates.  Simultaneous solutions to \myref{mod} and
\myref{ua} may require constraints on the $\G_c$-charged matter;
they are trivially solved if each $\Pi^\alpha$ is composed of $\Phi^B$'s
with the same Casimir $C_B$.  This is the case for $d=2$ and also for the
$d=3$ condensates in the FIQS $SO(10)$ model considered below, as well
as the toy $E_6$ and $SU(3)$ models considered in~\cite{bgw2}.}
\beq b'_c = {1\over8\pi^2}\(C_c - \sum_B C^B_c\), \qquad \sum_{\alpha
\in \Pi_d}b_c^\alpha = \sum_{B\in
\Pi_d}{C^B_c\over4d\pi^2}.\label{anomsol} \eeq
It is interesting to note that in this case, if $b^R_c\ne0$, the
contribution of $\Phi^{R,\R}$ to $b'_c$ exactly cancels the
contribution of $b_c^R$ in the expression \myref{beta} for $b_c$, in
conformity with the decoupling theorem: heavy states with
$M_R>\Lambda_c$ do not contribute to the running of the gauge
couplings.  However they also do not contribute to the anomaly
coefficients \myref{mod}, \myref{ua} and \myref{trace} in the
effective theory below $\Lambda_c$.  Therefore we should set
$b_c^R=0$. The contribution of $\Phi^R_c,\Phi^{\R}_c$ to the anomaly
at the string scale is still present in the form of a field-dependent
ultra-violet cut-off~\cite{gt} $M_{s R}$ in the standard loop integral,
but the infra-red cut-off should be the mass $M_R$, giving a net
contribution to the effective action
\bea \L_R &=& - {C^R_c\over16\pi^2}\superint {E\over R}U_c\ln(M_{s
R}/M_R) + {\rm h.c.},\nnn M_R &=& M_{s R} W_{R\R} = c_R(T)\prod_A C_A
e^{\hat\Theta A}M_{s R},\label{rmass}\eea
where $M_R$ is modular covariant (but not in general \ua\, covariant
when $C_A$ is held fixed) and $M_{s R}$ contains the contribution to the
anomalies; up to a constant factor of order one in Planck units
\beq
M^2_{s R} = e^{K - G^R - G^{\R} - 2\sum_a(q^a_R + q^a_{\R})V_a} \to 
e^{\hK - \hG^R - \hG^{\R} - 2\sum_a(q^a_R + q^a_{\R})h_a(L)},\eeq 
where the second expression is obtained after integrating out the
massive vectors with the field redefinitions of \mysec{rev}.  The
$\hat\Theta^A$ are the ``D-moduli'' that remain massless at the string
scale in the case when there are more scalar \vev's than broken \ua's.

In the next section we construct the effective potential under the
assumption that there are no mass terms of the form \myref{masspot} in
the strongly coupled sector and take $b^\alpha\ne0$ only if $d_\alpha
= 3$.  Note that the exponents $n^B_\alpha$ need not be integers,
because the effective superpotential \myref{pipot} contains terms
that, like the superpotential \myref{vyt} for $U_c$, are generated by
nonperturbative effects.  For example we will sometimes use as a
concrete example a model with a strongly coupled $SO(10)$ and matter
in three fundamental spinorial representations $\xi^B$; the lowest
dimension $SO(10)$-invariant matter-composite operators $O_\alpha$
have four factors of $\xi^B$.  No masses are generated for these
fields by \ua-breaking, and we require $\Pi^\alpha\sim
O_\alpha^{3\over4}$.

In \myapp{rm} we give the corrections to this construction when terms
like \myref{masspot} are present.  The sum rules \myref{qconds} and
\myref{pcond} given below are modified by the removal of the
$\Phi^R_c,\Phi^{\R}_c$ contributions.  This is compensated for by new
contributions from \myref{rmass} in such a way that, aside from the
usual renormalization group factor $\Lambda_c^2 \sim e^{-2/3b_c
g^2_s}$ that depends on the $\beta$-function factor for the massless
spectrum of the strongly coupled sector below the \ua-breaking scale,
the effective potential is determined by parameters defined in terms
of the modular weights and gauge charges of the full spectrum of the
effective theory at the string scale.

The superpotential \myref{pipot} is made modular invariant by
incorporating an appropriate $T^I$ dependence in $W_\alpha$ as
in~\cite{bgw2}. It is not \ua\, invariant; we follow the standard
approach for an effective theory with a broken symmetry: since the
symmetry breaking arises only from \vev's of $\G_c$ singlets, we first
construct a \ua-invariant superpotential by including appropriate
powers of these fields, and then replace them by their (T-moduli and
dilaton dependent) \vev's.  Then when we solve for the condensate
$vev$'s to get \myref{condeoms}, the effects of \ua-breaking appear in
$|u|^2$ at two-loop level through the superpotential couplings, as
they should.  The \vev's $\left<\Phi^A\right> = V^A_0(L,T^I)$ obtained
in \mysec{rev} assured vanishing D-terms at the \ua-breaking scale.
Once supersymmetry is broken at the condensation scale, one also
expects D-terms to be generated.  Therefore we replace those \vev's
by $V^A = V^A_0(L,T^I) + \del V^A$, with $\del V^A$, like
$U_c,\Pi^\alpha$ taken to be nonpropagating superfields to be
determined by solving the overall equations of motion.

\subsection{Solving for the vacuum at the supersymmetry-breaking scale}

Once supersymmetry is broken we cannot demand {\it a priori} that
$\lvev D_a\rvev=0$.  We may also generate F-terms associated with the
\ua-charged chiral fields that get \vev's. To include these, we
slightly modify the field redefinitions \myref{phi'0} and \myref{ugtr}
as follows:
\beq \Phi'^A = C_Ae^{\Theta'^A + \Delta^A}, \quad U'_a = U_a + h_a(L)
+ \sum_Bb_{aB}\Sigma'^B + \Del_a, \eeq
where
$\Delta_a$ and $\Del^A$ are vector and chiral superfields,
respectively with only constant scalar components:
\bea \delta_a &=& \l\Delta_a\r, \quad F_a = -{1\over4}\l D^2\Del_a\r,
\quad D_a = {1\over8}\l\Da\(\bD^2 - 8R\)\Dc\Del_a\r, \nnn \del^A &=&
\l\Delta^A\r, \quad F^A = -{1\over4}\l D^2\Del^A\r, \eea
and we take
\beq \lvev U_a\rvev = \lvev \Sigma'^A\rvev = 0 \eeq 
as before.  It is clear that we can make \ua\, gauge transformations
with constant chiral superfields $\Lambda_a$ to eliminate the scalar
and F components from $\Del_a$:
\beq \del_a = F_a = 0, \eeq
which just redefines $\del^A,F^A$.  The field redefinitions in
\mysec{rev} were chosen to eliminate terms linear in the heavy modes
$U$, neglecting terms in $\L_Q$ that are proportional to the squared
YM field strengths. When $\G_c$-charged fields condense, the term
quadratic in $\Wa_c$ becomes a contribution to the potential
proportional to $|u|^2$ with couplings to the $U$'s, including, in
particular, a coupling linear in the \ux\, gauge potential $U_X$.
Thus if $S_0(\tilde U)$, with $\tilde U = U + \Del$, represents the
effective action at the \ua-breaking scale, we require $\del S_0/\del
U|_{U=\Del = 0} = 0$.  At the condensate scale the effective action
becomes $S_c(\tilde U) = S_0(\tilde U) + \Del S(\tilde U)$, where
$\Del S$ is the condensate contribution.  The condition that there be
no terms linear in the heavy fields $U$ now reads $\del S_c/\del
U|_{U=0, \tilde U =\Del} = 0;$ it is automatically satisfied if we
minimize the effective theory with respect to $\Del$ keeping $U=0$.
However the shifts needed to go to unitary gauge are slightly
modified.  While we impose the constraints \myref{funct} on the
zeroth order vacuum values, here denoted by $k^A$:
\beq k^A = \CA e^{2\sum_a q^a_A h_a},\quad 
\sum_A q^a_A k^A = {\del_{aX}\dx\over2}L, \label{condk}\eeq
we use the true vacuum values $x^A$:
\beq x^A = k^A e^{\Delta^A + \bDel^A + 2\sum_a q^a_A\Delta_a} = k^A\[1
+ O(\Del)\],\eeq
in the conditions \myref{vltm} for going to unitary gauge.  We have
\bea\sum_A q^a_A x^A &=& {\del_{aX}\dx L\over2} + \sum_A q^a_A
 k^A\(\Del^A + \bDel^A\) + 2\sum_b N^{ab}\Del_b + O(\Del^2),
 \label{rels}\eea
using the notation introduced in \myref{nab}.  Now we have
a theory defined by
\bea K &=& k + \sum_A x^A + G, \nnn \tL &=& \hE\[2L s - \dx L\(\hG_X +
h_X + \Del_X\) + b L G - 3\] \equiv E\[2L S(L,\Del) + L\sum_I b_I g^I
- 3\],\nnn K' &=& k' + 2\sum_{A,a}h'_a q^a_A x^A = - 2L S' +
\dx\Del_X/2 + 2\sum_{a,A}q^a_A h'_a k^A\(\Del^A + \bDel^A\) +
O(\Del^2),\eea
where we used~\cite{GG02b} the relations,\footnote{A factor $x^Aq^b_A$
is missing from the next to last expression in Eq.(3.29)
of~\cite{GG02b}.} with $N_{ab}$ defined in \myref{nab},
\beq 2\sum_ah'_aN_{a b} = \del_{bX}{\dx\over2}, \quad 2h'_a = N^{-1}_{aX}
{\dx\over2}, \quad {\dx\over2}\sum_aN^{-1}_{aX}x^Aq^a_A =
2x^A\sum_aq^a_Ah'_a,\label{lderiv} \eeq
that follow from the $L$-derivative of (\ref{condk}), and prime
denotes differentiation with respect to $L$. Thus we have to perform a
further Weyl transformation to eliminate noncanonical Einstein terms
of order $\Del$.  When we eliminate the auxiliary fields in the
standard way we will obtain expressions for $D_a(\hel,x^A) \sim \Del$,
and the full scalar potential found below takes the form
\beq V = {s\over2}\sum_a D_a^2(\hel,x^A) +
{|u(\hel,t^I,x^A)|^2\over16}\[w(x^A) + v(\hel) + O(\del)\].\eeq
Since, as we shall see below, $\pp u/\pp\del^A \sim u$, $\pp
D_a/\pp\del^A,\pp w/\pp\del^A\sim 1$, the minimization equations for
$\del^A$ imply $D_a \sim \Del\sim |u|^2\ll 1$ in reduced
Planck mass units, and we need only keep terms up to order $\Del^2$ in
the Weyl transformation.  The details of this transformation are given
in \myapp{weyl}, following the procedure described in Appendix B of
\cite{GG02b}.  Working in the new Weyl basis, we obtain for the
nonderivative part of the bosonic Lagrangian
\bea e^{-1}\L_B &=& \sum_I{1 + b_I\hel\over(t^I + \t^I)^2}\bF^IF^I -
{1\over16\hel}
\[\pp_{\hel}\hK\u u - 4e^{\hK/2}\pp_{\hel}\hK\hel\(W\u + u\bW\)\] + 
\sum_{AB}\tK_{A\B}\bF^{\B}F^A \nonumber \\ & & +
{1\over9}\(\hel\pp_{\hel}\hK - 3\)\[\M M - {3\over4}\lbr\M\(b'_cu -
4We^{\hK/2}\) + {\rm h.c.}\rbr\] - {b'_c\over8}\pp_{\hel}\hK\u u
\nonumber \\ & & + \lbr{u\over4}\[\sum_\alpha
b_c^\alpha{F^\alpha\over\pi^\alpha} + \sum_I\[{b'_c- b_I\over2\re t^I}
- 2b_c^I\zeta(t^I)\]F^I - b'_c\sum_A\hK_AF^A\] + {\rm
h.c.}\rbr \nonumber \\ & & + {1\over8}L_c\(F_c - u\M + {\rm h.c.}\) +
{1\over2}\hS_A\(\u\bF^{\bar A} + u F^A\) + \sum_a\({s\over2}D^2_a +
\tK_aD^a\) \nonumber \\ & & + e^{\hK/2}\[\sum_IF^I\(W_I + \hK_IW\) +
\sum_\alpha F^\alpha W_\alpha + \sum_AF^A\(W_A + \hK_A W\)
 + {\rm
h.c.}\], \nnn L_c &=& 2\hS + b'_c\ln(e^{2-\hK}\u u) + \sum_\alpha
b_c^\alpha\ln(\pi^\alpha
 \bar{\pi}^\alpha) + \sum_I\[b_Ig^I -
2b_c^I\ln|\eta(t^I)|^2\],\label{bose}\eea
where hatted variables refer to the new Weyl basis in which the
Einstein term is canonically normalized, $\zeta(t) =
\pp\ln\eta(t)/\pp t$ and
\bea \hK(\hel,t,\del^A) &=& \hK|,\quad \hS(\hel,\del^A) = \hS|, \quad
\pp_{\hel}\hK = \pp \hK/\pp\hel, \quad \hS^A = \pp \hS/\pp\del^A,\quad
\tK = \hK + 2L\hS, \nonumber \\ \tK_a &=&
{1\over2}\l{\pp\tK\over\pp\Del_a}\r, \quad \tK_A =
{\pp\tK\over\pp\del^A},\quad \tK_{A\B} = {\pp\tK\over\pp
\del^A\bar{\del}^{\B}}, \quad W_A = {\pp W\over\pp\del^A},\quad etc.
\label{derivs}\eea
%
The equations of motion for the auxiliary fields give
\bea F^\alpha:&& \pi^\alpha e^{\hK/2}W_\alpha + {u\over4}b_c^\alpha =
0 =e^{\hK/2}W + {u\over4}(b_c-b'_c),\nonumber \\ F^c:&& \u u =
e^{-2b'_c/b_c}e^{\hK -2\hS/b_c}
\prod_\alpha|b^\alpha_c/4W_\alpha|^{-2b^\alpha_c/b_c} \prod_I(2\re
t^I)^{b_I/b_c}[|\eta(t^I)|^4]^{b_c^I/ b_c}, \nonumber \\ M:&& M =
{3\over4}ub_c = 3m_{\tG}, \nonumber \\ F^I:&& F^I = - {2\re t^I\over1
+ b_I\hel}{\u\over4}\[b_c - b_I + 2\re t^I\( \sum_\alpha
b_c^\alpha{\pp\over\pp\t^I}\ln\bW_\alpha - 2\bar{\zeta}(\t^I)
b_c^I\)\], \nonumber \\ F^A:&& F^A = -
{\u\over4}\sum_B\tK^{A\B}\[2\hS_{\B} - \sum_\alpha b_c^\alpha
{\pp\over\pp\bdel^{\B}}\ln\bW_\alpha - \hK_{\B}b_c\], \nonumber \\
D^a:&& D_a = - {1\over s}\tK_a,\label{aux}\eea
where $m_{\tG}$ is the gravitino mass.  Invariance under modular and
\ua\, transformations requires that the part of $W_\alpha$ that is
nonvanishing in the vacuum takes the form\footnote{There may be
additional factors of modular invariant holomorphic functions $f(T^I)$
that we are ignoring here.}
\beq W_\alpha = c_\alpha\prod_I[\eta(t^I)]^{2(q_I^\alpha +
p_I^\alpha - 1)} \prod_A(\phi^A)^{q_\alpha^A}, \quad \sum_Aq^a_A
q_\alpha^A = - q_\alpha^a, \quad p_I^\alpha = \sum_Aq_I^A
q_\alpha^A. \label{pdefs} \eeq
Note that once the $U(1)$'s are broken, the Lagrangian is still gauge
invariant (before one picks a gauge), but the symmetry is nonlinearly
realized.  In unitary-gauge the anomaly matching conditions read:
\bea \sum_\alpha b_c^\alpha q_I^\alpha &=& \sum_Bb_c^\alpha n^B_\alpha
q_I^B = \sum_B {C_c^B\over4\pi^2}q_I^B = b_I - b'_c -
b^I_c , \quad \sum_\alpha b_c^\alpha = b_c - b'_c,
\nonumber \\ \sum_\alpha b_c^\alpha q^a_\alpha &=& \sum_B b_c^\alpha
n^B_\alpha q^a_B = \sum_B {C_c^B\over4\pi^2}q^a_B =
-{1\over2}\dx\delta_{aX}, \quad b_I = b + \del b_I.
\label{qconds}\eea
Writing
\beq |\phi^A| = \sqrt{x^A}e^{-G^A/2 - \sum_a q^a_A
h_a},\label{physhift}\eeq
we have
\beq \prod_A|\phi^A|^{- 2\sum_\alpha b^\alpha_c q_\alpha^A} =
\exp\(-\sum_A p_A\ln x^A + \sum_I p_I g^I - \dx
h_X\),\label{phiprod}\eeq
where we define   
\beq p_I = \sum_\alpha b_c^\alpha p_I^\alpha, \quad p_A = \sum_\alpha
b_c^\alpha q^A_\alpha,\label{pdefs2}\eeq
and we used the second anomaly matching condition in \myref{qconds} 
which implies
\beq 2\sum_Aq_A^a p_A = 2\sum_{\alpha,A}b^\alpha_c q_\alpha^A q^a_A =
- 2\sum_\alpha b^\alpha_c q_\alpha^a = \dx\del_{aX}.\label{pcond} \eeq
Then we obtain
\bea \u u &=& e^{-2b'_c/b_c}e^{\kappa -2(\hS-\del s)/b_c}
\prod_\alpha|b^\alpha/4c_\alpha|^{-2b^\alpha/b_c} \prod_I\[2\re
t^I|\eta(t^I)|^4\]^{(b_I - b_c + p_I)/b_c}e^{-\sum_Ap_A\ln x^A /b_c},
\nonumber \\ F^I &=& - {2\re t^I\over1 + b_I\hel}{\u\over4}\(b_c - b_I
- p_I\)\[1 + 4\re t^I\bar{\zeta(t^I)}\], \nonumber \\ F^A &=& -
{\u\over4}\sum_B\tK^{A\B}\[2\hS_{\B} - p_B - \hK_{\B}b_c\],
\qquad  \kappa = \hK - G.\label{sols}\eea
Note that the factor $\myvev{e^{-2(\hS-\del s)/b_c}} =
\myvev{e^{-2s(\hel)/b_c}} + O(\del) = e^{-2/g_s^2b_c} + O(\del)$ has
the standard dependence on the $\beta$-function at the condensate
scale $\Lambda_c^6 = \myvev{u\u}$.  The auxiliary fields $F^A$ are
evaluated in \myapp{weyl}; the full potential takes the form
\bea V &=& {1\over2s}\sum_a\tK_a^2 + {|u|^2\over16} v(\hel,\del) +
\sum_I{1 + b_I\hel\over(t^I + \t^I)^2}\bF^IF^I,\nnn
v(\hel,\del) &=& {\pp_{\hel}\hK\over16\hel}(1 + b_c\hel)^2 -
3b_c^2.\label{pot1}\eea
In the remainder of this section we consider aspects of this potential
as well as the modular fermion masses.

\subsection{The moduli potential}

The potential is modular invariant, with a similar $t$-dependence as
in~\cite{bgw2}, so the moduli are still stabilized at self-dual points
$t_1 = 1, \; t_2 = e^{i\pi/6}$, with $\lvev F^I\rvev=0$. The T-moduli
masses are determined by the coefficients of $F^I$ in
\myref{sols}. Setting\footnote{The value of $m_t$ quoted
in~\cite{bgw3} is the average mass for $t = 1$ in the approximation
$\eta(1)\approx e^{-\pi/12}$.}
\beq t^I = {1\over\sqrt{2}}\(\tau^I + i a^I\),\eeq
we obtain
\bea m^I_{\tau,a} &=& {|(b_I + p_I -
b_c)u|\mu^I_{\tau,a}(t^I)|\over4(1 + b_I\ell)} = {|(b_I + p_I -
b_c)|\mu^I_{\tau,a}(t^I)|\over b_c(1 + b_I\ell)}m_{\tG},\nnn
\mu^I_{\tau}(t^I) &=& - 8\re t^I\[\zeta(t^I) + \re t^I\zeta'(t^I)\],
\quad \mu^I_a(t^I) = - 8\(\re t^I\)^2\zeta'(t^I), \nnn
\mu^I_{\tau}(t_1) &\approx& 2.56,\quad \mu^I_{a}(t_1) \approx
.56,\quad\mu^I_{\tau}(t_2) \approx 3.02,\quad \mu^I_{a}(t_2)
\approx 1.02.
\label{tmass}\eea
where $\tG$ is the gravitino, and here we neglect $O(\del)$
corrections. In the absence of an anomalous U(1), $p_I = 0,\;b_I=b$.
This was the case studied in~\cite{bgw2} using $b = b_{E_8}\approx
10b_c$, yielding a welcome mass hierarchy, $m_\tau/m_{\tG}\approx 30$,
$m_a/m_{\tG}\approx$ 6--10, thereby evading cosmological difficulties
that arise if there are many moduli degenerate with the gravitino.
However a large ratio $b/b_c$ is not generic to models with gauge
symmetry breaking by Wilson lines.  For example in the FIQS model,
\cite{fiqs} considered below, $b = b_c$, which would give massless
T-moduli without the additional contributions in \myref{tmass}.  Using
\myref{lderiv}, the contribution,
\beq \del b_I = -\dx\hG_X = -\dx\sum_A\hb_{XA}q^A_I = 
{\dx\over2}\sum_{A,a}N^{-1}_{aX}q^a_A x^A q^A_I, \eeq
reflects the shifts (\ref{shifts}) in the modular weights, as in
(\ref{delb}). We also have, using \myref{pcond} 
\beq p_I = \sum_{\alpha,A}b^\alpha_c
q^A_Iq_A^\alpha = \sum_A q^A_Ip^A = {\dx\over2}\sum_A q^A_Ip^A/\sum_B
q^B_Xp^B,\label{picond} \eeq
From the definition \myref{nab}, one generally expects $b_I/p_I>0$,
and we can get a hierarchy between the moduli masses and the gravitino
mass without requiring $b\gg b_c$ if $p_I\sim b_I\gg b_c$.  For
example, in minimal models with $n=m$ discussed at the end of
\mysec{redef}
\beq \del b_I = p_I, \quad b_I + p_I = b + \dx q^X_I + p_I = b +
2p_I.\label{minbi}\eeq
Specifically in the FIQS model introduced in the next subsection, with
$b=b_c$, $p_I\approx 2b$, the T-moduli masses \myref{tmass} are given
by
\beq {m^I_\tau\over m_{\tG}} \approx\cases{10\cr12\cr},\quad
{m^I_a\over m_{\tG}} \approx\cases{2\cr4\cr},\quad
{\rm for}\quad \myvev{t^I} = \cases{t_1\cr t_2\cr},\eeq
if $b_I\ell\ll 1$. 

\subsection{The dilaton potential}\label{dilpot}
From the results of \myapp{weyl}, the potential for $\ell$ at the
condensation scale is, evaluated at the moduli self-dual points,
\bea V &=& {1\over2s}\sum_a\tK_a^2 + {|u|^2\over16}v(\ell) +
O(\del|u|^2)\qquad v(\ell) = w + {k'\over\ell}(1 + b_c\ell)^2 - 3b_c^2
,\qquad \del\sim|u|^2, \nonumber \\ w &=& \sum_Aw(k^A),
\quad
 w(k^A) = b^2_ck^A + 2p_Ab_c + {p^2_A\over k^A} = {(p_A +
b_ck^A)^2\over k^A}.\label{pot}\eea
For this potential the positivity condition on $v$ found in
\cite{BGWa} is sufficient, but not necessary, since the other terms in
$V$ are positive-semi-definite. In fact we have to let $v$ go negative
to cancel the cosmological constant if $F,D\ne0$.  Since $D\ll F$, we
require $w+v\approx0$. Since $w(k^A)$ is minimized at $k^A = p_A/b_c$,
 \beq w \ge 4b_c p, \quad p =\sum_Ap_A = 
\sum_{A,\alpha}b_c^\alpha q_\alpha^A =
{1\over12\pi^2}\sum_{A,\alpha}C^A_c q_\alpha^A,\label{sump}\eeq
for the solution \myref{anomsol} of the anomaly constraints,
giving the constraint
\beq k'(1 + b_c\ell)\le \ell(3b_c^2 - 4b_cp), \label{kcond}\eeq 
at the vacuum. The right hand side of \myref{kcond} need not be
positive. A negative vacuum value for $k'$ would be unviable if $k$
were actually the dilaton K\"ahler potential, because the dilaton
metric would be proportional to $k'$, requiring $\lvev k'\rvev>0$.
However since the actual K\"ahler potential is $\tk$, we have the
weaker condition
\beq \tk'> 0, \quad k' > - \del k'= - \sum_A k'^A =  - \dx\ell h'_X 
.\label{kcond2}\eeq   
Together (\ref{kcond}) and (\ref{kcond2}) require
\beq \dx\ell h'_X + \(3b_c^2 - 4b_cp\)/\(1 + b_c\ell\) > 0.\eeq
If we neglect nonperturbative contributions, $k' = \ell^{-1}$, and
(\ref{kcond}) requires $\ell b_c> 1$, which is strong coupling since
$\ell = g_s^2/2$ in this case, so we still need to include
nonperturbative effects if we wish to stabilize the dilaton at weak
coupling.  However, unlike the models studied in
~\cite{BGWa,bgw2,casas}, these contributions are not needed to
stabilize the potential at very large $\ell$, because the D-term grows
with $\ell$ unless some $k^A\sim\ell$ also get large; but in this case
some $F$-terms become large, so $V \sim \ell$ in the large coupling
limit.  This allows more freedom in the parameterization of the
nonperturbative effects.  The vacuum value $\lvev \tk'\rvev$ is an
important parameter for phenomenology. A small value increases the
ratio $m_\ell/m_{\tG}$, where $m_\ell$ is the universal dilaton mass,
and suppresses the universal axion coupling; both are welcome features
for a viable modular cosmology.  However small $\lvev \tk'\rvev$ also
suppresses gaugino masses, which at tree level are the same as those
found in~\cite{bgw3} with the substitution $k'\to\tk'$:
$m_{\tg}/m_{\tG}\propto \lvev\tk'\rvev$.  This can become problematic
since the squark and slepton masses are generally of order $m_{\tG}$
or larger.

If the only flat direction at the string scale corresponds to a minimal
set $n=m$ of $n$ chiral multiplet \vev's that break $m$ \ua's, the
conditions \myref{condk} and \myref{pcond} have a unique solution:
\bea k^A &=& \ell p^A = {\ell\dx\over2}Q^A_X, \quad \del k' = p =
\sum_Ap_A, \nnn w &=& \ell^{-1}p(1 + b_c\ell)^2 = \ell^{-1}\del k'(1 +
b_c\ell)^2.\label{minsol}\eea
Then the vacuum condition \myref{kcond} determines the vacuum value
of the dilaton metric as
\beq \ell^{-1}\tk' = {3b^2_c\over(1+b_c\ell)^2},\label{vac0}\eeq
which is precisely the value found in \cite{bgw2}, resulting in
suppressed gaugino masses and axion coupling, and an enhanced dilaton
mass.  The latter two features are welcome, but there would be more
phenomenologically viable parameter space available if the gauginos
masses were, say, just a factor two larger. (Increasing the \vev\, of
$\tk'/\ell$ affects the dilaton and axion parameters only in
proportion to its square root).  However this would require adding a
{\it negative} contribution to the potential.  It is much easier to
find additional {\it positive} contributions.  This is simply a
consequence of the positivity of the vacuum energy in global
supersymmetry: negative contributions are necessarily connected to
higher dimension operators of local supersymmetry.  For example an
extra superpotential $W^\phi = f(T)\prod_A\Phi^A$ gives a negative
contribution: $V\ni - 3e^K|W^\phi|^2$, but since we need $\lvev
W^\phi\rvev\le 10^{-15}$, at least one $\Phi$ must have a very small
\vev, so there will be at least one larger F-term: $V\ni F^AF_A\gg
e^K|W^\phi|^2.$  As we note in the concluding section, loop
corrections from nonrenormalizable couplings with large coefficients
are not expected to significantly change this analysis.

If the vacuum is degenerate at the string scale
($n>m$), we may write
\beq k^A = \ell p_A + y^A, \quad \sum_A q^a_A y^A =
0,\label{ycond}\eeq
where the last equality assures that if \myref{pcond} is satisfied, so
is \myref{condk}.  Then at the condensation scale, the (approximate)
vacuum values will be those that minimize $w(k^A)$ with respect to the
$y^A$ subject to the condition in \myref{ycond}.  If $y^A=0\; \forall\;
A$, the dilaton potential is identical to the minimal case,
\myref{minsol}.

The conditions \myref{ycond} are most easily implemented by separating
out a minimal subset of the $\Phi^A$ with nonvanishing $p_A$:
\beq k^A,\;\; A = 1,\ldots n,\;\;\to\;\; (k^A,k^M), \quad
A = 1,\ldots, m,\quad M = 1,\ldots, n-m.\label{kakm}\eeq
Then defining $Q^A_a$ as in \myref{inft}, the constraint in 
\myref{ycond} reads
\beq y^A = - \sum_M\zeta^A_M y^M, \quad \zeta^A_M = \sum_a Q^A_a
q^M_a,\eeq
and we may take the $y^M$ as independent variables.  We have
(neglecting order $\del$ terms)
\beq {\pp V\over\pp y^M} = {\pp w\over\pp k^M} - \sum_A\zeta^A_M
{\pp w\over\pp k^A}= - (p_M/k^M)^2 + b_c^2 + \sum_A\zeta^A_M
\[(p_A/k^A)^2 - b_c^2\].\label{miny}\eeq
If $p_M = 0$, we require $k^M= y^m\ge0$, and
\beq \l{\pp V\over\pp y^M}\r_{y=0} = b_c^2 +
\sum_A\zeta^A_M\(\ell^{-2} - b_c^2\).\label{pm0}\eeq
If $\ell b^2_c<1$ and $\sum_A\zeta^A_M\ge 0$, the minimum indeed
corresponds to $y=0$.  However if $\sum_A\zeta^A_M<0$, the minimum
corresponds to a smaller $w$ with $y\ne0$.  For example if $y^M =
y'^A$, $q'^A = -q^A$, $\zeta^A_M = -\del^A_M$, the minimum occurs for
$y'^A = y^A = p_A\(\ell + 1/\sqrt{2}b_c\)$.  The matrix $N_{ab}$ is
scaled by a factor $(k^A + y^A)/2\ell p_A = 3 + \sqrt{2}/(b_c\ell)$
with respect to the minimal case. In the approximation $b_c\ell\ll 1$,
this gives
\bea w &\approx& \sqrt{2}b_c p(2+\sqrt{2}), \quad \del k' = \dx\ell h'_X
= {1\over 4}\dx^2 N_{XX}^{-1} \approx b_c p/\sqrt{2}, \nnn
\tk' &\approx& \del k - \ell w < 0. \eea
If $p_M\ne0$ we have instead of \myref{pm0}
\beq \l{\pp V\over\pp y^M}\r_{y=0} = 
\(\sum_A\zeta^A_M - 1\)\(\ell^{-2} - b_c^2\).\label{pm1}\eeq
In this case $y=0$ is the minimum if and only if $\sum_A\zeta^A_M =
1$, for example $y^M = y^A_i$, $i = 2,\ldots, N$, $n = Nm$, with $\vec
q_i^A = \vec q^A$, $\zeta^A_M = \del^A_M$.  If $\sum_A\zeta^A_M>0$,
the minimum will in general shift slightly from $y=0$.  For example
with a single \ux\, and two chiral superfields with $p_1 = p_2$,
The minimum occurs for $y_2 = -q_1y_1/q_2 = q_1/q_2 - 1.$  The dilaton
potential for these cases is not substantially different
from the minimal case.  On the other hand if $\sum_A\zeta^A_M>0$,
the situation is similar to the case with $p_M=0$: the minimum
occurs for larger $k^A$, with lower values for both the potential $w$ 
and $\del k'$ such that it becomes difficult to
maintain $\tk'>0$, {\it i.e.,} positivity of the dilaton metric.

To get an idea what values the various parameters might take, consider
the FIQS model described in section 4.2 of \cite{fiqs}, with the
$p_A$ along the ($S_1,S_2,S_3,S_6,S_8,Y_1$) sector.  In this sector
$Q^F_3 = 0, \; 6Y = {1\over3}Q^F_1 - {1\over2}Q^F_2 + Q^F_4 = 0$,
where, with the \ua-charge sign conventions used here, the properly
normalized charges are
\beq q'_a = - {1\over12\sqrt{3}} Q^F_a\times \cases{\sqrt{2}&for$\;a =
1,X$\cr\sqrt{3}&for$\;a = 2,6,7$\cr \sqrt{6}&for$\;a = 3,4,5$\cr}.\eeq
Then
\beq -Y = \sqrt{2\over3}q'_1 - q'_2 + \sqrt{2}q'_4.\eeq
We define the independent $\vec q$'s that are nonzero in this
sector and orthogonal to $\vec Y$ as $(\vec q^1,
\vec q^2,\vec q^5,\vec q^6,\vec q^7,\vec q^X)$, with
\bea q^1 &=& {1\over2\sqrt{11}}\(\sqrt{3}q'_1 + 4\sqrt{2}q'_2
+ 3q'_4\) = - {1\over12\sqrt{22}}\(Q^F_1 + 4Q^F_2 + 3Q^F_4\),
\nonumber \\ q^2 &=& {1\over2}\(\sqrt{3}q'_1 - 
q'_4\) = - {1\over12\sqrt{2}}\(Q^F_1 - Q^F_4\),
\nonumber \\ q^a &=& q'_a, \quad a = 5,6,7,X.\eea
Then the matrices defined in \myref{inft} are 
\bea {\bf q}^T &=& {1\over18\sqrt{2}}
\pmatrix{0&3\sqrt{11}&3\sqrt{11}&-3\sqrt{11}&-3\sqrt{11}&0\cr
12&-3&-3&-9&-9&12\cr0&6&-6&6&-6&0\cr
-6\sqrt{2}&0&6\sqrt{2}&0&-3\sqrt{2}&3\sqrt{2}\cr
0&-6\sqrt{2}&6\sqrt{2}&3\sqrt{2}&-3\sqrt{2}&0\cr
8\sqrt{3}&8\sqrt{3}&8\sqrt{3}&8\sqrt{3}&8\sqrt{3}&-4\sqrt{3}\cr}, \quad
\nonumber \\ Q^T &=& {\sqrt{2}\over6}
\pmatrix{3/\sqrt{11}&3&0&-4\sqrt{2}&2\sqrt{2}&\sqrt{3}\cr
6/\sqrt{11}&0&3&2\sqrt{2}&-4\sqrt{2}&\sqrt{3}\cr
6/\sqrt{11}&0&-3&2\sqrt{2}&2\sqrt{2}&\sqrt{3}\cr
-12/\sqrt{11}&0&6&2\sqrt{2}&2\sqrt{2}&\sqrt{3}\cr
-12/\sqrt{11}&0&-6&2\sqrt{2}&-4\sqrt{2}&\sqrt{3}\cr
-18/\sqrt{11}&6&0&8\sqrt{2}&-4\sqrt{2}&\sqrt{3}\cr}. \eea
Then we get $p = \dx\sqrt{3/2} = 6p_A$.  In this model the above
states come in degenerate groups\footnote{There is an additional
three-fold degeneracy for $Y_{1,2,3}$.} of 3, that we label by
$k^A_\alpha,\; \alpha = 1,2,3$ and set $k^A_\alpha = \ell p_A^\alpha$.
If $N$ of the $p^\alpha_A$ are nonvanishing we have $p =
\sqrt{3\over2}\dx = 6\sum_{\alpha=1}^N p^\alpha_A,$.  We can also
calculate the parameters relevant for T-moduli masses:
\beq p_I = \sum_Aq_I^A p_A = {2\over3} p + (p^1_6,p^2_6,p^3_6),
\quad \sum_\alpha p^\alpha_6 = {p\over6},\eeq
with
\beq \dx = {3\sqrt{6}\over4\pi^2} = \sqrt{6}b_c = \sqrt{6}b
\quad p = 3b. \eeq

In this model there are additional F-flat and D-flat directions
associated with ``invariant blocks'' ${\cal B}$ of fields such
that
\beq \sum_{A\in\cal B}q^a_A + \sum_{M\in\cal B}q_a^M = 0.\eeq
It is clear that if we choose $\Phi^M$ that form invariant
blocks with the $\Phi^A$, at least some $\zeta^A_M<0$.  The
conditions in \myref{ycond} are met for the choice 
\beq \Phi^A = (S_1,S_2,S_3,S_6,S_8,Y_1), \quad
\Phi^M = (S_4,S_7,Y_2,Y_3,S_5,S_9).\label{phiam}\eeq
The numerical solution to the minimization equations with just one set
of $k^A,p^A$ nonzero gives $\lvev \tk'\rvev<0$.  We expect this
problem to be generic. We therefore restrict here\footnote{If the
dilaton metric goes through zero, one should rewrite the theory in
terms of the canonically normalized field, in terms of which the zero
of the metric becomes a singularity in the potential.  It is not clear
that there might not be some viable region of parameter space in this
case.} the class of viable models to be the minimal
models,\footnote{We have also assumed a minimal K\"ahler potential for
matter fields, by which we mean that the matter field K\"ahler
potential is the minimal one consistent with modular invariance, as
given in \myref{kpt}; more general forms will be considered below.}
defined by models in which the potential $w(k^A)$ is minimized at
$k^A = \ell p_A$.  For example in the FIQS model if we set
$p^\alpha_A\ne0\;\forall\;\alpha$, condensation can occur only if
$k_\alpha^A\ne0\;\forall\;\alpha$, and it is likely that the directions
$k^M\ne 0$ are no longer F-flat~\cite{fiqs} so that the minimal
scenario described above is viable.  For these models the vacuum
conditions $\myvev{V} = \myvev{V'} = 0$ reduce to
\beq \tk'' = {\tk'(1 - b_c\ell)\over\ell(1 + b_c\ell)} =
{3b_c^2(1 - b_c\ell)\over(1 + b_c\ell)^3},\label{lvac}\eeq
and condition for a local minimum $\myvev{V'}>0$ reads
\beq \tk'''> {2\tk'(2 - 2b_c\ell + b_c^2\ell^2)\over\ell(1 +
b_c\ell)^2} = {6b_c^2(2 - 2b_c\ell + b_c^2\ell^2)\over(1 +
b_c\ell)^4}\equiv\mu_0^2.\label{lmin}\eeq
The dilaton mass
\beq m_\ell = \sqrt{2\over\tk'}\mu = \sqrt{2\over3}{\mu\over b_c^2}
\(1 + b_c\ell\)^2m_{\tG},\quad \mu^2 = \tk''' -
\mu^2_0,\label{dilmass}\eeq
can be considerably larger~\cite{bgw3} than the gravitino mass if
$\mu\sim1$.

If the K\"ahler potential has terms of order higher than quadratic in
fields with large \vev's, the minimal form assumed in \myref{kpt} may
not be valid. As shown in \myapp{nmapp}, if $K = k + G + f(x^A)$,
the dilaton potential takes the
form \myref{pot} with, neglecting $O(\del)$ corrections,
\bea w &=& \sum_{A B}K^{A B}\(p_A + b_c K_A\)\(p_B + b_c K_B\), \nnn
\del_{a X}\ell{\dx\over2} &=& \ell\sum_A q^a_A p_A = \sum_A q^a_A K_A
= \ell\sum_A q^a_A K'_A = 2\ell\sum_{A B b}q^a_A q^b_B h'_b K_{A
B},\nnn \del k' &=& 2\sum_{A a}q^a_A h'_a K_A = h'_X\del_X = - 2\del
s'.\label{newrels}\eea
For a minimal set we can use \myref{inft} to obtain from
the constraints in \myref{newrels}
\bea Q^A_X{\dx\over2} &=& p_A = 2\sum_{B b}K_{A B}q^b_B h'_b, \qquad
\sum_B K^{A B}p_B = 2\sum_b q^b_A h'_b,\nnn \sum_{A B} K^{A B}p_A p_B
&=& \dx\sum_{A b}Q^A_X q^b_A h'_b = \dx h'_X.\eea
Then since $K_A = \ell p_A$ for minimal models we get $w = \del k'$
and the dilaton potential is identical to that for the case of a minimal
K\"ahler potential. In the general case the F-term is
\beq \sum_{A B}F^A K_{A\B}\bF^{\B} = w - \del k' + O(\del)\ge0,\eeq
so a deviation from the minimal case can only give an additional
positive contribution to the vacuum energy, making it difficult
to maintain a positive dilaton metric with vanishing vacuum energy,
as discussed above for the minimal case.

\subsubsection{Modular fermion masses}
The mass matrix for the fermion superpartners of the dilaton and
moduli is given in \myref{ferms}.  In the FIQS model, with $b^I_c = 0$
and $p_I \approx 2b_c$ and $b_I\approx 3b_c$ nearly independent of
$I$, this reduces to the simpler form given in \myref{fiqsferms} and
\myref{fiqsferms2} in the case that all the moduli are stabilized at
the same self-dual point: $t_1=1$ or $t_2 = e^{i\pi/6}$. In this
approximation there are two linear combinations $\chi_b$ of the
``T-modulini'' that do not mix with the dilatino and that have 
approximately the same mass:
\beq |m_{\chi_b}(t_1)| \approx {8 + 6z\over1 + 3z}m_{\tG}, \quad 
|m_{\chi_b}(t_2)| \approx {10 + 6z\over1 + 3z}m_{\tG},\eeq
where $z = b_c\ell = .08\ell$ in the FIQS model.  The third modulino
$\chi_0$, which is approximately an equal admixture of the $\chi^I$,
mixes with the dilatino with via the mass matrix given in
\myref{fiqsferms2} with, in this model, 
\bea m_{\chi_0} &=& m_{\chi_b} + m' \approx m_{\chi_b} +
{m_{\tG}\over1+3z},\quad
m_{\chi_\ell\chi^I} = \sqrt{6}{3 + 7z + z^2 - z^3\over(1 + z)^2(1 +
2z)\sqrt{1 + 3z}}, \nnn m_{\chi_\ell} &\approx& -{1 - 13z + 24z^2 +
7z^3 + 35z^4 + 54z^5 + 6z^6\over3z(1 + z)^3(1 + 2z)}m_{\tG}
.\label{modferms}\eea
If $z=.11$, the mass eigenvalues are
\beq |m_1| = 14.2m_{\tG}, \;\; |m_2| = 8.7m_{\tG},\eeq
for $\myvev{t^I} = t_1$ or $t_2$; these numbers decrease monotonically
with $z$ for $z<1$ dropping to 6.1,.6 at $z=.6$.  Since the choice of
the self-dual point at which the moduli are stabilized has a minor
effect on the masses, we expect the same results to hold if they are
not all stabilized at the same point. All the masses in the
dilaton/T-moduli sector decrease somewhat as $z$ increases.

\mysection{Observable sector scalar masses}\label{obs}

Here we consider only the minimal models defined above.
Chiral fields $\Phi^M$ with vanishing \vev's have 
K\"ahler potential
\beq K(\Phi^M,\Ph^{\M}) = K_{(M)} = x^M(\ell) = e^{G'^M +
2\sum_aq^a_M h_a(\ell)}\left|\Phi^M\r^2, \quad \ell =
\ell(\hel,\del^A,x^N).\eeq
Terms bilinear in these fields appear in the potential only through
the functional $\hS$ and the effective K\"ahler potential $\tK$ in the
canonically normalized Weyl basis; at lowest order in $x^M,\del^A$:
\beq \tK_{(M)} = \hK_{(M)} + 2L\hS_{(M)} = K_{(M)} + 2L S_{(M)} =
x^M(\ell),\quad S_M = 0, \quad 2\hS_M = \pp_{\ell}x^M(\ell).\eeq
Since $\pp V/\pp\phi \propto \ph$ vanishes in the vacuum, the mass
matrix is diagonal:
\beq m^2_M = \({\pp\tK\over\pp\phi^M\pp\ph^{\M}}\)^{-1} {\pp
V\over\pp\phi^M\pp\ph^{\M}} = \({\pp
x^M\over\pp\phi^M\pp\ph^{\M}}\)^{-1} {\pp V\over\pp\phi^M\pp\ph^{\M}}
= {\pp V\over\pp x^M}.\label{m2M}\eeq
The complex scalar masses are evaluated in
\myapp{massapp}:
\beq m^2_M = m_{\tG}^2\(1 + \zeta_M{1 - z^2\over
z^2}\),\label{masses}\eeq
where
\beq \zeta_M = \sum_a q^a_M Q_a = \sum_{a,A}q^a_M Q^A_a,
\quad z = b_c\ell,\label{zetam}\eeq
$m_{\tG} = b_c|u|/4$ is the gravitino mass, and the matrix $Q^A_a$ is
defined in \myref{inft}.  Unless $|\zeta_M|< z^2,$ the D-term
contribution to the scalar masses strongly
dominates~\cite{kk}--\cite{dp} over the supergravity contribution
($\mu_M =1$ if $\zeta_M=0$) for weak coupling ($z\ll 1$), as has been
extensively discussed in the literature.  Moreover positivity of
Standard Model scalar squared masses with weak coupling would require
$\zeta^M>0\forall\; M$~\cite{bdu1}; this is not a generic feature of
orbifold compactifications (see, {\it e.g.} the FIQS example below).
Therefore models that are viable in the weak coupling regime {\it a
priori} require vanishing or very small values of $|\zeta_M|$ for the
standard model particles.

As a concrete example, in the FIQS model discussed above we have
\beq Q_a = \sum_A Q^A_a =
\(-{9\over\sqrt{22}},{3\over\sqrt{2}},0,4,-2,\sqrt{6}\).\eeq
In the same model the left-handed $SU(2)$ doublets $Q = Q_L$ have
\beq q^a_{Q} =
\({5\over2\sqrt{22}},{1\over2\sqrt{2}},0,0,0,0\),\quad
\zeta_{Q} = - {3\over11}.\label{QL}\eeq
There are two candidates $u_i$ for the left-handed anti-up quarks
$u^c$, two candidates $d_i$ for the left-handed anti-down quarks
$d^c$, five candidates $\ell_i$ for the left-handed leptons $\ell^+$,
four candidates $\tG_i$ for the Higgs doublet $H_d$, and five
candidates $G_i$ from which to choose the lepton doublets $L$ and the
Higgs doublet $H_u$. Each of these comes in a triplet of states with
identical gauge quantum numbers. To avoid flavor-changing neutral
currents we assume states with the same Standard Models gauge quantum
numbers also have the same \ua\, charges.  We obtain the smallest values
of $|\zeta_M|$ with the identification
\beq \(Q,u^c,d^c,\ell^+,L,H_u,H_d\)_L =
\(Q_L,u_2,d_2,\ell_i,G_j,G_k,\tG_{\rm 3\;or\; 4}\), \quad i,j,k 
\ne5. \label{ident}\eeq
With this choice we have for the left-handed anti-quarks:
\bea q^a_{u} &=& \({1\over\sqrt{22}},-{1\over3\sqrt{2}},0,
{1\over6},0,-{2\over3\sqrt{6}}\), \quad \zeta_{u} = -
{10\over11},\nnn q^a_{d} &=&
\(-{1\over2\sqrt{22}},{1\over6\sqrt{2}},{1\over3\sqrt{2}},
0,-{1\over3},-{2\over3\sqrt{6}}\), \quad \zeta_{d} = 
{5\over11}.  \eea
The fields $G_i,\ell_i,\; i=1,\ldots,4$, and $\tG_{3,4}$ have
different \ua\, charges, but degenerate values of $\zeta_M$:
\beq \zeta_{\ell^+} = - {7\over11}, \quad \zeta_{H_d} =
{2\over11}, \quad \zeta_{H_u}=\zeta_L = - {13\over11}.\eeq
In the FIQS model we have $b_c \approx .08$, so an acceptable mass
spectrum would require $\ell\approx 12$, much larger than the
classical value $\ell = g_s/2\approx1/4$.  Such a large value of
$\ell$ suggests strong coupling in the hidden sector, a conclusion
that will be revisited in Section 5.  Note that the ``best
fit''\cite{BGWsp,BGWcos} to the model of \cite{bgw2} requires a
smaller value $b_c\approx .03$, which would require still larger
$\ell$ and/or smaller values of $|\zeta_M|$.  However those analyses
may be modified in the class of models considered here.

These conclusions are not significantly modified if we take a
nonminimal form for the K\"ahler potential.  Taking inspiration from
the known form of the K\"ahler potential for the untwisted sector:
\beq K_{untw}(\Phi,\Ph) = -\sum_I\ln\(1 - \sum_A x^A_I\),
\label{untw}\eeq
we assume a K\"ahler potential of the form
\beq K = k + G + \sum_\alpha K^\alpha, \quad K^\alpha = - C_\alpha
\ln\[1 - C_\alpha^{-1}\(\sum_A x^A_\alpha + \sum_M
x^M_\alpha\)\].\label{newkahl}\eeq
The scalar masses in this case are given in \myref{nmmass}.  The
corrections due to the modification of the K\"ahler potential are
subleading in the weak coupling limit. The general expression is
rather complicated and we will just consider some illustrative
examples for minimal sets of \vev's. First assume that the fields with
large \vev's belong to the same set $\alpha,\; C_\alpha = C$.  In
the FIQS model, the quarks $Q_L$ are in the untwisted sector.  Then
from \myref{untw} not more than one $Q_L$ can belong to the set
$\alpha$; then none of them can if we require that they all have the
same mass so as to avoid flavor changing neutral currents.  If we
assume that all of the Standard Model chiral multiplets have $\beta\ne
\alpha$, we obtain ($\lambda = 1$ in the FIQS model)
\bea \mu^2_{M\beta} &=& {m^2_{M\beta}\over m^2_{\tG}} = 1 +\mmm
\zeta_M^\beta\({1\over z^2} - 1 + {3\lambda(1 + z)\[2C - 3\lambda
z)^2(1 - z)\]\over(1+2z)(1 + 3\lambda z)}\)\(1 -
{3\lambda^2z^2(1+z)^2\over(C + 3\lambda z)^2}\).\label{nm1} \eea
If instead we have three copies of a minimal set with $p_\alpha=p/3$,
we can assign each generation of standard model particles to one 
of these sets.  Then consistency with \myref{untw} for the 
untwisted sector particles ({\it e.g.,} $Q_L$ in the FIQS model)
requires $C_\alpha = 1$, and we obtain, {\it e.g.}, for $\lambda = 1$,
\bea \mu^2_{M\beta} &=& \zeta_M^\beta\({1\over z^2} -
{(1+z)\over(1+2z)}\) - {z(1+z)\over(1+2z)}.\label{nm2}
\eea
These corrections to the scalar masses relative to the ``minimal''
K\"ahler potential case \myref{masses} are negligible except for
$|\zeta_M/z|\sim1$.

There are many other possible parameterizations, since the K\"ahler
potential involving twisted sector fields is not known beyond the
quadratic terms.  It might be that a set $\alpha$ contains only fields
with the same modular weights, as in \myref{untw}.  Alternatively the
K\"ahler potential could be invariant under Heisenberg transformations
on the untwisted sector fields:
\bea K &=& k + G_{un} + f(X^A), \quad X^A =
|\Phi^A_{tw}|^2e^{G^A_{un}}, \quad G_{un} = G + K_{untw} = \sum_I
G^I_{un},\nnn G^A_{un} &=& \sum_I q^A_I G^I_{un}, \quad G^I_{un} = g^I
+ f^I,\label{heis}\eea
where the $\Phi^A_{tw}$ are twisted sector fields, and
$\exp{G_{un}^I}$ is a radius of compactification in string units.  It
has been argued~\cite{heis,lyth} that the effective supergravity
theory from the weakly coupled heterotic string may allow a viable
inflationary scenario if the K\"ahler potential is of this form.  In
this case, the contribution to untwisted sector squared masses
$m^2_{(M I)}$ from the terms evaluated in \myapp{nmapp} is somewhat
more complicated, but the corrections are still subleading with
respect to \myref{masses}. However under the ansatz \myref{heis} there
is an additional contribution to the untwisted sector masses from the
shift \myref{delgs} and the expression \myref{phiprod} with now $g^I$
replaced by $G^I$:
\beq \del\mu^2_{(M I)} = \[\(\del b_I + p_I\)^2/b_c^2\]\[1 + O(z)\],\eeq
with, {\it e.g.,} $(\del b_I + p_I)/b_c \approx 6$ in the FIQS model;
this would give a large positive contribution to $\mu_Q$.  (In the
FIQS model discussed in the previous section, the only untwisted
sector fields other than the $\G_c$-charged fields are the quark
doublets $Q$ and $u_2,\;\tG_1$, which are candidates for
$u^c_L,\;H_d$, respectively; they have larger values of $\mu_M$ than
the fields $u_2,\tG_{3\;{\rm or}\;4}$ that were identified with these
states in the previous section.) If the GS term is also Heisenberg
invariant, there is a further contribution
\beq \del\mu'^2_{(M I)} = \[\(b - b_c\)^2/b_c^2\]\[1 +
O(z)\].\label{gsshift}\eeq
If instead the full moduli + matter K\"ahler potential couples to the
GS term: $G\to K-k$ in \myref{oit}, all squared masses for the $x^M$
are shifted by \myref{gsshift}; this expressions vanishes identically
in the FIQS model but it is possible that effects such as these in
generic models could make all the masses positive.  Even so, the large
scalar/gaugino mass hierarchy remains problematic although one-loop
corrections~\cite{gnw} can significantly increase the gaugino masses
in the presence of scalar couplings in the GS term.

\mysection{Parameterizations of nonperturbative effects}\label{param}

It has been argued~\cite{shenk} on general grounds that the effective
supergravity Lagrangian from compactified string theory receives
corrections $\propto e^{-\beta/g_s}$, arising from string
nonperturbative effects.  This was indeed shown~\cite{eva} to occur in
an explicit compactification of the heterotic string.  Even in the
absence of these effects, one expects corrections $\propto
e^{-\beta/g_s^2}$ from both string and field theory~\cite{bd}
nonperturbative effects. 

To parameterize these effects, a simple form was assumed
in~\cite{BGWa,bgw2} for the functional\footnote{This is not the most
general ansatz since there can be different parameters in the
exponents.}
\beq f(L) = 2Ls(L) - 1 = \sum_na_nx^ne^{-x}, \quad x =
\beta/\sqrt{L}\label{fparam},\eeq
from which the functional $g'(L) = k'(L) -
L^{-1}$ was inferred using the condition \myref{canon}.  Retaining
just one or two terms in the sum \myref{fparam}, a fit to the vacuum
conditions $v = v' = 0$, $g^2_s \approx .5$, where found with order one
values of $\beta,a_{0,1}$.  However it might be more justified to
start with a simple form for the functional $g(L)$:
\beq g(x) = \sum_nA_nx^ne^{-x}, \label{gparam}\eeq
since this parameterizes the corrections to the actual K\"ahler
potential.  The conditions that the functions $g,f\to0$ in the weak
coupling limit $L\to0$ are met if one assumes both \myref{gparam} and
\myref{fparam}.  Then the constraint \myref{canon} relates the
coefficients by
\beq nA_n - A_{n-1} + (2 + n)a_n - a_{n-1}= 0.\label{arels}\eeq
If we also require that they are finite in the strong
coupling limit $x\to0$, we set $A_{n<0} = a_{n<0} = 0.$  Then
(\ref{arels}) is satisfied by 
\beq a_n = - {n\over n+2}A_n +
2\sum_{p=0}^{n-1}{(p+1)!\over(n+1)!}A_p.\eeq
This requires $a_0 = 0$, which could not be imposed in the model
of~\cite{BGWa,bgw2} because in those papers the nonperturbative terms
were needed to make the potential positive.  This is not the case for
the models considered here.

On the other hand, in the presence of nonperturbative corrections the
true string coupling constant is $g_s = 1/\sqrt{s(\ell)}\ne
\sqrt{2\ell}$, suggesting we should set $x= \beta\sqrt{s(L)}$ in
\myref{gparam} and \myref{fparam}.  Moreover, if it is the various
terms $\L_i$ in the {\it Lagrangian} $\L = \sum_i\L_i$ that receive
corrections of the form $L_i\to (1 + h_i)\L_i$, with $h_i(x)$ a
function of the form \myref{gparam}, one can show that this
corresponds to a correction to the K\"ahler potential of the form
\beq k = - \ln(2s) - \sum_i\alpha_i\ln(1 + h_i).\label{newparam}\eeq
All parameterizations of the nonperturbative corrections are
equivalent if these corrections are small, but in fact the vacuum
condition \myref{vac0} requires either large coupling or a significant
correction to the classical K\"ahler potential $k = \ln(\ell)$.  To
see this note that the condition
\beq k' + 2\ell s' = \tk' + 2\ell\ts' = 0 \eeq
implies
\beq {\pp k\over\pp s} =  {\pp\tk\over\pp\ts} = -2\ell,
\quad {\pp k\over\pp\ell} = {\pp s\over\pp\ell}{\pp k\over\pp s} = 
4\ell\({\pp^2k\over\pp s^2}\)^{-1}, \quad \kappa \equiv
{1\over\ell}{\pp k\over\pp\ell} = 4\({\pp^2k\over\pp s^2}\)^{-1},\eeq
with all relations holding for $k,s\leftrightarrow\tk,\ts$.  In the
classical limit we just have $k = - \ln(2s),$ $\kappa = \ell^{-2}$, so
the vacuum condition $\kappa = 3rb^2_c/(1 + \ell b_c)^2,$ $r\approx$
1--4, requires $\ell\sim b_c^{-1}/\sqrt{3r}$. The parameterization of
nonperturbative effects in~\cite{BGWa,bgw2} allowed a fit with
$f\sim1$, $\kappa \sim b_c^{-2}\ll \ell\sim 1$.  In that
parameterization $g_s = \sqrt{2\ell/(1+f)}$, $f>0$, so $g_s\ll \ell$
requires $f\gg1$, {\it i.e.,} nonperturbative effects must dominate
the dilaton potential at the vacuum, which is rather implausible in
the context of weak coupling.  A different parameterization might allow
$1/2s\ll\ell\sim b_c/\sqrt{3r}$ for moderate nonperturbative
contributions, and thus some suppression of the scalar masses found
in \mysec{obs}.

Suppose that for the observed value $s\approx 2$ of the coupling
constant, the sum in \myref{newparam} is dominated near the vacuum by
a single term:
\beq k = - \ln(2s) - \alpha\ln(1 + h).\label{hparam}\eeq
This gives, for $\alpha=1,$
\beq \ell = {1\over2s} + {h'\over2(1+h)}, \quad \kappa = 4
\({1\over s^2} - {h''\over(1+h)} + {h'^2\over(1+h)^2}\)^{-1},\eeq
where here prime means differentiation with respect to $s$.
If $h=\epsilon -1> -1,$ $h'>0$, one can get an enhancement of $\ell$
and a suppression of $\kappa$ for small $\epsilon$.  For example if we
take a monomial $h = - Ae^{-\sqrt{s}} <0$, we have $h'> 2(1+h)$ if
$3.5< A< 4.1$.  This gives $1.5\le\ell\le6.8$, $.51\ge \kappa\ge .02$
for $3.6\le A\le4$.  Alternatively if we take $h = -
A\sqrt{s}e^{-\sqrt{s}} <0$, we have $h'> 2(1+h)$ if $2.4< A< 2.9$, and
$1.6\le\ell\le5.9$, $.60\ge \kappa\ge .03$ for $2.5\le A\le2.8$.  

If canonical normalization of the Einstein term is imposed in the the
dual formulation for the dilaton in terms of a chiral superfield $S$,
the condition \myref{canon} arises from the solution to the equations
of motion for $L$ in the duality transformation:
\beq 2\re S = \int{d L\over2L}k'(L).\eeq
When the GS term is included, this is modified to read
\beq 2\re S + V_{GS} = \int{d L\over2L}k'(L), \quad V_{GS} = 
b G - \dx V_X.\eeq
When we shift $V_X$ from its ($L$-dependent) \vev, $S =
s(L)\to\ts(L)$, and $k(L)\to\tk(L)$, up to terms of order $\del$ and
terms quadratic in the heavy fields that we integrate out.  This
suggests that the parameterization \myref{hparam} should apply to
$\tk,\ts$ rather than to $k,s$.  The fact that the gauge coupling is
really $s$ rather than $\ts$ comes from a compensating term from field
theory quantum corrections that appear to be unrelated to the duality
transformation $L\leftrightarrow S + \S$.

The vacuum condition in the minimal models now reads
\beq  \tilde\kappa = \kappa + 3\ell^{-1}\lambda b_c = 3rb^2_c,
\eeq
which requires $\kappa<0$ unless $\lambda< rb_c\ell$, which for
example in the FIQS model would require $\ell> 13/r$; for the favored
value $b_c\approx .03$ in the analyses of~\cite{BGWsp,BGWcos}, this
constraint would be more stringent.  On the other hand the
parameterization \myref{hparam} of $k$ requires positive $\kappa$ in
order to enhance $\ell$ and suppress $|\kappa|$.  Therefore this
parameterization seems to be viable only if it applies to $\tk$.

Another important parameter of the effective low energy theory is the
scale $\mu_s$ of unification of the Standard Model gauge couplings.
The boundary condition found in \cite{gt}, that gives essentially the
standard result~\cite{kap} $\mu^2_s \approx g_s^2/2e$ in the
$\overline{MS}$ scheme, is changed due to modifications of the
K\"ahler potential from both string nonperturbative and \ua-breaking
effects. The string nonperturbative correction to $\mu^2_s$ was
found~\cite{bgw3} to be less than 1\% in the model of \cite{bgw2}.
In the present case we have new corrections as well as a wider range
of possible parameterizations for the string nonperturbative effects.

Making the appropriate modifications to Eq. (5.9) of~\cite{gt},
comparison with the renormalization group invariant~\cite{shif} gives,
following the analysis of~\cite{rulin}, the two-loop renormalization
group equations\footnote{Here we neglect moduli-dependent string
threshold corrections to the gauge kinetic terms.  The universal
terms~\cite{kkpr} drop out of the unification constraints.  The
nonuniversal terms \myref{lth} are absent in many quasi-realistic
orbifold models such as the FIQS model, and in any case are small when
the moduli are stabilized at their self-dual points as is the case
here. The factor $e$ relates~\cite{tatu} the scale $\lambda$ of the
external momentum ($-p^2 = \lambda^2$) to the scale $\mu$ of the
$\overline{MS}$ scheme: $\lambda^2 = e\mu^2$.}
\bea g^{-2}_a(\mu) &=& s -\frac{1}{16\pi^2}(C_a -
C^\chi_a)\tk(\ell)-\frac{2}{16\pi^2} \sum_M C^M_a\ln
\[1-P_M(\ell)\]\,\mmm + \frac{1}{16\pi^2}
(3C_a-C_a^\chi)\ln(e\mu^2) -\frac{2C_a}{16\pi^2}\ln g^2_a(\mu) -
\frac{2}{16\pi^2}\sum_M C^M_a\ln Z_M(\mu)\, ,\eee g_s^{-2} + {C_a\over
8\pi^2}\ln(e^{\tk}g_a^{-2}(\mu_s)) - \frac{1}{8 \pi^{2}} ( 3C_{a}
- C_{a}^\chi) \ln \frac{\mu_s}{\mu} + \frac{C_{a}}{8\pi^{2}} \ln
\frac{g_a^{2}(\mu_s)}{g_a^{2}(\mu)}\nonumber\\ & & + \frac{1}{8 \pi^{2}}
\sum_{M} C_a^{M} \ln \frac{Z_M(\mu_s)}{Z_M(\mu)}, \qquad\qquad
C^\chi_a = \sum_M C^M_a,
\label{rge}\eea
where $C^\chi_a$ is the chiral matter quadratic Casimir and the $Z_M$
are the renormalization factors for the matter fields $\Phi^M$.  In
writing the last equality we made the identifications
\beq \mu^2_s = e^{\tk-1},\quad s = g_s^{-2}, \quad
Z^{-1}_M(\mu_s) = 1 - P_M(\ell),\eeq
where the functions $P_M(\ell)$ arise from the dilaton-dependent term
in the reparameterization connection derived from the effective
K\"ahler metric $\tK = K + 2\hL S$; they depend on the \ua\, charges
$q^a_A$ and on the couplings of $\Phi^A$ in the GS terms.  If we
neglect two-loop running in the vicinity of coupling unification, this
contribution drops out, and we may set $g_s^{-2}(\mu^2_s) = s$ in the
second term on the RHS of \myref{rge}.  Then defining $\mu_{ab}$ by
\beq g_a(\mu_{ab}) = g_b(\mu_{ab}),\eeq
we find in this approximation 
\bea \mu_{ab} &\approx& e^{(\tk-1)/2}(e^{\tk}s)^{-\ep_{ab}} =
{g\lambda_{ab}\over2^{\ep_{ab}}\sqrt{e}},\nnn \lambda_{ab} &=&
(e^{\tk} 2s)^{\ep_{ab}},\quad \ep_{ab} = {1\over2} - {C_a - C_b\over
3C_a - C^\chi_a - 3C_b + C^\chi_b}.  \eea
The parameters $\lambda_{ab}$ measure the deviations from the
classical string theory prediction without string nonperturbative and
\ua-breaking effects: $\tk\to k\to \ln(2s)$.  Using the MSSM values for
the gauge and matter Casimirs:
\beq (C_a,C^\chi_a) = (0,6.6),\; (2,7),\; (3,6),\qquad a = 1,2,3,\eeq
gives 
\beq \ep_{12} = {1.5\over10.1}, \quad \ep_{23} = .25, \quad \ep_{31}
= {1.3\over6.6}.\eeq
Thus although some parameterizations of nonperturbative effects
considered above have $\mu_s$ considerably larger than its classical
value $g/\sqrt{2e}$, the effect on coupling unification is much less
significant.  Taking the most extreme of those cases (assuming, as
argued above, that the parameterization \myref{hparam} applies to
$\tk,\ts$)
\bea \tk &=& -\ln(2\ts) - \ln\(1 - 4e^{-\sqrt{\ts}}\), \quad \ts = s -
{\dx\over4}\ln\ell \approx s \approx 2,\nnn 2s e^{\tk} &\approx& \(1 -
4e^{-\sqrt{\ts}}\)^{-1}\approx 36,\label{extreme}\eea
we obtain
\beq \lambda_{12} = 1.66, \quad \lambda_{23} = 2.45,\quad \lambda_{31}
= 2.03.\label{lambda}\eeq
As is well known, the classical prediction $\lambda=1$ is
in contradiction with experiment since the measured unification scale is
lower than predicted by about an order of magnitude if the running is
due only to MSSM particles.  Orbifold models predict additional
particle content with masses somewhere between the electroweak scale
and the string scale, and many analyses~\cite{rulin,unif} have shown
that these masses can be adjusted to reconcile the theory with
experiment.  We expect that similar fits can be made in the present
case, even with slightly higher ``unification'' scales, as in 
\myref{lambda}.

The scale of condensation $\Lambda_c = |u\u|^{1\over6}$ and the
gravitino mass $m_{\tG} = b_cu/4$ are also governed by the factor
$e^{\kappa} = e^{\tk} + O(\del)$ in $u\u$ given in \myref{sols}; in
the ``extreme'' parameterization \myref{extreme}, $e^{\tk} \approx
9$. However there is a much stronger $\ell$-dependence that is
unrelated to nonperturbative effects; this is the factor
$\exp[-(\sum_A p_A\ln x^A)/b_c]$ in \myref{sols}.  In the minimal
models studied here that are subject to the constraints \myref{pcond},
we have $x^A = \ell p^A$, so this contribution grows as $\ell$
decreases, suggesting that these scenarios are more viable if $\ell$
is considerably larger than its classical value of $g^2_s/2$.  For
example, taking the FIQS model with $g_s^2 = .5,\; t^I = 1,\; c_\alpha
= 1$, gives a gravitino mass of 3000 TeV if $\ell = 6$, and this grows
dramatically as $\ell$ decreases.  Therefore a viable model in this
class requires a smaller $\beta$-function, as was also found in
studies~\cite{BGWsp,BGWcos} of models without an anomalous $U(1)$,
that were unable to accomodate the $SO(10)$ condensing gauge group of
the FIQS model.  

\mysection{D-moduli masses} 

It has been pointed out~\cite{dmod} that there is generally a large
degeneracy of the vacuum associated with \ua\, breaking, resulting in
many massless chiral multiplets, that we call D-moduli, between the
\ua-breaking and supersymmetry-breaking scales.  Moreover, in the
absence of superpotential couplings a number of these remain massless
even after supersymmetry breaking.  Here we show that couplings of the
D-moduli to the matter condensates {\it via} the superpotential
defined by \myref{pipot} and \myref{pdefs} is sufficient to lift the
degeneracy and give masses to the {\it real} parts of the D-moduli
scalars as well as the fermions, while the imaginary
parts of the scalars (``D-axions'') remain massless in the absence of
other superpotential couplings.  This remaining degeneracy may be at
least partially lifted by D-moduli couplings to other unconfined,
$\G_c$-neutral chiral supermultiplets.

Since our purpose in this section is only to establish that D-moduli
masses are generated, we will restrict our analysis to the simple case
in which there are $N$ minimal sets of chiral fields with the same
\ua\, charges so that $k^A = \ell p^A$, and we assume the minimal
K\"ahler potential \myref{kpt}.  Among the $n=Nm$ chiral fields with
$\myvev{\phi^A}\ne 0$, there are $n-m = (N-1)m$ chiral superfields
\beq D^i = (\sigma^i,a^i,\chi^i), \eeq  
that are the physical states orthogonal to the $m$ eaten Goldstone
bosons. The $D^i$ are defined in \myref{dadi} in terms of the chiral
superfields $\hTh^A$ introduced in \myref{ddef}. The relevant
Lagrangian for these fields is given in \myref{ld} and \myref{dferm}
of the appendix.  Their masses
\beq m_a = 0,\quad m_\sigma = {\sqrt{1 + z^2}\over1 - 4z}m_\chi =
{\sqrt{1 + z^2}\over z}m_{\tG}\label{dmass}\eeq
are generated by the F-terms associated with the superpotential
\myref{pipot}, \myref{pdefs} and therefore satisfy the sum rule
\beq m^2_a + m^2_\sigma = 2m^2_\chi\[1 + O(z)\],\eeq
where the $O(z)$ corrections vanish in the limit of vanishing
gravitino mass: $m_{\tG}\approx z m_\chi\to 0$ if $z\to0$.

For example in the FIQS model discussed in \mysec{dilpot}, there are
three identical sets $\Phi_\alpha^A$, $\alpha = 1,2,3$, defined as
$\Phi^A$ in \myref{phiam}. If one of these sets, say $\Phi^A_1$, has
no couplings to the condensates, $p^A_1=0$, then as shown in
\mysec{dilpot} the minimum of the potential will have
$\myvev{\phi^A_1} = 0$.  Then the 6 corresponding fermions $\chi^A_1$
will remain massless, and the complex scalars will acquire masses
given by \myref{masses} with $\zeta_{A_1} = 1$:
\beq m^2_{A_1} = {m^2_{\tG}\over z^2},\eeq
while those $\Phi^A_\alpha$ with nonvanishing $p^A_\alpha$ and
nonvanishing \vev's will have masses as in \myref{dmass}.  It was
argued that \mysec{dilpot} that it may be necessary for all the
$\Phi^A$ to have nonvanishing $p^A$ in order to stabilize the vacuum
against otherwise dangerous flat directions in the space of the
superfields $\Phi^M$ in \myref{phiam}.  In this case the $\phi^M$ as
well the scalar components of other superfields that form invariant
blocks~\cite{fiqs} with the $\Phi^A$ all acquire squared-mass terms of
the form \myref{masses} with
\beq \zeta_{S_4} = \zeta_{S_7} = \zeta_{Y_2} = \zeta_{Y_3} = -2, \quad
\zeta_{S_5} = \zeta_{S_9} = \zeta_{S_{10}} = \zeta_{S_{11}} = 1.
\eeq
There are three copies each of the superfields $S_i$ and nine copies
of the $Y_i$. Since $S_4$ and $Y_2$ have the same \uone\, charges, the
D-term potential for $(S_4,S_7,Y_2,Y_3)$ at the condensation scale has
an approximate\footnote{The symmetry is reduced to $[SO(6)]^8$ when
the differences in modular weights is taken into account, and could be
further reduced by the choice of K\"ahler potential; these effects
involve higher dimensional couplings and should generate very small
masses.}  $SO(24)\times SO(18)\times SO(6)$ invariance, resulting in
33 (approximately) massless (pseudo) Goldstone bosons when linear
combinations of these fields acquire \vev's, if there is no other
source for their masses.  The positive squared masses are safely large
for weak coupling, $z<< 1$.  However if we try to make the observable
sector viable in this model by substantially increasing $z$, we would
have an additional 27 scalars with masses uncomfortably close to the
gravitino mass.  In any case there would be 30 massless complex Weyl
fermions.  However there is a possible source of much larger masses
for at least some of these superfields.  Although terms trilinear in
the three fields that form each invariant block are
forbidden~\cite{fiqs}, supersymmetric masses of order
$\myvev{\phi^A}^4/m^3_P$ for $S_4,S_7,Y_2,Y_3$ would be generated if
the superpotential includes quadratic terms in the invariants
\beq (S_4S_3Y_1),\quad (S_7S_2Y_1), \quad (Y_2S_3Y_1), \quad
(Y_3S_6Y_1), \eeq 
thus possibly eliminating 3/5 of the troublesome light states
in this scenario.  Cosmological issues associated with massless and
TeV-scale fermions were discussed in \cite{dmod}.

\mysection{Conclusions and future directions}

We have studied a class of models based on the weakly coupled
heterotic string with an anomalous $U(1)$ and supersymmetry breaking
by condensation in a strongly coupled hidden gauge sector.  In
contrast to the models~\cite{BGWa,bgw2} previously considered without
an anomalous $U(1)$, dilaton stabilization is assured by the presence
of D-terms, but string nonperturbative corrections to the dilaton
K\"ahler potential are still needed to stabilize the dilaton at weak
coupling.  Several promising features of the earlier studies persist:
enhancement of the dilaton and T-moduli masses relative to the
gravitino mass, masslessness of the universal axion and a suppression
of its coupling constant, dilaton mediated supersymmetry breaking that
avoids potential problems with flavor changing neutral currents.

However some new difficulties arise unless the observable sector is
uncharged under the broken \ua's [or its charges are orthogonal to the
inverse charges of the fields with large \vev's: $\zeta_M = 0$ in
\myref{zetam}]. As noted in earlier
studies~\cite{kk}--\cite{dp},~\cite{bccm}, there is considerable
tension in maintaining {\it i}) a vanishing cosmological constant,
{\it ii}) a positive dilaton metric and {\it iii}) positive and
acceptably small scalar masses in the observable sector on the one
hand, while requiring {\it iv}) weak coupling and {\it v}) acceptably
large D-moduli/fermion masses on the other hand.  Some of this tension
might be attenuated by relaxing the requirement of a vanishing
cosmological constant.  However if one invokes an unknown mechanism to
cancel the cosmological constant there is no reason to assume that it
will not also contribute to scalar masses, making any predictions
meaningless.  Moreover, as discussed in the text, one needs a {\it
negative} contribution to the vacuum energy, which is very hard to
achieve, except by simply adding a constant to the superpotential.
Such a term could arise from the \vev\, of the three-form of
ten-dimensional supergravity:~\cite{drsw} $\myvev{\int
d\sigma^{lmn}H_{lmn}}\ne0$.  Such a contribution has been considered
in the past but was abandoned in the weak coupling context after the
realization that this \vev\, is quantized in Planck scale
units~\cite{rw}.  On the other hand it has been pointed out~\cite{ckn}
that quadratically divergent quantum corrections may induce a
significant contribution to the cosmological constant. These were
calculated in~\cite{uvdiv} for an arbitrary supergravity theory (with
at most two space-time derivatives at tree level), giving for the
leading terms in the number $N,N_G$ of chiral and gauge
supermultiplets, respectively:\footnote{The result quoted
in~\cite{clm} does not take into account the quadratically divergent
renormalization of the Einstein term.  The Weyl transformation
necessary to put the Einstein term in canonical form gives additional
corrections to the potential.}
\beq \myvev{V_{\rm quad}} \ni \[N_G\(\bigvev{V_{\rm tree}} -
m^2_{\tg}\) + N\(m^2_{\tG} - {g^2_s\over2}\sum_a
\bigvev{D^2_a}\)\]{\Lambda^2\over16\pi^2}.\label{quaddiv}\eeq
For the class of standard-like $Z_3$ orbifold models studied
in~\cite{Gie02a}, we have
\beq  N \ge 3N_G + 223,\eeq
suggesting, since $|D_a|\ll|m_{\tg}|<|m_{\tG}|$, that this
contribution, dominated by the term proportional to $N m_{\tG}^2$, is
always positive and of the same order as tree level terms if
$\Lambda\sim 1$ in reduced Planck units.  However, when the theory is
regulated in a way that preserves local supersymmetry, the cut-off
$\Lambda^2$ is replaced by
\beq \Lambda^2_{eff} = 2\sum_i\eta_i m_i^2\ln m_i^2, \eeq
where $\eta_i$ is the signature and $m_i$ is the mass of a
Pauli-Villars (PV) regulator superfield (and additional terms
quadratic in the PV masses are generated).  It was shown in Appendix C
of~\cite{casimir} that $\Lambda^2_{eff}$ has an indeterminate sign if
there are four or more terms in the sum over the PV fields that
regulate any one contribution to the quadratically divergent part of
the one-loop effective action.  Cancellation of all UV divergences in
realistic string-derived supergravity models requires~\cite{bhgp} at
least 5 PV chiral supermultiplets for each chiral supermultiplet of
the low energy theory and 51 PV chiral superfields for each light
gauge superfield, as well as gauge singlet PV chiral superfields and
Abelian PV vector superfields.  This proliferation of regulator fields
is not surprising, since the PV contributions parametrize those from
infinite towers of string and Kaluza-Klein modes of the underlying
string theory.  It might therefore seem reasonable to
include~\cite{ckn} an arbitrary constant of order $m^2_{\tG}$, which
need not be positive, in the effective potential at the condensation
scale.  However the terms in \myref{quaddiv}, together with their
supersymetric completion, respect supersymmetry only to one-loop
order; they are the $O(\epsilon = \Lambda^2/16\pi^2)$ corrections to
the potential due to a shift~\cite{pv1} $K\to K + \epsilon\Del K$ in
the K\"ahler potential.  If the coefficient of the correction is large
enough to be important, it must be retained to all orders in
evaluating the effective Lagrangian. Then it can be shown that the
effect on the vacuum energy is negligible~\cite{gnp}.

The generically large scalar-to-gaugino mass ratio might be reduced by
introducing more general forms of the K\"ahler potential for chiral
superfields, and/or including couplings of matter superfields in the
GS term.  Our current predictive power in this respect is
unfortunately limited by uncertainties in our knowledge of the
string-scale couplings.  In addition, the scalar masses can be reduced
relative to the gravitino mass by increasing the \vev\, of the dilaton
$\ell$.  Larger $\ell$ for fixed $g_s$ is also favored by requiring
that the gravitino mass and condensation scale be sufficiently
low. While this suggests strong coupling, we showed that with the
increased flexibility in the parameterization of string
nonperturbative effects in the presence of D-terms, $\ell$ can be
considerably larger than its classical value $g^2_s/2$ while
maintaining weak coupling: $g^2_s\approx1/2$. However either mechanism
for reducing the scalar-to-gaugino mass ratio also reduces the
D-moduli masses for fixed gaugino masses.

The weakly coupled heterotic string theory that we are using can be
obtained as the limit of zero separation between the hidden and observable
ten-dimensional boundaries of (suitably compactified generalizations of)
the Ho\v{r}ava-Witten (HW) scenario~\cite{hw}.
One might consider relaxing this strict weak coupling limit by allowing a
small separation between the ten-dimensional boundaries of the hidden and
observable effective supergravity theories.  Besides generating different
corrections to the coupling constants of these two sectors, it is conceivable
that some tuning of 11d moduli might allow for a suppression of the 
overlap factor $\zeta_M$ that governs observable sector masses.  However
we do not expect either of these effects to be significant unless we
approach the very strongly coupled HW limit, where we cannot use
perturbative results and instead have to appeal to 11-d supergravity to
extract data.  On the other hand, 11-d supergravity-based
calculations~\cite{Banks:1996ss}--\cite{Lukas:1997fg} show that one gets
an effective 4d supergravity theory very similar to those we are studying.
As a consequence, our results can easily be extended to
effective 4d descriptions of the strongly coupled heterotic string, since
the features of our effective field theory are general enough to
accommodate scenarios that occur in that context, to some level of
approximation.  A drawback to this approach is the greatly weakened
predictive power with respect to the case of the weakly coupled heterotic
string, since one does not have available the genus zero
conformal field theory calculations of the massless spectrum
and couplings.

Several other avenues for future investigation suggest themselves.  In
general left-handed fermions and anti-fermions have different
\ua-charges resulting in different masses for their scalar
superpartners; possible constraints on these mass differences from
precision electroweak data need to be investigated.
Analyses~\cite{BGWsp,BGWcos} of the model of \cite{bgw2} regarding a
viable vacuum at the electroweak symmetry breaking scale, neutralinos
as dark matter, a viable inflationary scenario and the Affleck-Dine
mechanism for baryogenesis need to be revisited with the FI D-term
included.  Indeed the inflationary scenario of~\cite{lyth} evoked such
a term for dilaton stabilization during inflation.  Including \ua\,
breaking provides new possibilities, such as D-axions as possible
candidates for quintessence and a see-saw mechanism for neutrino
masses if the right handed neutrinos acquire masses through Yukawa
couplings at the \ua-breaking scale.  There are also possible new
mechanisms for proton decay~\cite{GG02b}. We have used the FIQS
model~\cite{fiqs} as a benchmark to illustrate what might be
representative numerical results in a realistic model.  However this
model cannot reproduce the observed Standard Model Yukawa textures
\cite{Giedt:2000}, and in the present context gives implausibly large
values for $m_{\tG},$ and $\Lambda_c$, as well as an unacceptable
pattern of soft supersymmetry breaking scalar masses in the observable
sector.  The most extensively studied models~\cite{fiqs,Gie02a} have
$SU(3)\otimes SU(2)\otimes U(1)^5$ as the gauge group in the
observable sector. One might consider models in which the Standard
Model is embedded in a larger gauge group that is broken to the
Standard Model, at a scale considerably larger than the gravitino
mass, by the \vev\, of some field whose squared mass is driven
negative at the condensation scale.  This could give viable observable
sector scalar masses provided $\zeta_M\ll1$ for squarks and sleptons,
or if these particles are quasi-Goldstone bosons of an approximate
global symmetry that is broken at the same scale.

\vspace{0.20in}

\noindent {\bf \Large Acknowledgments}

\vspace{5pt}

\noindent 
We wish to thank Pierre Bin\'etruy, Brent Nelson and Erich Poppitz
for helpful discussions.  This work was
supported in part by the Director, Office of Science, Office of High
Energy and Nuclear Physics, Division of High Energy Physics of the
U.S. Department of Energy under Contract DE-AC03-76SF00098, in part by
the National Science Foundation under grants PHY-0098840 and
INT-9910077, and in part by the National Science and Engineering
Research Council of Canada.

\myappendix

\mysection{Weyl transformations}
\label{weyl}
In this appendix we work out the Weyl transformations necessary for
obtaining the potential given in \myref{bose}.

\subsection{General formalism}
We start with the Lagrangian defined by\footnote{Here we are assuming
nonvanishing \vev's only for fields with $O(\dx^{1\over2})$ \vev's at
the string scale.  We will consider additional $vev's$ of order $\Del$
in \myapp{addvev} below.}
\bea K &=& k + G + \sum_A\CA e^{(\Del^A + \bDel^{\bar A})}\prod_a
e^{2q_A^a(h_a + \Del_a)} = k + G + \sum_A x^A = \tk + G, 
\nnn S &=& s + {1\over2}\[\tG - \dx\(h_X + \Del_X\)\] = \ts +
\half\(\tG - \dx\Del_X\),\quad \tG = bG -\dx\hG_X,
\nnn \quad \tL &=& E\(2LS - 3\).\label{defth}\eea
   The Einstein term is canonical in zeroth order in $\Del$; to put
the remaining terms in canonical form we make a Weyl
transformation
\bea K &=& \hK + \Dk, \quad E = e^{-\Dk/3} \hE, \quad L = e^{\Dk/3}\hL, 
\nnn 2\hL \hS(\hL,g^I,\Del) &=& 2 \hL \left. S(L,g^I,\Del) \right|_{L =
e^{\Dk/3}\hL} + 3\(1 - e^{-\Dk/3}\)\eee 
2\hL S(L,g^I,\Del) - 2\hL\Del s.\label{wyl}\eea
The condition for a canonical Einstein term in the new Weyl basis is:
\bea 0 &=& \( {\p \hK \over \p \hL}\)_{\Del} + 2 \hL \( {\p \hS \over
\p \hL}\)_{\Del} \eee \({\pp L\over\pp\hL}\)_{\Del}\[K'(L) + 2\hL
S'(L)\] - \({\pp\Dk\over\pp\hL}\)_{\Del} - 2\hL\({\pp\Del
s\over\pp\hL}\)_{\Del}\eee e^{\Dk/3}\[1 + \hL\({\pp
\Dk\over3\pp\hL}\)_{\Del}\]\[K'(L) + 2\hL S'(L) + 2\Del s\edk\]
,\label{ech}\eea
where $F'(L)= \(\pp F/\pp L\)_{\Del}$, and the subscript on 
derivatives indicates the variable(s) held fixed, aside from the 
moduli that are held fixed throughout.  This gives
\bea 0 &=& K'(L) + 2\hL S'(L) + 2\Del s\edk\eee K'(L) + 2\hL S'(L) -
{3\over\hL}\(1 - \edk\)\edk\eee K'(L) + 2\hL S'(L) - {3\over L}\(1 -
\edk\) \eee K'(L) + 2\hL S'(L) - {3\over L}\(1 - {\hL\over
L}\).\label{wyl2}\eea
Defining $\pp_\alpha F = \pp F/\pp \Del^\alpha$,
$\alpha = a,A,\bar A$, the $\Del$ derivatives of the effective
K\"ahler potential $\tK = \hK + 2\hL\hS$ satisfy
\bea \tK_\alpha &=& \(\pp_\alpha\tK\)_{\hL} = \(\pp_\alpha K\)_L +
2\hL\(\pp_\alpha S\)_L + \(\pp_\alpha L\)_{\hL}
\[K'(L) + 2\hL S'(L)\] - \(\pp_\alpha\[\Dk 
+ 2\hL\Del s\]\)_{\hL}\eee \(\pp_\alpha K\)_L + 2\hL\(\pp_\alpha S\)_L
+ \(\pp_\alpha\Dk/3\)_{\hL} \lbr L\[K'(L) + 2\hL S'(L)\] - 3 +
3\edk\rbr \eee \(\pp_\alpha K\)_L + 2\hL\(\pp_\alpha S\)_L.\eea
The D-terms in the scalar potential are determined in \myref{aux}  
by the scalar component of $\tK_a:$
\bea \l\tK_a\r &=& 2\sum_{A}q^a_A x^A(\ell,\del_A,\bdel_{\bar A}) -
\del_X\hel\del_{aX}\eee \del_X\del_{aX}\(\ell - \hel\) + 2\sum_A q^a_A
k^A \(\del^A + \bar\del^A\) + O(\del^2)\sim\del,\label{tkal} \eea
where $\del_{a X}$ is the Kronecker delta-function.  Since $(\pp_A
S)_L=0$, for $\alpha = A$ we have simply
\bea \l\tK_A\r &=& x^A(\ell,\del^A,\bdel^{\bar A})\label{tka2}.\eea 
%
 To evaluate the F-terms, we need the K\"ahler metric $\l\tK_{A\B}\r$.
In the following we define
\bea x^A &\equiv& x^A(\ell,\del_A,\bdel_{\bar A}) =
k^A(\ell)e^{\del^A + \bdel^{\bar A}}, \quad x'^A = \(\pp
x^A/\pp\ell\)_{\del},\quad etc., \nnn \ell &=& e^{\Dk/3}\hel,\quad
\ell_A = (\pp_A\ell)_{\hel}, \quad \lhel = (\pp_{\hel}\ell)_{\del}.
\label{xa}\eea
Differentiation of the $\theta=0$ component of \myref{wyl2} 
gives\footnote{In \cite{GG02a,GG02b} we parametrized the Weyl
transformation by the function $\alpha(\ell) = 3\dx c(\ell)/\ell$.}
\bea 0 &=& \lhel\[K''(\ell) + 2\hel\ts''(\ell) + 3/\ell^2 -
6\hel/\ell^3\] + 2\ts'(\ell) + {3/\ell^2}\eee \ell_A\[K''(\ell) +
2\hel\ts''(\ell) + 3/\ell^2 - 6\hel/\ell^3\] + x'^A,\nnn \ell_A &=&
x'^A{\lhel\ell^2\over3 - \ell\tk'(\ell)} =
\lhel c(\ell)x'^A,\label{lhl}\eea
since $\tk'(\ell) = - 2\ell\ts'(\ell)$. Then we obtain 
\beq \l\tK_{A\B}\r = \(\pp_{\B}\pp_A\tK\)_{\hel} =
\[\pp_{\B}x^A(\ell,\del^A,\bdel^{\bar A})\]_{\hel} = 
\del_{AB}x^A + c\lhel x'^Ax'^B,
\label{kab}\eeq
and the inverse metric $\tK^{A\B}$ is
\bea \tK^{A \B} &=& {\del^{AB}\over x^A} - {x'^Ax'^B\over
x^Ax^B}\[(\lhel c)^{-1} + \sum_C{(x'^C)^2\over
x^C}\]^{-1}\label{tkinv}.\eea
%
%
%
Above the condensate scale, quadratic terms in the $\Del_\alpha$
appear only through the combination of functionals $\tK = \hK +
2\hL\hS$.  When a gaugino condensate potential is added, quadratic
terms in $\Del_A$ appear through a different linear combination of
$\hK,\hS$.  Thus we need to evaluate these separately.  Since
$\(\pp_A S\)_L = 0$, we have
\bea \hK_A &=& \(\pp_A\hK\)_{\hL} = \tK_A - 2\hL\hS_A,\nnn \hS_A &=&
\(\pp_A\hS\)_{\hL} = \(\pp_A L\)_{\hL} S'(L) - \(\pp_A\Del s\)_{\hL},
\quad \Del s = 3\(L^{-1} - \hL^{-1}\)/2,\nnn \l\hS_A\r &=&
\ell_A\[\ts'(\ell) + 3\ell^{-2}/2\] = \ell_A\[3 - \ell\tk'(\ell)
\]/2\ell^2 = {\lhel\over 2}x'^A.\label{kasa}\eea
To study the dilaton potential we can drop terms of order $\del$
and set $x^A = k^A,;\hel = \ell$.  In this approximation the auxiliary
field $F^A$ in \myref{sols} is just:
\bea F_A &=& \sum_B\tK_{A\B}\bF^{\B} = -{\u\over4}\[\(k'^ A\(1 +
b_c\hel\) - p_A - b_c k^A\) + O(\del)\].\eea
It follows from the definitions \myref{tka2} and \myref{xa} and the
constraints \myref{tkal} and \myref{pcond} that
\bea \sum_Ap_A k'^A/k^A &=& \ell^{-1}\sum_A k'^A = \sum_A(k'^A)^2/k^A =
\dx h'_X(\ell),\label{prel}\eea
and therefore that 
\beq  \sum_A F_A x'^A/x^A = O(\del\u)\label{faxp}\eeq
is negligible in this approximation, giving
\bea \sum_A F_A F^A &=& {|u|^2\over16}\[w(k) - \(1 + b_c\hel\)^2 \dx
h'_X(\ell) + O(\del)\], \nnn w(x) &=& \sum_A{\(p_A + x^A b_c\)^2\over
x^A}.\label{sums}\eea
Setting the moduli at their self-dual points $F^I=0$,
the full potential for the dilaton is
\bea V &=& {|u|^2\over16}v(\ell) + O(\del^2,\del|u|), \quad v(\ell) =
w[k(\ell)] + \ell^{-1}k'(\ell)(1 + b_c\ell)^2 -
3b_c^2.\label{apot}\eea

To evaluate the \vev's of $\del$ and the D-terms $\tK_a$, we need only
retain terms linear in the coefficient of $|u|^2$, since
$\tK_a\sim\del\sim|u|^2$.  
It follows from \myref{faxp} that
\beq \sum_A F_A F^A = \sum_A|F_A|^2/x^A + O(\del^2),\eeq
Taking the $\ell$ derivative of \myref{tkal} gives
\beq 2\sum_A q^a_A x'_A = \tK'_a + \lhel^{-1}\dx\del_{a X} = \tK'_a +
\dx\del_{a X}\lhel^{-1}.\eeq
Then again using \myref{tkal} and \myref{pcond} we have
\bea \sum_A|F_A|^2/x^A &=& w(x) + 2(1+ b_c\hel)\sum_A\[(1+
b_c\hel)x'^A h'_a q^a_A - 2 p_A h'_a q^a_A - b_c x'^A\]\eee w(x) - (1+
b_c\hel)^2h'_X\dx\lhel\mmm + (1+ b_c\hel)\lhel\sum_a h'_a\[\lhel(1+
b_c\hel)\tK'_a - 2b_c\tK_a\].\label{fterm} \eea
To obtain the remaining contribution to $v(\hel,\del)$ we use the lowest
component of \myref{wyl2} which reads
\beq \hel - \ell = -\tc\sum_a h'_a(\ell)\tK_a,\qquad \tc(\ell) =
{\ell^2\over 3 - \ell k'(\ell)} = {1\over c^{-1} + \dx h'_X(\ell)},
\label{tilc}\eeq
%
\bea \hK'(\ell) &=&  k'(\ell) + \sum_a h'_a\tK_a + \dx h'_X\hel + 
3\pp_{\ell}\ln\(1 - \ell^{-1}\tc\sum_a h'_a\tK_a\).\label{hkp} \eea
In addition, from \myref{lhl} or \myref{tilc},
\bea \lhel &=& \[1 - \pp_\ell\(\tc\sum_a h'_a\tK_a\)\]^{-1}
= 1 + \pp_\ell\(\tc\sum_a h'_a\tK_a\) + O(\del^2),\label{lhel}\eea
so
\bea \lhel\hK'(\ell) &=& k'(\ell) + \(k'(\ell) - 3\ell^{-1}\)
\pp_{\ell}\(\tc\sum_a h'_a\tK_a\) + \sum_a h'_a\tK_a\(1 +
3\ell^{-2}\tc\) + \lhel\dx
h'_X\hel + O(\del^2)\eee k'(\ell) - {\ell\over\tc}
\pp_{\ell}\(\tc\sum_a h'_a\tK_a\) + \sum_a h'_a\tK_a\(2 + \tc
\ell^{-1}k'\) + \lhel\dx h'_X\hel +
O(\del^2),\label{lhelhkp}\eea
where in the last equality we used the definition of $\tc$ in
\myref{tilc}.  Terms in $\tK'_a$ and
terms proportional to $\dx h'_X$ cancel between \myref{fterm}
and \myref{lhelhkp} in the contribution $v(\hel,\del)$ to the
potential \myref{pot1}:
\bea v(\hel,\del) &=& \hel^{-1}\(1 + b_c\hel^2\)^2\lhel\hK'(\ell) +
\sum_A|F^A|^2/x^A - 3b_c^2 + O(\del^2)\eee\hel^{-1}\(1 + b_c\hel\)
\sum_a\tK_a\lbr h'_a\[\(1 + b_c\hel\)\(1 + {\tc k'\over\ell} -
{\ell\tc'\over\tc}\) - 2b_c\hel\] - \ell h''_a\rbr\mmm + w(x) + \(1 +
b_c\hel^2\)^2\hel^{-1}k'(\ell) - 3b_c^2 +
O(\del^2).\label{velldel}\eea

We may further expand \myref{velldel} using \myref{tilc} to write
\beq k'(\ell) = k'(\hel) + \tc\sum_a h'_a\tK_a k'' + O(\del^2), \qquad
w[x(\ell)] = w[x(\hel)] + \tc\sum_a h'_a\tK_a w' + O(\del^2).\eeq
Then using the minimization equation for $\hel$ and vanishing of the
vacuum energy
\bea v'(\myvev{\hel}) &\equiv& v'(\myvev{\hel},0) = \l\[w' + \(1 +
b_c\hel\)^2\hel^{-1}k''(\hel) - \hel^{-2}\(1 -
b_c^2\hel^2\)k'(\hel)\]\r_{\hel = \myvev{\hel}} = O(\del), \nnn
v(\myvev{\hel}) &\equiv& v(\myvev{\hel},0) = \l\[w + \(1 +
b_c\hel\)^2\hel^{-1}k'(\hel) - 3b_c^2\]\r_{\hel = \myvev{\hel}} =
O(\del),\label{vac}\eea
gives 
\bea v(\myvev{\hel},\del) &=& w[x\myvev{\hel})] - w[k\myvev{\hel})] +
\sum_a\tK_a A_a(\myvev{\hel}) + O(\del^2),\label{vheldel} \eea
where
\bea A_a(\ell) &=& \ell^{-1}\(1 + b_c\ell\) \lbr h'_a\[1 - b_c\ell +
2{\tc k'\over\ell} - \(1 + b_c\ell\){\ell\tc'\over\tc}\] - \ell
h''_a\rbr.\label{tkp} \eea
%
%
%
Then, with the dilaton at its vacuum value, the potential for $\del$
is
\bea V(\myvev{\hel},\del) &=& {1\over2s(\ell)}\sum_a\tK_a^2 +
{|u|^2\over16}v(\myvev{\hel},\del), \qquad v(\myvev{\hel},\del) =
\sum_A\(\del^A + \bar\del^A\)v_A(\myvev{\hel}) + O(\del^2),\nnn
v_A(\ell) &=& w_A(\ell) + \sum_a A_a(\ell)\tK_{a A},\qquad w_A(\ell) =
b_c^2k^A(\ell) - {p^2_A\over k^A(\ell)},\label{newpot}\eea
with, using \myref{tkal} and \myref{tilc},
\bea \tK_{a A} &=& (\pp_A\tK_a)_{\hel} = 2q^a_A k^A + \dx\del_{a
X}\tc\sum_b h'_b\tK_{b A} + O(\del)\eee 2q^a_A k^A + \del_{a
X}{\tc\dx k'^A\over1 - \tc\dx h'_X} = 2q^a_A k^A + \del_{a
X}c\dx k'^A.\label{tkaA}\eea
The vacuum conditions, in addition to \myref{vac}, are [using
\myref{lderiv}]
\bea {|u|^2\over16}v_A(\hel) &=& -{2\over s(\hel)}k^A(\hel)\sum_a
q_A^a\[\tK_a + c h'_a\dx\tK_X\] + O(\del^2), \quad
\tK_a\sim \del\sim |u|^2, \nnn \sum_A v_A(\hel) &=& -
{16c\dx\hel\over|u|^2 \tc s(\hel)}\tK_X + O(\del^2),\nnn \sum_A q^a_A
v_A(\hel) &=& -{32\over|u|^2 s(\hel)}\(\sum_b N^{ab}\tK_b +
{c\over4}\del^a_X\dx^2\tK_X\) + O(\del^2),\label{dterms}\eea
which may be  by inverted to evaluate the D-terms:
\beq \tK_X = -{|u|^2\tc s\over16c\dx\hel}\sum_A v_A, \quad \tK_a = -
{|u|^2 s\over32}\sum_{A,b}N^{-1}_{ab}q^b_A v_A - c\dx h'_a\tK_X,
\label{tkxa}\eeq
where we used \myref{lderiv}.  Consistency of these equations
for $a = X$ requires
\beq \sum_A v_A = 2\ell\sum_{A,a}q^a_A h'_a v_A,\label{consist}\eeq
which is automatically satisfied for the minimal models with $n=m$
discussed at the end of \mysec{redef}, since [see \myref{inft} and
\myref{minrel} below] $2\ell\sum_a q^a_A h'_a = \sum_{a,B}Q^B_a q^a_A
= 1$ in these models.
In general, we have from \myref{newpot}
\beq \sum_A v_A = \sum_A w_A + \dx\ell{c\over\tc}A_X(\ell).\eeq
Using the definition of $N_{a b}$ in \myref{nab} gives
\beq 2\sum_A q^b_A\tK_{a A} = 4\(N_{ab} + \del_{a X}c\dx\sum_c h'_c
N_{b c}\).\label{tkrel} \eeq
Then using the second equality in \myref{lderiv} gives
\bea 2\ell\sum_{A b}q^b_A h'_b\tK_{a A} &=& {c\over\tc}\ell\dx\del_{a
X}, \nnn 2\ell\sum_{A b}q^b_A h'_b v_A &=& 2\ell\sum_{A b}q^b_A h'_b
w_A + \dx\ell{c\over\tc}A_X(\ell),\eea
and the consistency condition \myref{consist} becomes
\beq \sum_A w_A = 2\ell\sum_{A,a}q^a_A h'_a w_A.\label{extra}\eeq
Making the same decomposition as in \myref{kakm}, the minimization
conditions \myref{miny} give 
\bea w_A &=& k^M\sum_A\zeta^A_M w_A/k^A,\nnn
\sum_A w_A + \sum_M w_M &=& \sum_A w_A\(1 + \sum_M\zeta^A_M k^M/k^A\)
= {\dx\ell\over2}\sum_A w_A Q^A_X/k^A, \eea
where the last equality follows from the conditions \myref{funct}.
In addition 
\bea &&2\ell\sum_a h'_a\(\sum_A q^a_A w_A + \sum_M q^a_M w_M\) =
2\ell\sum_a h'_{a A}w_A\(q^a_A + \sum_M q^a_M\zeta^A_M k^M/k^A\)\ddd = 
2\ell\sum_a h'_{a A}w_A/k^A\sum_b Q^A_b N_{b a} = {\ell\dx\over2}
\sum_A  w_A Q^A_X/k^A,\eea
where we again used the definition \myref{nab} of $N_{ab}$ and the
second equality in \myref{lderiv}.

\subsection{Minimal models}
For these models we have, at leading order in $\del$,
\bea k^A(\ell) &=& \ell p_A = {\dx\over2}Q^A_X\ell, \quad h'_a =
{1\over2\ell}Q_a = - \ell h''_a, \quad Q_a = \sum_A Q^A_a\nnn \del k'
&=& \sum_A k'^A = \dx h'_X\ell = p = \sum_A p_A, \quad \del k = \sum_A
k^A = \dx h'_X\ell^2, \nnn w(k) &=& \ell^{-1}p(1 + b_c\ell)^2 = \dx
h'_X(1 + b_c\ell)^2,\nnn A_a &=& A h'_a, \qquad {\ell A(\ell)\over1 +
b_c\ell} = -3b_c\ell + {\tc k'\over\ell}\(1 - b_c\ell\) - \tc
k''.\label{minrel}\eea
To simplify notation we also define
\beq z = b_c\myvev{\ell}, \quad \myvev{\ell^2h'_X}\dx =
Q_X\dx\myvev{\ell}/2 = 3\lambda z.\eeq
For example, $\lambda = 1$ in the FIQS model considered in the text.
Then at the vacuum, at leading order in $\del$, using \myref{nab} and
\myref{inft} we obtain from \myref{tkxa}
\bea \tK_a &=& {|u|^2s\over32\ell^2}Q_a\[{\tc\over c}(1 - z^2) + a(z)\], \qquad
a(z) = - \bigvev{\ell A(\ell)} = (1 + z)\(\bigvev{\ell\tc w'} - 3z\).\nnn
\bigvev{\ell\tc w'} &=& -{\lambda z(1-z^2)(1 + z)^2\over(1 + z)^2(1 +
\lambda z) - z^2}, \qquad {\tc\over c} = {1 + 2z\over(1 + z)^2(1 +
\lambda z) - z^2}.\label{tka}\eea
These results are unchanged if we replace a minimal set $k^A = \ell
p_A$ with $N$ minimal sets $k^i = \ell p^i_A$ with the same $U(1)$
charges such that $\sum_{i=1}^N p^i_A = p_A$.

\subsection{Additional $O(\del)$ \vev's}\label{addvev}

The parameterization of the K\"ahler potential in \myref{defth} is
valid only if $\CA\sim \del^0$.  Once we allow the D-terms to be
nonvanishing, $\tK_a|\sim \del$, we might expect additional order
$\del$ terms to occur from \vev's of fields $\Phi^M$ that are \uone\,
charged, but lie in F-flat directions, and have $p_M = 0$. Since these
fields acquire \vev's at the condensation scale where $t^I$ and $\ell$
are also determined, modular invariance, the linearity condition and
local supersymmetry are broken, and the formalism of
\cite{GG02a,GG02b} does not apply.  The exact treatment of the
minimization equations involves mixing among all the ``light'' fields
($\phi^M,t^I,\ell$).  However, for the purpose of determining the
parameters that define the vacuum, we can set to zero any field that
vanishes in the vacuum.  This allows us to parameterize the
contributions in a fashion analogous to the $\Phi^A$ with large \vev's
such that the above results still hold.  For the fields $\Phi^M,$
define
\bea x^M &=& \epsilon^M e^{(\Del^M + \bDel^{\M})}\prod_a e^{2q_M^a(h_a
+ \Del_a)} = \lvev e^{G^M}|\Phi^M|^2\rvev \prod_a e^{2q_M^a(h_a +
\Del_a)}, \nnn \epsilon^M&\sim& \del,\quad \Del^M| = \del^M \sim 1,
\quad k^M = 0.\eea
Then the results given above as expansions in $\del$ are unchanged.

\subsection{$O(\del^2)$ terms}

In order to determine if the extrema found above are true minimal, we
need to evaluate the terms quadratic in $\del$.  The same terms are needed
to evaluate masses of D-moduli in nonminimal cases.
Writing
\beq V = {1\over2s(\ell)}\sum_a\tK_a^2 + {|u|^2\over16}\[v(\hel) +
\sum_A\del^A v_A(\hel) + \half\sum_{A B}\del^A\del^B v_{AB}(\hel) +
O(\del^3)\], \eeq
and recalling that the vacuum conditions require $v(\hel),\;
\pp_{\hel}v(\hel),\; \tK_a\sim\del\sim |u|^2$,
we have
\bea V_{\hel\hel} &=& \mu^2_{\hel\hel} = {|u|^2\over16}v''(\ell) +
O(\del^2), \nnn
V_{\hel A} &=& \mu^2_{\hel A} = {1\over
s}\sum_a\[\tK_a\pp_{\hel}\tK_{aA} + \tK_{aA}\(\tK'_a
- {s'\over s}\tK_a\)\] + {|u|^2\over16}\(\pp_{\hel}v_A +
v_A\pp_{\hel}\ln|u|^2\) + O(\del^2),
\nnn V_{AB} &=& M^2_{AB} + \mu^2_{AB}, \quad M^2_{AB} = 
{1\over s}\sum_a\tK_{aB}\tK_{aA},
\nnn \mu^2_{AB} &=& {1\over s}\sum_a\tK_a\tK_{aAB} + {|u|^2\over16}
\(v_{AB} + v_A\pp_B\ln|u|^2 + v_A\pp_B\ln|u|^2\) + O(\del^2).
\label{2deriv}\eea
The matrix $V_{AB}$ is the mass matrix for the real fields $\Sigma^A$
introduced in \myref{jjrt}. In the minimal case, its properly
normalized eigenvalues are the squared masses of the \ua\, vector
bosons which are positive. In this case the only requirement
for the vacuum to be a local minimum is $v''(\ell)>0$; since
$M^2\sim \del^0$ and $\mu^2\sim\del$,
$\Det V =  v''\Det M^2 + O(\del^2)$, and similarly for any submatrices
on the diagonal.

For the general case, we can write the mass terms for the eaten
Goldstone bosons $\hat\Sigma_a$ and the D-moduli $\hSig^A$
defined in \myref{baA}--\myref{newu}:
\beq 
\Sigma^A = \hSig^A - 2\sum_a q_A^a\hSig_a, \quad
\sum_A q^A_a x^A\hSig^A = 0. \label{dmod}\eeq
We have 
\beq \tK_{aA} = 2x^A\(q^a_A + c\del_{aX}\dx\sum_b q^b_A h'_b\) =
2x^Aq^a_A + c\del_{aX}\dx x'^A, \quad
\sum_A\tK_{aA}\hSig^A = 0,\eeq
so the relevant squared-mass matrix for the light $\hel,\hSig^A$
sector is $\mu^2$, and mixing of these states with $\hSig_a$ is
negligible. From the last equality in \myref{dmod}, which also
implies 
\beq \sum_A x'^A\hSig^A = 0,\quad x^A = k^A + O(\del),\eeq
we may drop terms proportional to $q^a_A k^A,q^a_B k^B, k'^{A,B}$
in $\mu_{\hel A}$ and $\mu_{AB}$. Thus we may drop terms
proportional to $\tK_{aA}$.
This means in particular that in the expansion of $v(\hel,\del)$
to order $\del^2$ we may drop all terms of containing $\tK_a\tK_b$
or $\tK_a\tK'_b$.  Then there are no new relevant $O(\del^2)$
terms in the expression \myref{hkp} for $\hK'$. Since $\lhel\dx h'_X$
cancels between \myref{fterm} and \myref{lhelhkp}, the only relevant
additional terms in \myref{lhelhkp} are, using \myref{lhel},
\beq \[\lhel\hK'(\ell)\]_{\del^2} = \(k' -
3\ell^{-1}\)(\tc\sum_ah'_a\tK'_a)^2 = -
\ell\tc(\sum_a h'_a\tK'_a)^2.\label{del2}\eeq
From \myref{fterm} we get a contribution 
\beq \[\sum_A F_A|^2/x^A\]_{\del^2} = {|u|^2\over16}\[w_{\del^2} +
2\tc\(1 + b_c\ell\)^2(\sum_a h'_a\tK'_a)^2\].\label{del2f}\eeq
Finally, there is a contribution from the second term in \myref{tkinv}:
\beq F_A F^A\ni - \tc(\sum_A F_A x'^A/x^A)^2 + O(\del^3|u|^2) =
- {|u|^2\over16}\tc\(1 + b_c\ell\)^2(\sum_a h'_a\tK'_a)^2 +
O(\del^3|u|^2).\label{del2f2}\eeq
Further expansion of \myref{velldel} in $\ell - \hel$ gives only
terms that are higher order in $\tK_a$.  Projection
onto the D-moduli sector also gives
\bea 
 v_A &\to& w_A \qquad \tK'_{aA} \to
2k'^A\(q^a_A + c\dx\del_{aX}\sum_b q^b_A h'_b\),\nnn
\tK_{a AB} &=& \del_{AB}\tK_{a A} + \ell_B\pp_{\ell}\tK_{a A}\to 0,
\eea
since [see \myref{lhl}] $\ell_B \propto x'^B$.  Then collecting
\myref{del2}, \myref{del2f} and \myref{del2f2}, terms quadratic
in $\tK'_a$ also cancel, and $v_{A B}$ reduces to
\beq v_{A B} = w_{A B},\eeq
while the contributions to $\pp_{\hel}v_A$ are just the $\ell$-derivatives
of $v_A$ in \myref{newpot} with
\beq\pp_{\hel}\[A_a(\hel)\tK_{aA}\] \to A_a(\hel)\tK'_{aA} .\eeq
From \myref{sols}, \myref{kasa}, the first
condition in \myref{prel} and the expression \myref{delk} for $\del
k'$, we have
\bea \pp_{\hel}\ln|u|^2 &=& \tk' - 2s'/b_c - \sum_A p_A k'^A/b_c k^A =
\del k' + k'(b_c\ell -1)/b_c\ell - \dx h'_X/b_c \eee \tk'(b_c\ell
-1)/b_c\ell, \nnn
\pp_A\ln|u|^2 &=& - p_A/b_c + \hK_A - 2\hS_A/b_c +
\ell_A\pp_{\hel}\ln|u|^2 \to - p_A/b_c + k^A.\eea
Then using the vacuum conditions \myref{vac} and the 
results \myref{tkxa} we obtain the matrix elements of $\mu$.

These expressions simplify further for the case of $N$ copies of a
minimal set with vacuum values that satisfy $k^A = \ell p^A$,
$q^a_A k'^A = \ell^{-1}q^a_A k^A$. In this case we may also drop
$\tK'_a$ when evaluating the squared mass matrix for
$\hel,\hSig^A$.  Then the above expressions reduce to
\bea v_A&\to& - p^2_A/k^A + b^2_c k_A = -\ell^{-2}k^A\(1 - z^2\), \nnn
\pp_{\hel}v_A &\to& k'_A\[(p_A/k^A)^2 + b_c^2\] = \ell^{-2} k'^A\(1 +
z^2\)\to 0,\nnn v_{AB}&\to& \del_{AB}\ell^{-2} k^A\(1 + z^2\), \eea
and, using the vacuum value \myref{vac0} for $\tk'$, the relevant
elements of $\mu$ in \myref{2deriv} reduce to
\bea \mu^2_{\hel\hel} &=& {|u|^2\over16}v''(\ell),\quad \mu^2_{\hel
A} = {|u|^2\over16}\ell^{-2}k^A{(1 - z)^2\over 1+z},\nnn \mu^2_{AB}&=&
{|u|^2\over16}\[\del_{AB}\ell^{-2} k^A\(1 + z^2\) + {2\over
z}\ell^{-2}k^A k^B(1 - z)^2(1+z)\].  \eea
In the case under consideration, the condition \myref{sigcond} becomes
\bea 0 &=& \sum_{A,\alpha}q^a_A k^A_\alpha\hSig^A_\alpha =
\sum_{A,\alpha,a}Q^B_a q^a_A k^A_\alpha\hSig^A_\alpha \eee \sum_\alpha
k^B_\alpha\hSig^B_\alpha\quad \forall\; B,\label{dcond}  \eea
and there is no mixing of the dilaton with the D-moduli.  Defining
chiral fields $D^A$ as in \myref{ddef} and setting to zero the $m$
eaten Goldstone modes $\hSig_a$, the mass term for the
D-moduli\footnote{The prime on $\Sigma$ refers to the field
redefinitions make in \mysec{redef} and does not denote
differentiation with respect to $\ell$.} $\sigma'^A = \l\Sigma'^A\r$
or $d^A = \l D^A\r$ is given by
\beq \L_{m_D} = - \half\sigma'^A\mu^2\sigma'^B = - \half m^2_{\tG}
{1+z^2\over z^2}\sum_{A=1}^nk^A(d^A + \bar d^{\bA})^2.\eeq
The K\"ahler potential for the D-moduli is\footnote{The
second term in Eq. (3.33) of~\cite{GG02b} is missing a factor
1/2}
\beq K(D,\bar D) = \half\sum_{A=1}^nk^A(D^A + \bar
D^{\bA})^2.\label{dkahl}\eeq
When we reexpress the $d^A$ in terms of an orthonormal set
$D^i$ subject to the constraint \myref{dcond}:
\beq D^A = \sum_{i=1}^{n-m}c^A_i D^i, \quad
\sum_A k^A c^A_i = 0,\label{dadi} \eeq
the K\"ahler metric and the squared mass matrix 
\beq K_{i\bj} = \sum_A c^A_i k^A c^A_{j}, \quad \mu_{ij}^2 =
m^2_{\tG} {1+z^2\over z^2}\sum_A c^A_i k^A c^A_{j},\eeq
are diagonalized by the same unitary transformation:
\beq D^i\to D^i = U^i_j D^j, \quad d^i = \l D^i\r = N_d(\sigma^i + i
a^i),\label{diprime}\eeq
where the normalization constant $N_d$ is chosen to make the kinetic
energy term canonically normalized. Then the Lagrangian quadratic in
the scalar D-moduli reads
\beq \L_D = \half\sum_i\[\pp_\mu\sigma^i\pp^\mu\sigma^i + 
\pp_\mu a^i\pp^\mu a^i - m^2_{\tG} {1+z^2\over z^2}(\sigma^i)^2\].
\label{ld}\eeq

\subsection{Scalar masses}\label{massapp}

For fields $\Phi^M$ with vanishing \vev's and K\"ahler potential
\beq K = \sum_M x^M, \quad x^M = e^{G'^M + 2\sum_aq^a_M
h_a}|\Phi^M|^2,\eeq
referring to \myref{m2M}, the complex scalars have masses
\beq m^2_M = {\pp V\over\pp x^M} = V_M = {|u|^2\over16}v_M
+ {1\over s}\sum_a\tK_a\tK_{aM}, \eeq
where everywhere the subscript $M$ denotes differentiation
with respect to $x^M$. We have 
\beq \tK_M = \hK_M + 2L\hS_M = K_M + 2L S_M = K_M = 1.\eeq
Since $x^M$ and $\del$ appear in the functionals $\tK$ and $\hS$ in
the same way, the terms linear in $x^M$ can be directly extracted from
the formulate for those linear in $\del_A$. The $x^M$ derivatives are
obtained from the $\del^A$ derivatives by the replacements
\beq K_A = k^A\to K_M = 1,\quad K'_A = k'^A\to K'_M = {x'^M\over x^M} =
2\sum_a h'_a q^a_M \equiv 2h'_M, \quad p_A \to0.\label{subs}\eeq
Thus we obtain from \myref{newpot}, to zeroth order in $\del$:
\bea v_M &=& b^2_c + \sum_a A_a(\ell)\tK_{aM},\qquad \tK_{aM} = 2q^a_M
+ 2c\del_{aX}\dx h'_M.\label{vmrels}\eea
Using \myref{tkxa} and \myref{tkrel} to solve for the D-terms
in the vacuum gives
\beq \tK_a = -{|u|^2s\over16}\(\sum_A w_A B_{A a}+ A_a\), \qquad
B_{A a}(\ell) = \half\sum_b N^{-1}_{a b}q^b_A - {\tc h'_a\over\ell}.
\label{tkagen}\eeq
The functions $A_a(\ell)$ drop out of the scalar masses:
\beq m^2_M = {|u|^2\over16}\[b_c^2 - \sum_{A a}w_A B_{A a}\tK_{a
M}\].\eeq
For the minimal models \myref{tkagen} reduces to \myref{tka}, and 
\bea v_M &=& b_c^2 - \ell^{-1}a(z)\sum_ah'_a\tK_{aM} = b^2_c - 2h'_M{c
a(z)\over\tc}, \nnn \sum_a\tK_a\tK_{aM} &=& {|u|^2s\over8\ell}h'_M\[1
- z^2 + {c a(z)\over\tc}\],\eea
giving
\bea m^2_M &=& {|u|^2s\over16\ell}\[\ell b_c^2 + h'_M(1 - z^2)\] =
\(\zeta_M{1 - z^2\over z^2} + 1\)m^2_{\tG},\nnn \zeta_M &=& \sum_{a,A}
q^a_M Q^A_a.\label{minmass}\eea

\subsection{Nonminimal K\"ahler potential for matter}\label{nmapp}
First consider the toy model with just one charged superfield $\Phi$.
Following~\cite{GG02a} set 
\beq \del K \equiv K(\Phi,\Ph) = f(x),\quad x = e^{G_q +
2qV}|\Phi|^2 = e^{G_q + 2q V'} = e^{2qU '}.\eeq 
We require $\lvev D_X\rvev= 0$ at the Planck scale, where 
\beq D_X = qK_\Phi\Phi - {\dx L\over2} = qxf'(x) - {\dx L\over2}, 
\quad \lvev D_X\rvev_L = \lvev qe^{2qh}f'(qe^{2qh}) - {\dx L\over2},
\rvev_L\eeq 
where 
\beq U' = h(L) + U, \quad \lvev U\rvev = 0.\eeq 
Then we have 
\bea \del k &=& f(e^{2qh}), \quad \del s = - {\dx h\over2}, \nonumber \\
\del k' &=& 2qh'e^{2qh}f'(e^{2qh}) = h'\dx L, \quad \del 
s' = - {\dx h'\over2}, \quad \del k' + 2L\del s' = 0.\eea
So the Einstein condition is again satisfied for $U=0$.  Next consider 
the terms linear in $U$: 
\beq \del K = f(x) = \del k + U\left.{\pp f\over\pp U}\right|_{U=0},
\quad \left.{\pp f\over\pp U}\right|_{U=0} = 2qe^{2qh}f'(e^{2qh}) =
\dx L, \quad \del S = \del s - {\dx\over2}U.\eeq
The linear terms are the same as in~\cite{GG02a}, and are removed
by the same Weyl transformation.

Next consider the general case.  Allowing for $O(|u|^2)$ D-terms, we
have
\bea \del K &=& f(x^A), \quad D_a = \sum_Aq^a_Ax^Af_A(x) - {\dx
L\over2} \del_{Xa}, \nnn f_A &=& {\pp f\over\pp x^A}, \quad x^A =
|\Phi^A|^2e^{G^A + 2\sum_aq^a_AV_a},\nonumber \\ \lvev
x^A\rvev_{(L,T)} &=& |C_A|^2e^{d_A + \bar d_A}\prod_ae^{2q^a_A(h_a(L)
+ \Del_a)} = k^Ae^{\tilde\Del_A}, \quad \tilde\Del_A = d_A + \bar d_A
+ 2\sum_a q^a_A\Del_a = O(|u|^2), \nonumber \\ 0 &=& \sum_A
q^a_A k^Af_A(k) - {\dx L\over2}\del_{Xa}, \quad \del k = f(k^A) +
O(|u|^2), \nonumber \\ \del k'(k) &=& 2\sum_{A,b}q_A^b h'_b k^Af_A(k)
= \dx L h'_X = -2L\del s'.\eea
To evaluate the condensate-induced potential, we set $x^A = k^A
e^{\tilde\Del_A}$ and expand in $\tilde\Del_A$ as before:
\bea \tK_A &=& K_A = \left.{\pp f\over\pp d_A}\right| =
x^Af_A(x), \quad \tK_a = 2\sum_A q^a_A K_A - \hel\dx\del_{a X},
\nonumber \\ \hK_A &=& \tK_A - 2\hL\hS_A, \quad \hS_A =
{\lhel\over2}K'_A, \quad \tK_{A\B} = K_{A\B} + c\lhel K'_A K'_B,\nnn
K_{A\B} &=& K_{AB} = K_A\del_{AB} + x^Ax^Bf_{AB}, \quad f_{AB} =
{\pp^2f\over\pp x^A\pp x^B},\nnn
F^A &=& - {\u\over4}\tK^{A\B}\[(1 + b_c\hel)\lhel K'_B - p_B
-K_Bb_c\].\eea
To obtain the potential we need the inverse $\tK^{A\B}$ of
$\tK_{A\B}$; defining $K^{A\B}$ to be the inverse of $K_{A\B}$, we
obtain
\beq \tK^{A\B} = K^{A B} - \sum_{C,D}K^{A C}K'_C K^{B
D}K'_D\[(c\lhel)^{-1} + \sum_{E,F}K^{E F}K'_E K'_F\]^{-1}.\eeq
The $K_A(k)$ satisfy the same constraints as $k^A$ in the case of a
minimal K\"ahler potential, although now $K_A(x)\ne x^A$ in
general. The potential is the same as before except for the
replacements $x^A\to K_A$, $\del^{AB}/x^A \to K^{AB}$. In particular,
the relations
\bea 2\sum_A q_A^a K_A &=& \dx\del_{aX}\hel + \tK_a, \quad K'_A =
2\sum_{B,b}h'_b q^b_B K_{AB}, \nnn 2\sum_A q_A^a K'_A &=& 4\sum_{A B
b}q_A^a q^b_B h'_b K_{A B} = \lhel^{-1}\dx\del_{aX} + \tK'_a, \nnn
\del k' &=& 2\sum_{A,a}h'_aq^a_A K_A = h'_X\dx\hel + \sum_a
h'_a\tK_a,\label{nonminrels}\eea
give
\bea \sum_{A B}K^{A\B}K'_A K'_B &=& 2\sum_{A,a}h'_a q^a_A K'_A = h'_X\dx
+ \sum_a h'_a\tK'_a,\nnn \sum_{A B}K^{A\B}K'_A K_B &=& 2\sum_{A,a}h'_a
q^a_A K_A = h'_X\dx\hel + \sum_a h'_a\tK_a,\nnn \sum_{AB}K^{AB}K'_A
p_B &=& 2\sum_{A,a}h'_a q^a_A p_A = h'_X\dx.\label{nmrels} \eea
Thus
\beq \sum_A F_A K^{A\B}K'_B = O(\u\del),\eeq
and we obtain the result in \myref{apot} with
\beq w = K^{A\B}\(p_A + K_A b_c\)\(p_B + K_B b_c\),\eeq
which in minimal models reduces to
\beq w = K^{A\B}\(1 + b_c\ell\)^2p_A p_B.\label{neww}\eeq
In this case we can invert the equations in \myref{nonminrels} to
obtain, dropping order $\del$ terms,
\beq 2K_A = 2\ell p_A = 2\ell K'_A = \dx Q^A_X\ell, \quad \sum_B
K^{A\B}p_B = 2\sum_a h'_a q^a_A, \quad \sum_{A B}K^{A\B}p_A p_B = \dx
h'_X.\label{nmmin}\eeq
Then in these models
\beq w = (1 + z)^2\dx h'_A + O(\del),\qquad \sum_{A B}F_A\tK^{A\B}\bF_B
= \sum_{A B}F_A K^{A\B}\bF_B + O(\del^2) = O(\del),\eeq
as before, and the dilaton potential is unchanged with respect to the
case of a minimal K\"ahler potential for matter.  Using \myref{nmrels}
for the general case, the expression for $\sum_{A B}F_A K^{A\B}\bF_B$
is the same as the right hand side of \myref{fterm}, with $w(x)$ now
given by \myref{neww}, and we obtain \myref{velldel}--\myref{tkp} with
the same substitution.  The results \myref{newpot}--\myref{tkxa} and
\myref{tkagen} are modified as follows:
\bea w_A &=& 2b_c\(p_A + b_c K_A\) - K^{D\bar C}K_{\bar C A E}
K^{E\B}\(p_D + b_c K_D\)\(p_B + b_c K_B\),\nnn
\tK_{aA} &=& 2\sum_B q^a_B K_{A B} + c\del_{aX}\dx K'_A + O(\del)
,\nnn
{|u|^2\over16}v_A(\hel) &=& -{2\over s(\hel)}\sum_{a,B}K_{AB}
q_B^a\[\tK_a + c h'_a\dx\tK_X\] + O(\del^2), \nnn \sum_{A B}K_B K^{B
A}q^a_B v_A(\hel) &=& -{32\over|u|^2 s(\hel)}\(\sum_b N^{ab}\tK_b +
{c\over4}\del^a_X\dx^2\tK_X\) + O(\del^2),\nnn \sum_{A B}K_B
K^{B A}v_A(\hel) &=& - {16c\dx\hel\over|u|^2 \tc s(\hel)}\tK_X +
O(\del^2),\qquad N^{a b} = \sum_A q^a_A q^b_A K_A(k),\nonumber \eea
\bea \tK_a &=& - {|u|^2 s\over32}\sum_{A B b}K_B K^{B
A}N^{-1}_{ab}q^b_B v_A - {c\over4}\dx^2N_{a X}^{-1}\tK_X = -
{|u|^2s\over16} \(\sum_A B_{A a}w_A + A_a\),\nnn \tK_X &=& -{|u|^2\tc
s\over16c\dx\ell}\sum_{A B}K_B K^{B A}v_A, \quad B_{A a} = \sum_{B}K_B
K^{B A}\(\sum_b N_{a b}^{-1}q^b_B - {\tc\over2\ell}\dx N_{a X}^{-1}\),
\label{nmka}\eea
where we used, instead of \myref{tkrel},
\beq \sum_{AB}K_B K^{B A}q^a_B\tK_{A a} = 2\sum_b N_{a b}\(1 +
c\dx\del_{a X} h'_b\),\quad \sum_{AB}K_B K^{B A}\tK_{A a} =
{c\over\tc}\dx\del_{a X}. \eeq 
With a K\"ahler potential of the form \myref{newkahl}
\bea K^\alpha_A &=& x^A_\alpha e^{K^\alpha/C_\alpha}, \quad
K^\alpha_{A B} = K^\alpha_A\del_{A B} + C_\alpha^{-1}K^\alpha_A
K^\alpha_B, \nnn K_\alpha^{AB} &=& {1\over K^\alpha_A}\del^{AB} -
{1\over C_\alpha + \sum_CK^\alpha_C},\nnn
w^\alpha_A &=& b_c^2 K^\alpha_A - (p^\alpha_A)^2/K^\alpha_A -
C^{-1}_\alpha\(p^\alpha_A + b_c K^\alpha_A\)w_\alpha,\nnn w &=&
\sum_\alpha w_\alpha, \quad w_\alpha = \sum_A{1\over
K^\alpha_A}(p^\alpha_A + b_c K^\alpha_A)^2 - 
{\[\sum_A(p^\alpha_A + b_c K^\alpha_A)\]^2\over C_\alpha + \sum_B
K_B^\alpha}.\label{nmkahl}\eea
For a minimal set \myref{nmmin} we have $N^{-1}_{a b} = \sum_A Q^A_a
Q^A_b/K_A$, $N_{a X}^{-1} = 2Q_a/\dx\ell$, and we obtain
\bea w_\alpha &=& {C_\alpha p_\alpha(1+z)^2\over\ell\(C_\alpha + \ell
p_\alpha\)}, \qquad p_\alpha = \sum_A p^\alpha_A =
Q_X^\alpha{\dx\over2}\equiv{3z\lambda_\alpha\over\ell},\qquad
Q_a^\alpha = \sum_A Q^{A\alpha}_a,\nnn w^\alpha_A &=& \sum_B
K^\alpha_{A B}{\pp w^\alpha\over\pp K^\alpha_B} = \ell^{-2}\sum_B
K^\alpha_{A B}g_\alpha, \qquad \sum_A B_{A a}w_A =
{1\over\ell^2}\sum_\alpha g_\alpha\(Q^\alpha_a - {\tc\over\ell}Q_a
p_\alpha\)\nnn g_\alpha &=& z^2 - 1 + 3z^2(1+z)\lambda_\alpha{3
\lambda_\alpha(1-z) - 2C_\alpha\over\(C_\alpha +
3z\lambda_\alpha\)^2}.\eea
Finally we can solve for $h'_a$ using
\bea K^\alpha_A &=& {k^A_\alpha\over1 - C^{-1}_\alpha\sum_B
k^B_\alpha} = \ell p^\alpha_A,\qquad k^A_\alpha = {\ell
p^\alpha_A\over1 + C^{-1}_\alpha\ell p^\alpha_A},\nnn h'_a &=&
{1\over2}\sum_{A\alpha}Q^{A\alpha}_a{k'^A_\alpha\over k^A_\alpha} =
{1\over2\ell}\sum_\alpha{Q^\alpha_a\over1 + C^{-1}_\alpha\ell
p_\alpha}, \qquad h'_X = {1\over\ell\dx}\sum_\alpha{C_\alpha
p_\alpha\over C_\alpha + \ell p_\alpha}.\eea
Then at the vacuum
\beq c = {\ell^2\over3 - \ell k' - \dx\ell^2 h'_X} = {\ell^2\over3 +
(\ell^2w - 3z)/(1+z)^2 - \dx\ell^2 h'_X} =
{\ell^2(1+z)^2\over3(1+2z)}.\eeq
is unchanged with respect to the case of a minimal K\"ahler potential.
When scalars $\Phi^M_\beta$ with vanishing \vev's are included in the
K\"ahler potential \myref{newkahl}, we have (in order $\del^0$)
\beq \tK^\beta_M = \hK^\beta_M + 2L\hS^\beta_M = K^\beta_M =
\(\pp^\beta_M K^\beta\)_{\hel} = {1\over1 - C_{\beta}^{-1}\sum_A
 k^A_\beta} = 1 + C^{-1}_{\beta}\ell p_\beta,\label{km2}
\eeq
The scalar masses are now given by
\bea m^2_{M\beta} &=& (K^\beta_M)^{-1}{\pp V\over\pp x_\beta^M} =
(K^\beta_M)^{-1}V^\beta_M, \nnn V^\beta_M &=& {|u|^2\over16}v^\beta_M
+ {1\over s}\sum_a\tK_a\tK_{a,M\beta},\qquad
v^\beta_M = w^\beta_M + \sum_a A_a\tK_{a,M\beta}. \eea
As before the terms the terms proportional to $\sum_a\tK_a\tK_{a M}$
cancel out, and we obtain 
\beq m_{M\beta}^2 = (K^\beta_M)^{-1}{|u|^2\over16}\(w^\beta_M - \sum_{A\beta
a}w^\beta_A B_{A\beta a}\tK_{a M}\).\eeq
We now have
\bea w^\beta_M &=& 2b_c^2K^\beta_M - K_\beta^{D\bar C}K^\beta_{\bar C
M E}K_\beta^{E\B}\(p^\beta_D + b_c K^\beta_D\)\(p^\beta_B + b_c
K^\beta_B\)\eee \sum_B K^\alpha_{M B}{\pp w^\alpha\over\pp K^\alpha_B}
= \ell^{-2}K^\beta_M\[1 + g_\beta\(1 + C^{-1}_\beta\ell p_\beta\)\] +
O(x_M),\nnn \tK_{a,M\beta} &=& 2\sum_B q^a_{B\beta}K^\beta_{M B} +
c\del_{aX}\dx K'^\beta_M + O(\del,x_M)\eee
2K^\beta_M\sum_b\(q^b_{M\beta} + C^{-1}_\beta\sum_A
q^b_{A\beta}K^\beta_A\)\(\del_{a b} + c\del_{a X}\dx h'_b\),\eea
where in the first equalities the sums are over all chiral multiplets,
and in the second equalities we specialized to the K\"ahler potential
\myref{newkahl}.  For the minimal case we obtain
\bea {m^2_{M\beta}\over m^2_{\tG}} &=& {1\over z^2}\[1 + g_\beta\(1 +
C^{-1}_\beta\ell p_\beta\)\]\mmm - {1\over z^2}
\sum_\alpha\(\zeta^\alpha_{M\beta} + C_\beta^{-1}\ell\sum_A
p^\beta_A\zeta^\alpha_{A\beta}\)\sum_\gamma \[\del_{\alpha\gamma} +
{3c z\lambda_\gamma C_\alpha\over\ell^2\(C_\alpha + 3z\lambda_\alpha\)}\]
\(g_\gamma - {3z\tc\over\ell^2}\sum_\del\lambda_\del g_\del\), 
\nonumber \eea \bea \zeta_{M \beta}^\alpha &=& \sum_a
q^a_{M \beta}Q^{\alpha}_a, \quad\zeta_M^\beta = \sum_\alpha\zeta_{M
\beta}^\alpha, \quad\zeta_{A\beta}^\alpha = \sum_a
q^a_{A\beta}Q^{\alpha}_a.\label{nmmass} \eea
It straightforward to check that \myref{nmmass} reduces to
\myref{minmass} for $C_\alpha\to\infty,\;3z\lambda_\alpha\to \ell^2\dx
h'_X$. In the examples considered in the text, each sector $\alpha$
with some $k^\alpha_A\ne0$ includes a complete minimal set with
charges $q^a_{A\alpha} = q^a_A$, so $\zeta^\alpha_{A\beta} = 1$.

\mysection{Fermion masses}\label{ferm}

The general expression for the Yukawa couplings of the fermionic
superpartners of the D-moduli
\beq \chi^A = {1\over\sqrt{2}}\l\D D^A\r \eeq
is somewhat complicated and involves
the reparameterization connection derived from the effective
K\"ahler potential $\tK$:
\beq \Gamma^A_{B C} = \tK^{A D}\tK_{D B C}\label{dcon}.\eeq
However, due to the condition \myref{dcond}, in the simple model
considered here, the Yukawa term simplifies considerably since we can
drop terms proportional to $\sum_A k^A\chi^A$, and $k^A = \ell p^A =
\ell k'^A,$ $k''^A = 0$. Therefore the connection \myref{dcon} drops
out, as do terms proportional to $\sum_A W_A\chi^A$ and
$\sum_A\hK_A\chi^A$.  Defining
\beq \chi^\alpha = {1\over\sqrt{2}}\l\D \Pi^\alpha\r,\quad
\chi_c = {1\over\sqrt{2}}\l\D U\r, \quad
\chi_\ell = \sqrt{2}\l\D L\r, \quad
\chi^I = {1\over\sqrt{2}}\l\D T^I\r, \eeq
the Yukawa couplings take the form\footnote{A factor $u_{(a)}$ is
is missing from the second and third term on the right hand side of
 (D.2) in~\cite{joel}.}~\cite{joel} (in the gauge $\sigma_m\psi^m=0$
 for the gravitino)
\bea \L_Y &=& - \half\sum_{A B}\(\chi^A\chi^B\)\lbr e^{\hK/2}\(W_{A B}
+ \hK_{A B}W\) + {u\over4}\[2\hS_{A B}\(1 - 2b_c\ell\) - b'_c\hK_{A
B}\]\rbr\mmm + \sum_\alpha\lbr\(\chi^\alpha\chi^\alpha\)
{u\over8}{b^\alpha_c\over(\pi^\alpha)^2} -
e^{\hK/2}\[\sum_I\(\chi^\alpha\chi^I\)\(W_{\alpha I} + \hK_I
W_\alpha\) + \sum_A\(\chi^\alpha\chi^A\)W_{\alpha A}\]\rbr\mmm -
e^{\hK/2}\lbr\half\sum_{I J}\(\chi^I\chi^J\)\[W_{IJ} + \(2W_I\hK_J +
\hK_I\hK_J W\)\(1 - \del_{IJ}\)\] + \sum_{IA}\(\chi^A\chi^I\)W_{A
I}\rbr\mmm + {1\over4}\sum_I\(\chi^I\chi_c\)\[(b'_c - b_I)\hK_I +
2b^I_c\zeta(t^I)\] + {1\over16\ell}\[2(1 + b'_c\ell)\tk' + \tk''\ell +
{(\tk')^2\ell\over\tk'\ell - 3}\]\(\chi_c\chi_\ell\) \mmm +
{u\over8}\sum_I\(\chi^I\chi^I\)\(\pp_I - 2\hK_I\)\lbr\[b'_c -
b_I(1-2b_c\ell)\]\hK_I + 2b^I_c\zeta(t^I)\rbr\mmm -
{1\over8}\(\chi_\ell\chi_\ell\)\lbr{u\over4\ell^2(3 -
\ell\tk')}\[3\tk' - \ell(\tk')^2(1 - b_c\ell) - \ell\tk''\(3 -
2\ell\tk'\)(1 + b_c\ell)\]\rbr \mmm\qquad\qquad\qquad\l + \(\tk'' +
\tk'^2\)e^{\hK/2}W - {b'_c u\over4}\tk''\rbr -
\sum_\alpha{b^\alpha_c\over4\pi^\alpha}\(\chi^\alpha\chi_c\) -
{b'_c\over8u}\(\chi_c\chi_c\)\mmm - \half e^{\hK/2}\tk'\[\sum_\alpha
W_\alpha\(\chi_\ell\chi^\alpha\) + \sum_I\(W_I + \hK_I
W\)\(\chi_\ell\chi^I\)\] + {\rm h.c.},\label{LY}\eea
where we used
\beq \Gamma^I_{II} = 2K_I = 2\hK_I = - (\re t^I)^{-1},\quad
\hK_{IJ} = \del_{IJ}\hK_I^2,\label{KI} \eeq
and we dropped terms proportional to $F^I$ which vanishes in the vacuum.
Eliminating the static fields $\chi^\alpha,\chi_c$ by their equations
of motion gives 
\bea \chi^\alpha &=& {\pi^\alpha\over u}\chi_c + {4(\pi^\alpha)^2\over
 u b^\alpha_c}e^{\hK/2}\[\sum_I\chi^I\(W_{\alpha I} + \hK_I W_\alpha\)
 + \sum_A\chi^A W_{\alpha A} + {\tk'\over2}\chi_\ell W_\alpha\],\nnn 
\chi_c &=& {u\over4b'_c\ell}\[2(1 + b'_c\ell)\tk' + \tk''\ell +
{(\tk')^2\ell\over\tk'\ell - 3}\]\chi_\ell\mmm + {u\over
b'_c}\sum_I\[\hK_I(b'_c - b_I) + 2b_c^I\zeta(t^I)\]\chi^I -
\sum_\alpha{u b^\alpha_c\over b'_c\pi^\alpha}\chi^\alpha \eee
\(1 - {b_c\over b'_c}\)\chi_c + {u\over4b_c'\ell}\(2\tk'\(b_c\ell +1\)
+ \tk''\ell + {(\tk')^2\ell\over\tk'\ell - 3}\)\chi_\ell\mmm + {u\over
b'_c}\sum_I\chi^I\[(b_c - b_I)\hK_I + 2\(b_I + p_I -
b_c\)\zeta(t^I)\],\label{statchi} \eea
where we used the equation of motion for $F^\alpha$ in \myref{aux},
the constraints \myref{qconds} and the definitions \myref{pdefs2};
in particular
\beq \sum_{A,\alpha}b^\alpha_c q^A_\alpha\chi^A = \sum_A p_A\chi^A 
= 0. \label{pchi}\eeq
Evaluating \myref{statchi} at the moduli vacuum values
\beq F^I = 0 = 1 + 4\re t^I\zeta(t^I), \quad \hK_I = 2\zeta(t^I),
\label{modvac}\eeq
gives
\bea 
X^\alpha &\equiv& e^{\hK/2}W_\alpha\chi^\alpha = {b^\alpha_c
u\over4}\[2\sum_K\chi^I\zeta(t^I)\(q^\alpha_I + p^\alpha_I\) 
+ \sum_A\chi^A q_\alpha^A + {\tk'\over2}\chi_\ell -
{\chi_c\over u} \]\eee {b^\alpha_c u\over4}\[2\sum_I\chi^I
\zeta(t^I)\(q^\alpha_I + p^\alpha_I - {p_I\over b_c}\) +
\sum_A\chi^A q_\alpha^A - {1\over4b_c\ell}\(2\tk' 
+ \tk''\ell + {(\tk')^2\ell\over\tk'\ell - 3}\)\chi_\ell\], \nnn
\chi_c &=& {u\over4b_c\ell}\[2\tk'\(b_c\ell +1\)
+ \tk''\ell + {(\tk')^2\ell\over\tk'\ell - 3}\]\chi_\ell + 
{2u\over b_c}\sum_I\chi^I p_I\zeta(t^I). \eea
Using \myref{qconds} and \myref{pchi}, we have
\bea e^{\hK/2}W_{A B} &=& - {u\over4}\sum_\alpha b^\alpha_c q^\alpha_A
q^\alpha_B, \quad e^{\hK/2}W_{A I} = - {u\over2}\sum_\alpha b^\alpha_c
q^\alpha_A\(q^\alpha_I + p^\alpha_I\) ,\nnn e^{\hK/2}W_{I J} &=& -
u\zeta(t^I)\zeta(t^J)\[\sum_\alpha b^\alpha_c \(q^\alpha_I +
p^\alpha_I\)\(q^\alpha_J + p^\alpha_J\) - 2b_I - 2p_I + 2b_c^I + b_c +
b'_c\]\ddd\qquad\qquad\qquad\qquad - {u\over2}\zeta'(t^I)\(b_I + p_I
-b_c - b^I_c\)\del_{I J}.\eea
Then defining
\bea \L_\alpha &\equiv& -\half\sum_\alpha
X^\alpha\lbr2\sum_I\chi^I\zeta(t^I)\(q^\alpha_I + p^\alpha_I\) +
\sum_A \chi^A q^\alpha_A + \chi_\ell{\tk'\over2}\rbr \eee
e^{\hK/2}\[\half\sum_{IJ}\(\chi^I\chi^J\)W_{IJ} +
\half\sum_{AB}\(\chi^A\chi^B\)W_{AB} +
\sum_{IA}\(\chi^I\chi^A\)W_{IA}\]\mmm - {u\over2b_c}
\sum_{IJ}\(\chi^I\chi^J\)\zeta(t^I)\zeta(t^J)\[\(b_I + p_I - b^I_c\)
\(2b_c - p_J\) - b^2_c - b'_c\(b_c - p_J\)\]\mmm +
{u\over4}\sum_I\(\chi^I\chi^I\)\zeta'(t^I)\(b_I + p_I -b_c -
b^I_c\)\mmm + {u\over16b_c\ell}\sum_{I}\zeta(t^I)\lbr(b_c-b'_c)\(2\tk'
+ \tk''\ell + {(\tk')^2\ell\over\tk'\ell - 3}\)\right.\ddd
\qquad\qquad\qquad\qquad\l - 2\ell\tk'\[b_c\(b_I
- b'_c - b_c^I\) + p_I b'_c\]\rbr\(\chi^I\chi_\ell\)\mmm +
{u\tk'\over64b_c\ell}(b_c-b'_c)\(2\tk' + \tk''\ell +
{(\tk')^2\ell\over\tk'\ell - 3}\)\(\chi_\ell\chi_\ell\)\nnn
\L_c &\equiv& {1\over8}\lbr\sum_I\(\chi^I\chi_c\)\[(b'_c - b_I)\hK_I +
2b^I_c\zeta(t^I)\] + {b'_c\tk'\over2}\(\chi_c\chi_\ell\)\rbr\eee
{u p_I\over2b_c}\sum_{IJ}\(\chi^I\chi^J\)\(b^I_c - b_I + b'_c\)
\zeta(t^I)\zeta(t^J) \mmm +
{u\over16b_c\ell}\sum_I\(\chi^I\chi_\ell\)\lbr\[2(1 + b_c\ell)\tk' +
\tk''\ell + {(\tk')^2\ell\over\tk'\ell - 3}\]\(b^I_c - b_I + b'_c\) +
2\ell b'_c p_I\rbr\zeta(t^I),\mmm
+
{ub'_c\tk'\over64b_c\ell}\[2(1 + b_c\ell)\tk' + \tk''\ell +
{(\tk')^2\ell\over\tk'\ell - 3}\]\(\chi_\ell\chi_\ell\) \eea
we see that the second derivatives of the superpotential $W$ drop out
of the Yukawa coupling:
\beq \L_Y = \L\(\chi^A,\chi^I,\chi_\ell\) + \L_\alpha +
\L_c + {\rm h.c.} = \L(\chi^A) + \L\(\chi^I,\chi_\ell\) + {\rm h.c.}. \eeq
Referring to \myref{kab} and \myref{kasa}, the D-fermion masses are
determined by 
\beq \L(\chi^A) = - {u\over8}\sum_A\(\chi^A\chi^A\) \[k'^A\(1 -
3b_c\ell\) - b_c k^A\] = - {m_{\tG}\over2z}\sum_A\(\chi^A k^A\chi^A\)
\(1 - 4z\), \eeq
Here we evaluated (at leading order in $\del$) $\l S_{A B}\r$ from
\myref{kasa} using \myref{lhel} to obtain
\beq \lhela = 2c\sum_a h'_a q^a_A k'^A +
c\sum_a\tK_{aA}\[\tc^{-1}\tc' h'_a + h''_a\] = 
c k''^A + c' k'^A = (\pp_{\hel}\ell_A)_A, \label{solve}\eeq
where the last equality follows directly from \myref{lhl}.  When we
make the fermion field redefinitions analogous to
\myref{dadi}--\myref{diprime}, the Lagrangian quadratic in D-fermions
reads
\beq \L_{f_D} = \half\sum_A\bar\chi^{\bar i}\(i\notd -
m_{\tG}{1-4z\over z}\)\chi^i,\label{dferm}\eeq
where $z = b_c\ell$ and here $\chi^i = C(\bar\chi^{\bar i})^T =
 C(\overline{\chi^i})^T$ is a four-component Weyl fermion.  For the
 dilatino and T-modulini, we have, using \myref{modvac} in \myref{LY},
\bea \L(\chi^I,\chi_\ell) &=& -{u\over64\ell^2}\[2\tk' - \ell\tk'^2\(1 +
2b_c\ell\) - \ell^2\tk''\(2b_c\ell + 3\tk' + 2b_c\ell\tk'\)\right.
\mmm \qquad\qquad\l + {\ell\over3 - \tk'\ell}\lbr\ell\tk'\(2b_c\tk' +
4(\tk')^2 + \tk''(1 + b_c\ell\)\rbr\]\(\chi_\ell\chi_\ell\)\mmm +
{u\over2b_c}\sum_{I J}\(\chi^I\chi^J\)p_I p_J\zeta(t^I)\zeta(t^J)\mmm
+ {u\over4}\sum_I \(\chi^I\chi^I\) \lbr\(b_I + p_I - b_c\)\zeta'(t^I)
-2\zeta^2(t^I)\[b_I\(1 + 2b_c\ell\) + p_I + b_c\]\rbr\mmm +
{u\over16b_c\ell}\sum_{I}\zeta(t^I)(b_c + b_c^I -b_I)\(2\tk' +
\tk''\ell + {(\tk')^2\ell\over\tk'\ell - 3}\)\(\chi^I\chi_\ell\) .\eea
To determine the fermion masses we evaluate this expression at the
vacuum values $v'(\ell) = v(\ell) =0$.  Using \myref{KI},
\myref{modvac}, and the vacuum conditions \myref{vac0}, \myref{lvac},
we obtain
\bea \L(\chi^I,\chi_\ell) &=& {u\over4}\sum_I \(\chi^I\chi^I\)
\[\(b_I + p_I - b_c\)\zeta'(t^I) - {p_I + b_c + b_I(1 + 2b_c\ell)\over
8(\re t^I)^2}\] \mmm - {u\over32}\sum_{I J}\(\chi^I\chi^J\){p_I p_J
\over(\re t^I)^2}- {3z u(3 + 7z + z^2 -
z^3)\over64\ell(1+z)^3(1 + 2z)} \sum_I\(\chi^I\chi_\ell\){b_c + b_c^I
- b_I\over\re t^I}\mmm - {z^2(1 - 13z + 24z^2 + 7z^3 + 35z^4 + 54z^5
+ 6z^6)u\over64\ell^3(1
+ z)^5(1 + 2z)}\(\chi_\ell\chi_\ell\)
.\eea
The corresponding kinetic energy terms~\cite{joel} are given in 
Dirac notation in terms of 4-component Majorana spinors by
\bea \L_{KE}(\chi^I,\chi_\ell) &=& {i\tk'\over16\ell}\bar\chi_\ell\notd
\chi_\ell + {i\over2}\sum_I(1 + b_I\ell)\hK_I^2\bar\chi^I\notd\chi^I\eee
{3iz^2\over16(1 + z)^2\ell^2}\bar\chi_\ell\notd
\chi_\ell + \sum_I{i(1 + b_I\ell)\over8(\re t^I)^2}\bar\chi^I\notd\chi^I,
\eea
and the mass matrix in terms of the canonically normalized fields
$\chi^I_N$ is
\bea m_{\chi^I\chi^J} &=& {u\over4(1 + b_I\ell)}\lbr\del_{IJ}\[8(\re
t^I)^2(b_I + p_I - b_c)\zeta'(t^I) - p_I - b_c - b_I(1 + 2b_c\ell)\] +
p_I p_J\rbr,\nnn 
m_{\chi_\ell} &=& -b_c u{(1 - 13z + 24z^2 + 7z^3 + 35z^4 +54z^5 +
6z^6)\over12z(1 + z)^3(1 + 2z)},\nnn
m_{\chi_\ell\chi^I} &=& -{\sqrt{3\over2}}{u\over4}{(3 + 7z +
z^2 - z^3)\(b_c + b_c^I - b_I\)\over(1 + z)^2(1 + 2z)\sqrt{1 +
b_I\ell}}.\label{ferms}\eea
In the FIQS model used in the text as an illustrative example, $b_c^I
= 0$, and $b_I = b_c + p_I$. Since $b_I,p_I$ are nearly independent of
$I$, the mixing simplifies considerably in the case that all three
moduli are stabilized at the same self-dual point.  In this case the
eigenstates of $m_{\chi^I\chi^J}$ are
\beq \chi_0 = \sum_I\chi^I_N/\sqrt{3},\eeq
and two orthogonal combinations $\chi_b$ that have the same mass:
\beq m_{\chi_b} = u{8(\re t^I)^2p_I\zeta'(t^I) - (p_I + b_c)(1 +
b_c\ell)\over2(1 + b_I\ell)}.\label{fiqsferms}\eeq
Only $\chi_0$ mixes with the dilaton, with mass matrix
\beq m = \pmatrix{m_{\chi_\ell}&\sqrt{3}m_{\chi_\ell\chi^I}\cr
\sqrt{3}m_{\chi_\ell\chi^I}&m_{\chi_b} + m'\cr}, \quad
m' = {3u p_I^2\over4(1 + b_I\ell)}.\label{fiqsferms2}\eeq

\mysection{Massive $\G_c$-charged chiral multiplets}\label{rm}

If some $\G_c$-charged chiral multiplets acquire masses at the
\ua-breaking scale as in \myref{rmass}, they do not contribute to the
anomalies of the effective theory below the condensation scale, and
the anomaly matching conditions \myref{qconds}, \myref{pcond} and
\myref{picond} are modified to read
\bea \sum_\alpha b_c^\alpha q_I^\alpha &=&  b_I - b'_c -b^I_c +
 \sum_R {C_c^R\over4\pi^2}\(1 - q_I^R - q_I^{\R}\), \nonumber \\
 \sum_\alpha b_c^\alpha q^a_\alpha &=&  -{1\over2}\dx\delta_{aX} -
 \sum_R{C_c^R\over4\pi^2}\(q^a_R + q^a_{\R}\),\nnn
 2\sum_Aq_A^a p_A &=& \dx\del_{aX} + \sum_R{C_c^R\over2\pi^2}\(q^a_R +
 q^a_{\R}\),\label{pcond2}\eea
where $b_c,b'_c$ are defined as in \mysec{cond} in terms of the
Casimirs of the massless spectrum of the strongly coupled gauge group.
Then \myref{phiprod} becomes
\beq \prod_A|\phi^A|^{- 2\sum_\alpha b^\alpha_c q_\alpha^A} =
\exp\[-\sum_A p_A\ln x^A + \sum_I p_I g^I - \dx
h_X - \sum_{R,a}{C_c^R\over2\pi^2}\(q^a_R +
 q^a_{\R}\)h_a\],\label{phiprod2}\eeq
The contribution of \myref{rmass}
\bea \L_R &=& - \sum_R{C^R_c\over16\pi^2}\superint {E\over
R}U_c\ln\lbr c_R\prod_I\[\eta(T^I)\]^{2\(q_I^R + q_I^{\R} + p^R_I -
1\)}\prod_A\(\phi^A\)^{q^A_R}\rbr + {\rm h.c.},\label{rmass2}\eea
gives an additional contribution to the bosonic Lagrangian
\myref{bose}:
\bea e^{-1}\L_B^R &=& - u\sum_R{C^R_c\over8\pi^2}\[2\sum_I\(q_I^R +
q_I^{\R} + p^R_I -1\)\zeta(t^I)F^I + q^A_R F^A\] + {1\over8}L^R_c\(F_c
- u\M\) + {\rm h.c.},\nnn L^R_c &=& \sum_R{C^R_c\over2\pi^2}\[\ln
|c_R|^2 + 2\sum_I\(q_I^R + q_I^{\R} + p^R_I
-1\)\ln|\eta(t^I)|^2\right.\ddd \l + \sum_A q^A_R\(\ln x^A - G^A -
\sum_a q^a_A h_a\)\],\label{bose2}\eea
where we used \myref{physhift}.  Gauge invariance and modular
covariance of the superpotential \myref{masspot} imply
\beq \sum_A q^A_R q^a_A = - q^a_R - q^a_{\R},\qquad
p^R_I =  \sum_A q^A_R q^A_I.\eeq
Then defining 
\beq p^r_I \equiv \sum_R{C^R_c\over4\pi^2}p^R_I =
\sum_A p^r_A q^A_I, \qquad p^r_A \equiv
\sum_R{C^R_c\over8\pi^2}q^A_R q^a_A ,\eeq
the equations of motion \myref{sols} for
the auxiliary fields $F_c,F^I$ and $F^A$ are modified to read
\bea \u u &=& e^{-2b'_c/b_c}e^{\kappa -2(\hS-\del s)/b_c}
\prod_\alpha|b^\alpha/4c_\alpha|^{-2b^\alpha/b_c} \prod_I\[2\re
t^I|\eta(t^I)|^4\]^{(b_I - b_c + \hp_I)/b_c}e^{-\sum_A\hp_A\ln x^A /b_c},
\nonumber \\ F^I &=& - {2\re t^I\over1 + b_I\hel}{\u\over4}\(b_c - b_I
- \hp_I\)\[1 + 4\re t^I\bar{\zeta(t^I)}\], \nonumber \\ F^A &=& -
{\u\over4}\tK^{A\B}\[2\hS_{\B} - \hp_B - \hK_{\B}b_c\],
\label{sols2}\eea
where 
\beq \hp_I = p_I - p^r_I = \sum_A\hp_A q^A_I, \qquad
\hp_A = p_A - p^r_A, \qquad \sum_A\hp_A q^a_A = \dx\del_{aX}.
\eeq
Therefore the effective potential is determined by parameters defined
in terms of the modular weights and gauge charges of the full spectrum
of the effective theory at the string scale, except for the
renormalization group factor $\Lambda_c^2 \sim e^{-2/3b_c g^2_s}$ that
depends on the $\beta$-function factor for the massless spectrum of
the strongly coupled sector below the \ua-breaking scale.

\end{document}